\documentclass[lettersize,journal]{IEEEtran}

\usepackage{cite}

\usepackage{longtable}
\usepackage{enumitem}
\usepackage{amsfonts}
\usepackage{enumitem,kantlipsum}
\usepackage{textcomp}
\usepackage{stfloats}
\usepackage{url}
\usepackage{caption}
\usepackage{graphics}
\usepackage{verbatim}
\usepackage{pifont}
\usepackage{float}
\usepackage{amsmath}
\usepackage{cancel}
\usepackage{amsmath}
\usepackage{graphicx}
\usepackage{wrapfig}
\usepackage{float}
\usepackage{lscape}
\usepackage{ltxtable}
\usepackage{hhline}
\usepackage{caption}
\usepackage{wasysym}
\usepackage{xurl}
\usepackage{subcaption}
\usepackage{algorithm} 
\usepackage{color} 
\usepackage[export]{adjustbox}
\usepackage{colortbl}
\usepackage{algpseudocode}
\usepackage{verbatim}
\usepackage{lipsum}
\usepackage{tikz}
\usetikzlibrary{shapes,arrows}
\usetikzlibrary{intersections}
\usetikzlibrary{automata, arrows.meta, positioning}
\usetikzlibrary{decorations.pathreplacing}
\usepackage{amsthm}
\usepackage{booktabs}
\usepackage{amsmath}
\usepackage{algorithm}
\usepackage{inputenc}
\usepackage{adjustbox}
\usepackage{url} 
\usepackage{multirow}
\usepackage{color,soul}
\usepackage{lipsum}
\usepackage{mathtools}
\usepackage{subcaption}
\usepackage{cuted}
\usepackage{wasysym}
\usepackage{pgfplotstable}
\usepackage{mdframed}
\usepackage{glossaries}
\usepackage[normalem]{ulem}
\usetikzlibrary{arrows.meta}
\usepackage{smartdiagram}
\definecolor{lemon}{rgb}{1.0, 1.0, 0.13}
\usetikzlibrary{arrows.meta}
\usepackage{smartdiagram}
\usetikzlibrary{positioning, fit, calc, shapes, arrows,trees}

\definecolor{columbiablue}{rgb}{0.61, 0.87, 1.0}

\usepackage{algpseudocode}

\definecolor{BulletsColor}{rgb}{0, 0, 0.9}
\newlist{myBullets}{itemize}{1}

\setlist[myBullets]{
 label={\textbullet},
 leftmargin=*,
 topsep=0ex,
 partopsep=0ex,
 parsep=0ex,
 itemsep=0ex,
}
\newcounter{module}

\tikzset{%
 thick arrow/.style={
 -{Triangle[angle=120:1pt 1]},
 line width=0.8cm, 
 draw=blue!20
 },
 arrow label/.style={
 text=black,
 align=center
 },
 set mark/.style={
 insert path={
 node [midway, arrow label, node contents=#1]
 }
 }
}

\usepackage{amsmath}
\usepackage{acronym}
\usepackage{glossaries}
\usepackage[cmintegrals]{newtxmath}

\usepackage{hhline}
\definecolor{columbiablue}{rgb}{0.61, 0.87, 1.0}

\tikzset{%
 thick arrow/.style={
 -{Triangle[angle=120:1pt 1]},
 line width=0.8cm, 
 draw=blue!20
 },
 arrow label/.style={
 text=black,
 align=center
 },
 set mark/.style={
 insert path={
 node [midway, arrow label, node contents=#1]
 }
 }
}
\newcommand\deleted{\bgroup\markoverwith{\textcolor{red}{\rule[0.5ex]{2pt}{0.4pt}}}\ULon}

\usetikzlibrary{pgfplots.groupplots}

\newacronym{FL}{FL}{Federated Learning}%
\newacronym{BCFL}{BCFL}{Blockchain-based Federated Learning}%
\newacronym{HFL}{HFL}{Horizontal Federated Learning}%
\newacronym{VFL}{VFL}{Vertical Federated Learning}%
\newacronym{SMPC}{SMPC}{Secure Multi-Party Computation}%
\newacronym{FTL}{FTL}{Federated Transfer Learning}%
\newacronym{HIPAA}{HIPAA}{Health Insurance Portability and Accountability Act}%
\newacronym{GDPR}{GDPR}{General Data Protection Regulation}%
\newacronym{AI}{AI}{Artificial Intelligence}%
\newacronym{BC}{BC}{Blockchain}%
\newacronym{EHRs}{EHRs}{Electronic Health Records}%
\newacronym{CT}{CT}{Computed Tomography}%
\newacronym{MRI}{MRI}{Magnetic Resonance Imaging}%
\newacronym{FL-BETS}{FL-BETS}{FL-based BC-Enabled Task Scheduling}%
\newacronym{PET}{PET}{Positron Emission Tomography}%
\newacronym{DICOM}{DICOM}{Digital Imaging and Communications in Medicine}%
\newacronym{CAD}{CAD}{Computer-Aided Detection}%
\newacronym{SNPs}{SNPs}{Single Nucleotide Polymorphisms}%
\newacronym{CNVs}{CNVs}{Copy Number Variations}%
\newacronym{RNA-Seq}{RNA-Seq}{RNA Sequencing}%
\newacronym{ECG}{ECG}{Electrocardiogram}%
\newacronym{EEG}{EEG}{Electroencephalogram}%
\newacronym{EMG}{EMG}{Electromyography}%
\newacronym{CGM}{CGM}{Continuous Glucose Monitoring}%
\newacronym{PGD}{PGD}{Patient-Generated Data}%
\newacronym{NLP}{NLP}{Natural Language Processing}%
\newacronym{MAR}{MAR}{Medication Administration Records}%
\newacronym{CBC}{CBC}{Complete Blood Count}%
\newacronym{ESR}{ESR}{Erythrocyte Sedimentation Rate}%
\newacronym{EMRs}{EMRs}{Electronic Medical Records}%
\newacronym{PROs}{PROs}{Patient Reported Outcomes}%
\newacronym{CNNs}{CNNs}{Convolutional Neural Networks}%
\newacronym{PCA}{PCA}{Principal Component Analysis}%
\newacronym{t-SNE}{t-SNE}{t-distributed Stochastic Neighbor Embedding}%
\newacronym{RL}{RL}{Reinforcement Learning}%
\newacronym{IoT}{IoT}{Internet of Things}%
\newacronym{DApps}{DApps}{Decentralized Applications}%
\newacronym{P2P}{P2P}{Peer-to-Peer}%
\newacronym{PoW}{PoW}{Proof of Work}%
\newacronym{PoS}{PoS}{Proof of Stake}%
\newacronym{IBC}{IBC}{Inter-Blockchain Communication}%
\newacronym{DPoS}{DPoS}{Delegated Proof of Stake}%
\newacronym{BFT-based consensus}{BFT-based consensus}{Byzantine Fault Tolerant-based consensus}%
\newacronym{BFT}{BFT}{Byzantine Fault Tolerance}%

\newacronym{IBFT}{IBFT}{Improved Byzantine Fault Tolerance}%
\newacronym{pBFT}{pBFT}{Practical Byzantine Fault Tolerance}%
\newacronym{PoC}{PoC}{Proof of Capacity}%
\newacronym{PoST}{PoST}{Proof of Space and Time}%
\newacronym{DAG}{DAG}{Direct Acyclic Graph Tangle}%
\newacronym{PoB}{PoB}{Proof of Burn}%
\newacronym{PoA}{PoA}{Proof of Authority}
\newacronym{DeFi}{DeFi}{Decentralized Finance}%
\newacronym{IoB}{IoB}{Internet of Blockchains}%
\newacronym{MNIST}{MNIST}{Modified National Institute of Standards and Technology}%
\newacronym{IoMT}{IoMT}{Internet of Medical Things}%
\newacronym{GANs}{GANs}{Generative Adversarial Networks}%
\newacronym{CPS}{CPS}{Cyber-Physical Systems}%
\newacronym{HAM10000}{HAM10000}{Human Against Machine with 10000 training samples}%
\newacronym{ML}{ML}{Machine Learning}%

\begin{document}

\title{Integration of Federated Learning and Blockchain in Healthcare: A Tutorial }

\author{Yahya~Shahsavari,
 Oussama~A.~Dambri,
 Yaser~Baseri,
 Abdelhakim~Senhaji~Hafid, 
 and~Dimitrios~Makrakis

\IEEEcompsocitemizethanks{\IEEEcompsocthanksitem Yahya~Shahsavari, Yaser~Baseri and Abdelhakim~Senhaji~Hafid are with Montreal Blockchain Lab, University of Montreal, Montreal, Canada.
Emails: yahya.shahsavari@umontreal.ca; yaser.baseri@umontreal.ca; ahafid@umontreal.ca \protect
\IEEEcompsocthanksitem Oussama~A.~Dambri and Dimitrios~Makrakis are with the School of Electrical Engineering and Computer Science, University of Ottawa, Ottawa, Canada. E-mails: doussama@uottawa.ca; dmakraki@uottawa.ca \protect} 
\thanks{}}


\maketitle
\begin{abstract}

Wearable devices and medical sensors revolutionize health monitoring, raising concerns about data privacy in \gls{ML} for healthcare. This tutorial explores \gls{FL} and \gls{BC} integration, offering a secure and privacy-preserving approach to healthcare analytics. \gls{FL} enables decentralized model training on local devices at healthcare institutions, keeping patient data localized. This facilitates collaborative model development without compromising privacy. However, \gls{FL} introduces vulnerabilities. \gls{BC}, with its tamper-proof ledger and smart contracts, provides a robust framework for secure collaborative learning in \gls{FL}. After presenting a taxonomy for the various types of data used in \gls{ML} in medical applications, and a concise review of \gls{ML} techniques for healthcare use cases, this tutorial explores three integration architectures for balancing decentralization, scalability, and reliability in healthcare data. Furthermore, it investigates how \gls{BCFL} enhances data security and collaboration in disease prediction, medical image analysis, patient monitoring, and drug discovery.
By providing a tutorial on \gls{FL}, blockchain, and their integration, along with a review of \gls{BCFL} applications, this paper serves as a valuable resource for researchers and practitioners seeking to leverage these technologies for secure and privacy-preserving healthcare \gls{ML}. It aims to accelerate advancements in secure and collaborative healthcare analytics, ultimately improving patient outcomes.
\end{abstract}


\IEEEpeerreviewmaketitle

\section{Introduction}\label{Sec:intro}

In recent years, the rapid proliferation of wearable devices and medical sensors has unlocked remarkable possibilities for continuous health monitoring, facilitating real-time tracking of various health parameters~\cite{seneviratne2017survey,ometov2021survey}. This diverse array of devices, ranging from commonly used smartwatches to highly specialized medical sensors, has significantly expanded the scope of data available for healthcare applications~\cite{kang2022wearing}. The voluminous and continuous streams of data generated by these devices hold immense potential for advancing personalized healthcare and contributing to medical research~\cite{haghi2017wearable,vijayan2021review,wu2019wearable}. However, the centralized nature of medical data gathering, where data is often collected, stored, and analyzed in centralized databases or cloud servers, presents a significant challenge to privacy~\cite{datta2018survey,cilliers2020wearable,arias2015privacy}. This centralization poses a risk as it consolidates sensitive health information in a single location, making it an attractive target for potential breaches or unauthorized access. Concentrating such valuable and private data in one repository increases the potential impact of a security breach, jeopardizing the confidentiality and privacy of individuals' health information~\cite{cilliers2020wearable}.

Medical data privacy can be compromised when used for \gls{ML}, due to the centralized nature of data gathering. Concentrating sensitive health information in centralized databases or cloud servers poses a significant risk~\cite{qayyum2020secure}. If these centralized repositories become targets for security breaches or unauthorized access, the potential impact is heightened. A security breach on such a consolidated repository becomes more attractive to malicious actors as it grants access to a wealth of valuable and private data from a diverse array of wearable devices and medical sensors~\cite{cilliers2020wearable,kim2015reliability}. The centralized storage and analysis of voluminous and continuous streams of health data increase the surface area for potential privacy threats, jeopardizing the confidentiality and privacy of individuals' health information. This centralization introduces a single point of failure, making it more challenging to ensure the security and integrity of the data, thereby creating a potential avenue for compromising privacy when medical data is utilized for \gls{ML} purposes. Consequently, addressing the inherent privacy challenges associated with centralized medical data gathering becomes crucial to harnessing the benefits of wearable devices while safeguarding the sensitive nature of health-related information.

\gls{FL}, as an innovative decentralized learning framework, enables model training across multiple local devices while keeping raw data localized~\cite{rieke2020future,zhang2021survey,li2020review}. In the context of medical data, where patient confidentiality is paramount, \gls{FL} emerges as a promising solution to address inherent privacy challenges~\cite{rieke2020future,pfitzner2021federated,antunes2022federated,xu2021federated}.

One of the key advantages of \gls{FL} lies in its ability to facilitate collaborative model training without the need for centralized data aggregation. In conventional \gls{ML} models, centralizing medical datasets raises concerns regarding unauthorized access, data breaches, and the potential for patient re-identification. \gls{FL} distributes the model training process across individual devices, ensuring that raw data remains on the local device, and only model updates are shared. This decentralized approach substantially reduces the risk of privacy breaches, as sensitive patient information is not exposed beyond the confines of the local device.

\begin{table}[!hbpt]
\centering \footnotesize
\caption{Table of Acronyms}
\label{tab:abbreviation}
\resizebox{\columnwidth}{!}
{%
\begin{tabular}{|l|l|}
 \hline
 \textbf{Acronym} & \textbf{Explanation} \\
 \hline
 \glsentryname{AI} & \glsentrylong{AI}\\ \hline
 \glsentryname{BFT} & \glsentrylong{BFT}\\ \hline
 \glsentryname{BFT-based consensus} & \glsentrylong{BFT-based consensus}\\ \hline
 \glsentryname{BC} & \glsentrylong{BC}\\ \hline
 \glsentryname{BCFL} & \glsentrylong{BCFL}\\ \hline
 \glsentryname{CAD} & \glsentrylong{CAD}\\ \hline
 \glsentryname{CBC} & \glsentrylong{CBC}\\ \hline
 \glsentryname{CNVs} & \glsentrylong{CNVs}\\ \hline
  \glsentryname{CNNs} & \glsentrylong{CNNs}\\ \hline
 \glsentryname{CGM} & \glsentrylong{CGM}\\ \hline
 \glsentryname{CT} & \glsentrylong{CT}\\ \hline
 \glsentryname{CPS} & \glsentrylong{CPS}\\ \hline
 \glsentryname{DApps} & \glsentrylong{DApps}\\ \hline
 \glsentryname{DAG} & \glsentrylong{DAG}\\ \hline
 \glsentryname{DeFi} & \glsentrylong{DeFi}\\ \hline
 \glsentryname{DICOM} & \glsentrylong{DICOM}\\ \hline
 \glsentryname{DPoS} & \glsentrylong{DPoS}\\ \hline
 \glsentryname{ECG} & \glsentrylong{ECG}\\ \hline
  \glsentryname{EEG} & \glsentrylong{EEG}\\ \hline
 \glsentryname{EMG} & \glsentrylong{EMG}\\ \hline
 \glsentryname{EMRs} & \glsentrylong{EMRs}\\ \hline
 \glsentryname{EHRs} & \glsentrylong{EHRs}\\ \hline
 \glsentryname{ESR} & \glsentrylong{ESR}\\ \hline
 \glsentryname{FTL} & \glsentrylong{FTL}\\ \hline
 \glsentryname{FL} & \glsentrylong{FL}\\ \hline
 \glsentryname{FL-BETS} & \glsentrylong{FL-BETS}\\ \hline
 \glsentryname{GANs} & \glsentrylong{GANs}\\ \hline
 \glsentryname{GDPR} & \glsentrylong{GDPR}\\ \hline
 \glsentryname{HIPAA} & \glsentrylong{HIPAA}\\ \hline
 \glsentryname{HAM10000} & \glsentrylong{HAM10000}\\ \hline
  \glsentryname{IBFT} & \glsentrylong{IBFT}\\ \hline
 \glsentryname{IoB} & \glsentrylong{IoB}\\ \hline
 \glsentryname{IoMT} & \glsentrylong{IoMT}\\ \hline
 \glsentryname{IoT} & \glsentrylong{IoT}\\ \hline
 \glsentryname{IBC} & \glsentrylong{IBC}\\ \hline
 \glsentryname{ML} & \glsentrylong{ML}\\ \hline
 \glsentryname{MAR} & \glsentrylong{MAR}\\ \hline
  \glsentryname{MNIST} & \glsentrylong{MNIST}\\ \hline
 \glsentryname{MRI} & \glsentrylong{MRI}\\ \hline
 \glsentryname{NLP} & \glsentrylong{NLP}\\ \hline
 \glsentryname{PET} & \glsentrylong{PET}\\ \hline
 \glsentryname{PCA} & \glsentrylong{PCA}\\ \hline
 \glsentryname{PGD} & \glsentrylong{PGD}\\ \hline
 \glsentryname{PoA} & \glsentrylong{PoA}\\ \hline
 \glsentryname{PoB} & \glsentrylong{PoB}\\ \hline
 \glsentryname{PoC} & \glsentrylong{PoC}\\ \hline
 \glsentryname{PoS} & \glsentrylong{PoS}\\ \hline
 \glsentryname{PoST} & \glsentrylong{PoST}\\ \hline
 \glsentryname{PoW} & \glsentrylong{PoW}\\ \hline
 \glsentryname{PROs} & \glsentrylong{PROs}\\ \hline
 \glsentryname{RL} & \glsentrylong{RL}\\ \hline
 \glsentryname{RNA-Seq} & \glsentrylong{RNA-Seq}\\ \hline
 \glsentryname{SNPs} & \glsentrylong{SNPs}\\ \hline
 \glsentryname{SMPC} & \glsentrylong{SMPC}\\ \hline
 \glsentryname{t-SNE} & \glsentrylong{t-SNE}\\ \hline
 \glsentryname{VFL} & \glsentrylong{VFL}\\
 \hline
 \end{tabular}}
\end{table}

Furthermore, \gls{FL} empowers healthcare institutions to collaborate on model development without the necessity of sharing raw patient data. This collaborative learning paradigm allows models to be trained on diverse datasets from different sources, fostering the creation of robust and generalizable models. Importantly, by enabling the aggregation of insights from various healthcare providers, \gls{FL} promotes the development of models that are representative of broader populations, thereby enhancing the efficacy and applicability of medical \gls{AI} solutions.

Although \gls{FL} preserves the confidentiality of sensitive data by training models on decentralized devices, it introduces new vulnerabilities, such as the risk of model poisoning~\cite{fang2020local,cao2019understanding}, compromises to data integrity~\cite{lyu2020threats,bouacida2021vulnerabilities,lyu2022privacy}, and decentralized communication errors~\cite{ye2022decentralized}. \gls{BC}, with its decentralized and tamper-resistant nature, coupled with smart contracts for protocol execution, offers a robust framework to combat malicious activities, ensure data integrity, and enhance the overall security of the collaborative learning process~\cite{nguyen2021federated,nguyen2021federated}. By synergizing the strengths of \gls{FL} and \gls{BC} technology, we can strike a balance between privacy and security, fostering a trustworthy and resilient environment for the advancement of \gls{ML} in sensitive domains such as healthcare.



This paper significantly contributes to the field of decentralized \gls{ML} by specifically focusing on the integration of blockchain technology with \gls{FL} in the healthcare domain. Our contributions are as follows:

\begin{enumerate}[label=\textbf{\arabic*.}]
\item \textbf{Motivations for Blockchain-\gls{FL} Integration in Healthcare:} We analyze the need for blockchain to address data privacy, security, and trust concerns in healthcare-oriented \gls{FL} systems. This analysis emphasizes the unique challenges and opportunities presented by healthcare data, setting the stage for understanding the value proposition of blockchain integration.

\item \textbf{Exploring Integration Architectures for Healthcare \gls{FL}:} We discuss three distinct integration architectures – fully coupled, semi-coupled, and loosely coupled – specifically within the context of healthcare applications. This analysis provides valuable insights for researchers and practitioners designing and implementing \gls{BCFL} systems for healthcare data, considering factors like patient privacy, regulatory compliance, and data security.

\item \textbf{Impact of Blockchain-based Federated Learning (BCFL) on Healthcare:} We delve into the emerging paradigm of \gls{FL} and showcase its potential to revolutionize healthcare analytics. By demonstrating how \gls{FL} frameworks can preserve patient privacy while enabling collaborative model training across healthcare institutions, this section highlights the transformative impact of \gls{FL} on various healthcare domains, including disease prediction, medical image analysis, patient monitoring, and drug discovery.

\item \textbf{Research Gaps and Future Directions in Healthcare \gls{FL}:} Recognizing limitations in existing surveys, we identify key research gaps specifically at the intersection of blockchain, \gls{FL}, and healthcare. We advocate for comprehensive future research endeavors encompassing regulatory, technical, and adoption aspects to guide further advancements in \gls{FL} for effective healthcare solutions.

\item \textbf{Synthesis for Advancing Decentralized Healthcare Machine Learning:} By synthesizing insights from various sections of the paper, including motivations, integration architectures, \gls{FL} applications, and research gaps, we offer a comprehensive perspective on the contribution of blockchain to \gls{FL} in the context of healthcare. This synthesis aims to foster collaboration, innovation, and adoption of \gls{FL} specifically for healthcare applications, ultimately accelerating progress in decentralized \gls{ML} for improved healthcare outcomes.

\end{enumerate}


The remainder of the paper is structured as follows. Section~\ref{sec:MedData} provides a review of medical data and their applications in \gls{ML}. In Section~\ref{sec:MedAI}, we offer a brief tutorial on the application of \gls{ML} and \gls{FL} in medical use cases. Section~\ref{sec:MedBC} offers a brief background on blockchain technology necessary for understanding this paper. The integration of blockchain and \gls{FL} in healthcare is discussed in Section~\ref{sec:MedIngr}. Section~\ref{sec:relatedWork} presents related works on the integration of blockchain and \gls{FL} in healthcare as well as currently existing surveys. Finally, Section~\ref{sec:conclusion} concludes the paper.

\section{Medical Data and Applications in Machine Learning}\label{sec:MedData}
Medical data encompasses a diverse array of information crucial for understanding, diagnosing, and treating various health conditions. This data, ranging from patient demographics and medical history to diagnostic images and genomic sequences, holds immense potential for advancing healthcare through \gls{ML} applications. By harnessing the power of \gls{ML} algorithms, medical data can be analyzed to extract valuable insights, predict patient outcomes, personalize treatments, and optimize healthcare delivery. However, the utilization of medical data for \gls{ML} requires careful consideration of data storage and management practices to ensure compliance with privacy regulations, maintain data integrity, and facilitate seamless access for research and clinical purposes. This section explores the different types of medical data and their applications in \gls{ML}.
\subsection{Electronic Health Records}
\gls{EHRs}~\cite{rahman2015electronic,yadav2018mining,hoerbst2010electronic,simon2007physicians} are digital versions of patients' paper charts in a healthcare setting. \gls{EHRs} contain a patient's medical history, diagnoses, medications, treatment plans, immunization dates, allergies, medical images (e.g., radiology images), and laboratory test results. They are designed to provide a comprehensive and up-to-date view of a patient's health information, allowing healthcare providers to make informed decisions about a patient's care. The key features of \gls{EHRs} include:

\begin{itemize}
\item \textbf{Digital Format:} \gls{EHRs} replace traditional paper records with electronic versions, making it easier to store, access, and share health information.

\item \textbf{Interoperability:} \gls{EHRs} are designed to be interoperable, meaning that they can be shared among different healthcare providers and organizations. This facilitates better coordination of care, especially when patients see multiple healthcare providers.

\item \textbf{Real-Time Access:} Authorized healthcare professionals can access a patient's EHR in real-time, providing quick access to critical health information, which is crucial, especially during emergency situations.

\item \textbf{Patient Engagement:} \gls{EHRs} often include features that allow patients to access their health information, schedule appointments, communicate with healthcare providers, and participate more actively in their own care.

\item \textbf{Decision Support:} \gls{EHRs} may include tools that provide decision support for healthcare providers, such as alerts for potential drug interactions, reminders for preventive screenings, and clinical guidelines.

\item \textbf{Data Security and Privacy:} \gls{EHRs} are designed with security measures to protect the confidentiality and privacy of patient information. Access to \gls{EHRs} is typically restricted to authorized personnel~\cite{jin2009patient,blobel2004authorisation,fernandez2013security,tang2019efficient}.
\end{itemize}

\gls{EHRs} serve as a rich source of structured and unstructured data that can be leveraged for \gls{ML} applications in healthcare. Since \gls{EHRs} contain comprehensive information about patients, \gls{ML} algorithms can analyze this wealth of information to extract meaningful insights, predict patient outcomes, identify patterns, and improve clinical decision-making~\cite{arbet2021lessons, wu2020statistics, deo2016learning}. In the context of ML, EHR data can be used for a variety of tasks, including:

\begin{itemize}
\item \textbf{Predictive Analytics:} \gls{ML} models trained on EHR data can predict the likelihood of various medical events, such as hospital readmission, disease onset or progression, adverse drug reactions, and mortality rates. These predictions enable proactive interventions and personalized care strategies~\cite{wu2010prediction, johnston2019using, dev2022predictive, muniasamy2020deep}.

\item \textbf{Disease Identification and Diagnosis:} \gls{ML} algorithms can analyze EHR data to assist in the early detection and diagnosis of diseases. By identifying subtle patterns and anomalies in patient data, these models can aid healthcare providers in making accurate and timely diagnoses~\cite{sajda2006machine, singh2021diagnosing, li2020heart, mall2022implementation, ahsan2022machine, alanazi2022identification, battineni2020applications}.

\item \textbf{Treatment Recommendations:} \gls{ML} models trained on EHR data can suggest personalized treatment plans based on patients' medical history, demographic information, genetic profiles, and responses to previous interventions. These recommendations help optimize treatment outcomes and minimize adverse effects~\cite{chekroud2021promise, bica2021real, wong2018using, moehring2021development, chu2020treatment, komal2019drug}.
\end{itemize}

In summary, \gls{EHRs} serve as a valuable resource for \gls{ML} and \gls{FL} applications in healthcare, enabling predictive analytics, disease identification, personalized treatment recommendations, and collaborative model training, while ensuring patient privacy and data sovereignty.

\subsection{Medical Imaging Data}
Medical imaging data encompasses diverse visual representations of the internal structures of the human body, intended for clinical scrutiny and medical intervention. These visual depictions are acquired utilizing distinct imaging modalities~\cite{lalitha2022medical, abhisheka2023recent, elangovan2016medical} and play a pivotal role in the diagnosis and ongoing assessment of diverse medical conditions. Key components of medical imaging data are as follows:

\begin{itemize}
\item \textbf{Patient Information:} consists of basic information about the patient, including their name, identification number, age, gender, and other relevant demographic details as well as information about the patient's medical history, current symptoms, and any relevant details that might impact the interpretation of the images.
\item \textbf{Imaging Modalities:} Each imaging modality utilizes distinct principles and technologies to generate specific types of images~\cite{doi2006diagnostic}. Different imaging modalities offer complementary information, and the choice of modality depends on the specific clinical question, the type of tissue or organ under investigation, and the desired level of detail~\cite{khoon2016survey,santhi2022survey}. Common imaging modalities include:
 \begin{enumerate}
\item \textit{X-ray:} uses ionizing radiation to produce images of bones and some soft tissues~\cite{huda2015x,mustapha2021comparative,aaslund2010detectors}.
\item \textit{Computed Tomography(CT):} combines X-rays with computer technology to create detailed cross-sectional images of the body~\cite{brooks1993computed,sluimer2006computer,withers2021x,kalender2006x,shrimpton2005doses}.
\item \textit{Magnetic Resonance Imaging (MRI):} Uses strong magnetic fields and radio waves to generate detailed images of soft tissues, organs, and joints~\cite{katti2011magnetic,khoo1997magnetic,fatahi2015magnetic,tirotta201519f,prasad2005making,mori1999diffusion}.
\item \textit{Ultrasound:} Utilizes sound waves to create images of internal structures, commonly used for examining the abdomen, pelvis, and fetus during pregnancy~\cite{avola2021ultrasound,huang2017review,whittaker2011ultrasound,ellis2020exploring,ortiz2012ultrasound,fenster2001three,lees2001ultrasound,schellpfeffer2013ultrasound}.
\item \textit{Nuclear Medicine:} Involves the injection of radioactive substances to visualize the functioning of organs and tissues~\cite{hricak2021medical,schoder2003pet,blankenberg2002nuclear,vaz2020nuclear,van2003nuclear,prandini2006nuclear}.
\item \textit{\gls{PET}:} Combines nuclear medicine with computed tomography to provide functional information about tissues and organs, particularly in cancer diagnosis and staging\cite{bailey2005positron,tai2004applications,gallamini2014positron,wagner1998brief,townsend2008positron}.
\item \textit{Fluoroscopy:} Involves continuous X-ray imaging to capture real-time moving images, commonly used for procedures such as angiography and gastrointestinal studies~\cite{jones2014medical,maher1986digital,krohmer1989radiography,rudin1992region,mahesh2001fluoroscopy,hoheisel2006review}.
\end{enumerate}
\item \textbf{Image Files:} Medical data~\cite{ulrich2022understanding} including metadata~\cite{sweet2013electronic, mclean2008electronic,vardaki2009statistical} such as patient information, acquisition parameters, and image details are being stored, shared, and transmitted regarding a set of technical standards~\cite{de2015data} referred to as \gls{DICOM}~\cite{Lim2006}. 

\item \textbf{Annotations and Measurements:} Annotations in medical data involve adding context, insights, or labels to enhance the interpretation of information, while measurements quantify various aspects of physiological parameters, imaging features, and laboratory results, providing objective data for diagnosis and treatment decisions. Both annotations and measurements are essential for comprehensive and accurate medical analysis and decision-making~\cite{gur2017towards,patel2018annotation,seifert2010semantic,aljabri2022towards}.
\item \textbf{Image Series:} In medical imaging, an image series refers to a sequence of images acquired during a single imaging study or examination. This sequence of images is typically obtained to capture different views, slices, or time points of a particular anatomical region or physiological process~\cite{hajnal2001medical}.
\item \textbf{Metadata:} Metadata in the context of medical imaging refers to additional information that accompanies the actual image data. This information is crucial for understanding, interpreting, and managing medical images~\cite{ulrich2022understanding}. Key components of metadata in the medical context are acquisition parameters (i.e. exposure settings, magnetic field strength, etc.), image acquisition time, patient information (i.e. patient-related data, including name, ID, age, gender, clinical history, etc.), study information (e.g. study ID, modality, etc.), image details (e.g. orientation and position, slice thickness, etc.), device information (e.g. make and model of the scanner or machine, software version), dose parameters ( e.g. amount of radiation exposure), and image processing history (i.e. reconstruction algorithm, and image reconstruction method)~\cite{chileshe2023large}.
\item \textbf{Reports:} In the context of medical imaging, reports refer to narrative documents generated by radiologists summarizing their findings based on the analysis of the medical images. These reports are often used by referring physicians for treatment planning.
\end{itemize}

Medical imaging data has a wide range of applications in \gls{ML} within the healthcare industry~\cite{razzak2018deep}. Some key applications include:

\begin{itemize}
 \item \textbf{Disease Diagnosis and Classification:} \gls{ML} algorithms can analyze medical imaging data to assist in the diagnosis and classification of various diseases and conditions, such as cancer, neurological disorders, cardiovascular diseases, and musculoskeletal 
 disorders~\cite{ker2017deep,yadav2019deep,rana2023machine,latif2019medical,chan2020deep,castiglioni2021ai}.
 \item \textbf{\gls{CAD}:} \gls{CAD} systems use \gls{ML} algorithms to assist radiologists and clinicians in detecting abnormalities in medical images, such as tumors, lesions, fractures, and other anomalies. These systems can help improve diagnostic accuracy and efficiency~\cite{jeyaraj2019computer,chan2020computer}.
 \item \textbf{Medical Image Segmentation:} Segmentation involves partitioning an image into multiple segments to identify and analyze specific regions of interest, such as organs, tissues, or abnormalities. \gls{ML} algorithms can be trained to perform image segmentation tasks, which are valuable for tasks such as tumor delineation, organ segmentation, and identifying abnormalities within tissues or organs~\cite{wang2022medical,lai2015deep,wang2018interactive}.
 \item \textbf{Image-Based Biomarker Discovery:} \gls{ML} algorithms can analyze medical imaging data to identify imaging biomarkers associated with specific diseases or treatment responses. These biomarkers can be valuable for predicting disease prognosis, assessing treatment efficacy, and guiding personalized medicine approaches~\cite{echle2021deep,savadjiev2019image}.
 \item \textbf{Treatment Planning and Monitoring:} Medical imaging data can be used to develop personalized treatment plans for patients based on their unique anatomical characteristics and disease progression. \gls{ML} models can analyze imaging data to predict treatment outcomes, monitor disease progression over time, and optimize treatment strategies for individual patients~\cite{xu2019deep}.
 \item \textbf{Image Reconstruction and Enhancement:} \gls{ML} techniques, such as deep learning-based image reconstruction and enhancement algorithms, can improve the quality of medical images acquired from various imaging modalities (e.g., \gls{MRI}, \gls{CT}, ultrasound). These techniques can help reduce imaging artifacts, enhance image resolution, and improve overall image quality for better interpretation by healthcare professionals~\cite{wang2018image,zhang2020review,ahishakiye2021survey}.
 \item \textbf{Image Registration and Fusion:} Image registration and fusion techniques align and combine multiple medical images from different modalities or time points to provide comprehensive information for diagnosis and treatment planning. \gls{ML} algorithms can be used to perform automatic image registration and fusion tasks, enabling more accurate visualization and analysis of complex medical data.
 \item \textbf{Drug Discovery and Development:} Medical imaging data can also be used in drug discovery and development processes. \gls{ML} models can analyze imaging data to evaluate the effects of experimental drugs or therapies on disease progression, identify potential drug targets, and optimize drug delivery methods~\cite{vamathevan2019applications}.

\end{itemize}

\subsection{Genomic Data}
Genomic data pertains to information concerning the organization and operation of the genome within an organism~\cite{louie2007data,he2017big}. Genomic data harnessed in healthcare \gls{ML} pertains to the comprehensive information extracted from an individual's genome, encompassing the entirety of their DNA and genetic composition~\cite{libbrecht2015machine,chafai2023emerging,diao2018biomedical,wu2018deep,lin2017machine,khan2023bioinformatics,whalen2022navigating}. This invaluable dataset plays a pivotal role in healthcare applications, offering profound insights into genetic predispositions, treatment responses, and overall health~\cite{akgun2015privacy}. 
Key components of Genomic data in medical use cases can be summarized as follows:

\begin{itemize}
 \item \textbf{Genome Sequences:} The genome sequence represents the complete set of genetic material (DNA) present in an organism, including all of its genes and non-coding regions. It consists of a sequence of nucleotides (A, T, C, and G) that encode genetic information.
 \item \textbf{Genes:} Genes are specific sequences of DNA that encode instructions for the synthesis of proteins, which are essential for various biological processes in the body. Genomic data includes information about the location, structure, and function of genes within the genome.
 \item \textbf{\gls{SNPs}:} \gls{SNPs} are variations in a single nucleotide base at a specific position in the genome that occur with relatively high frequency within a population. These variations can influence traits, susceptibility to diseases, and drug responses. Genomic data includes information about \gls{SNPs} and their association with phenotypic traits and diseases~\cite{leache2017utility}.
 \item \textbf{\gls{CNVs}:} \gls{CNVs} are genomic alterations that involve changes in the number of copies of specific DNA segments, ranging in size from kilobases to megabases. \gls{CNVs} can affect gene dosage and expression levels, leading to phenotypic variations and disease susceptibility. Genomic data includes information about \gls{CNVs} and their association with various diseases and traits~\cite{zarrei2015copy}.
 \item \textbf{Gene Expression Profiles:} Gene expression refers to the process by which genetic information is used to synthesize proteins. Genomic data includes information about gene expression profiles, which can provide insights into the activity levels of different genes in specific tissues or under different conditions. Gene expression data is often generated using technologies such as microarrays or \gls{RNA-Seq}~\cite{audic1997significance}.
 \item \textbf{Epigenetic Modifications:} Epigenetic modifications are heritable changes in gene expression that do not involve alterations in the DNA sequence itself. These modifications, such as DNA methylation and histone modifications, can influence gene activity and phenotype without changing the underlying DNA sequence. Genomic data includes information about epigenetic modifications and their role in regulating gene expression and disease susceptibility~\cite{portela2010epigenetic}.
 \item \textbf{Genetic Variation Databases:} Genetic variation databases compile and organize genomic data from various sources, including population studies, disease databases, and clinical datasets. These databases provide comprehensive repositories of genetic variants, annotations, and associated phenotypic information, facilitating research on genetic variation and its implications for human health~\cite{sherry2001dbsnp}.
\end{itemize}

These components of genomic data are essential for understanding the genetic basis of diseases, identifying genetic risk factors, and developing personalized approaches to healthcare and medicine. Genomic data analysis plays a crucial role in advancing our understanding of genetics, genomics, and their implications for human health and disease.
Utilization of genomic data in healthcare \gls{ML} can be described as follows:

\begin{itemize}
 \item \textbf{Disease prediction and Risk Stratification:} \gls{ML} algorithms scrutinize genomic data to discern patterns and variations linked to specific diseases. This analytical approach allows for the assessment of the risk associated with certain health conditions~\cite{gulamali2022machine,okser2013genetic,tseng2020development,kruppa2012risk,quazi2022artificial}.

 \item \textbf{Personalized Medicine:} genomic data forms the basis for the development of personalized treatment plans~\cite{west2006embracing,ginsburg2009genomic}. \gls{ML} models, through predictive analytics, anticipate individual responses to different medications, facilitating tailored and optimized treatment strategies~\cite{vadapalli2022artificial,quazi2022artificial,khan2023bioinformatics,maceachern2021machine,wu2018deep,chafai2023emerging}.

 \item \textbf{Drug Discovery and Computational genomics:} By leveraging \gls{ML}, researchers can analyze genomic data to expedite drug discovery and development processes. This entails identifying potential drug targets and comprehending the genetic underpinnings of diseases, ultimately leading to more targeted and efficacious therapeutic solutions\cite{gupta2021artificial,carracedo2021review,ramsundar2019deep,dana2018deep}.

 \item \textbf{Genomic Counseling:} \gls{ML} algorithms play a crucial role in delivering personalized genetic counseling by deciphering complex genomic data. This assists healthcare professionals in effectively communicating intricate genetic details, including health risks and familial implications~\cite{gordon2018future, kearney2020artificial}.

 \item \textbf{Early Detection and Diagnostics:} The integration of genomic data and \gls{ML} algorithms facilitates the early detection and diagnosis of diseases. These algorithms can discern subtle genetic variations indicative of specific conditions, enabling timely intervention and enhanced prognostic outcomes~\cite{dias2019artificial,benning2022advances,venugopalan2021multimodal,tao2020machine}.

 \item \textbf{Research and Population Health Informatics:} Aggregated and anonymized genomic data, subjected to \gls{ML} methodologies, advances our understanding of the genetic foundations of diseases at a population level. This knowledge informs public health initiatives, epidemiological studies, and the identification of genetic factors influencing disease prevalence~\cite{mullen2021race,shameer2018machine}.

 \item \textbf{Genomic Sequencing and Computational Analysis:} \gls{ML} serves a crucial role in the interpretation of extensive genomic datasets generated through advanced sequencing technologies. These algorithms play a pivotal role in identifying genetic mutations, variations, and other pertinent information impacting individual health~\cite{leung2015machine,schmidt2021deep,cortes2022computational}.
\end{itemize}

\subsection{Biometric Data}
Biometric data in the medical context refers to unique physical or behavioral characteristics that can be used for identifying and monitoring individuals. This data is increasingly important in healthcare for patient identification, access control to medical records, personalized treatments, and monitoring of health conditions. The components of biometric data in medical data can include, but are not limited to:

\begin{itemize}
 \item \textbf{Physiological Biometrics:} Consists of data types such as fingerprints data, facial recognition data, iris recognition data, Ear recognition data, retina scans, DNA, Smell recognition, and hand geometry data~\cite{dargan2020comprehensive}.
 \item \textbf{Behavioral Biometrics:} Consists of data types as follows: voice recognition data, gait analysis data, and typing dynamics data~\cite{dargan2020comprehensive}. 
 \item \textbf{Health-Related Biometrics:} Consists of data types such as heart rate, blood pressure, blood Oxygen levels, \gls{ECG} patterns~\cite{pinto2018evolution}, and brain wave patterns (i.e. \gls{EEG} patterns)~\cite{paranjape2001electroencephalogram}.
 \item \textbf{Emerging Technologies:} There are some recently emerging technologies for extracting biometric data such as:
 \begin{enumerate}
 \item \textit{Microbial Biometrics:} Analyzing the unique microbiome of an individual.
 \item \textit{Olfactory Biometrics:} Unique body odors can potentially be used for the identification and detection of diseases.
 \end{enumerate}

\end{itemize}
Biometric data integrated into healthcare \gls{ML} encompasses quantifiable physiological or behavioral attributes~\cite{dargan2020comprehensive,abdulrahman2023comprehensive,boulgouris2009biometrics} unique to individuals. This data serves the purpose of accurate identification~\cite{kisku2019design,fatima2019biometric,mason2020investigation,okoh2015biometrics}, secure access~\cite{kaul2020secure,hei2011biometric,barka2022implementation}, and health assessment~\cite{mohsin2018real,choi2022hybrid} within healthcare settings. The primary application of biometric data in healthcare \gls{ML} can be succinctly outlined as follows:

\begin{itemize}
 \item \textbf{Biometric Identification and Access Control:} Biometric data, encompassing features like fingerprints or facial characteristics, is pivotal for precise identification and access control in healthcare environments. Implementation of biometric authentication ensures heightened security measures, governing access to sensitive areas, \gls{EHRs}, and medical devices. \gls{ML} algorithms play a crucial role in processing and analyzing these biometric patterns to enable swift and accurate verification~\cite{kumar2019deep,sengar2020multimodal}.

 \item \textbf{Patient Identification and Record Matching:} employing biometric identifiers such as fingerprints or iris scans ensures precise linkage of patients to their health records, minimizing the risk of identification errors~\cite{piera2020patient,leonard2008realization,jonas2014patient,sohn2020clinical}. \gls{ML} algorithms enhance the accuracy of patient matching, mitigating potential medical errors and elevating the overall quality of patient care~\cite{fatimah2022biometric,labati2019deep,prakash2023deep}.

 \item \textbf{Biometric Monitoring for Health Assessment:} Continuous monitoring of biometric data—like heart rate, blood pressure, or \gls{ECG} signals—via wearable devices or sensors facilitates real-time health assessment~\cite{mohsin2018real}. \gls{ML} algorithms analyze dynamic biometric data, enabling early detection of anomalies or health issues. This capability supports timely intervention and personalized healthcare management~\cite{ismail2022recent,fei2021machine}.

 \item \textbf{Behavioral Biometrics for Mental Health Monitoring:} Behavioral biometrics, encompassing patterns like typing or voice modulation, contribute to mental health assessment and behavioral changes detection~\cite{gomes2023survey,xefteris2016behavioral,patel2018mental,killoran2023can}. \gls{ML} models discern patterns indicative of mental health conditions, aiding healthcare providers in delivering targeted interventions and support~\cite{garcia2018mental}.

 \item \textbf{Biometric Data in Clinical Trials:} Biometric data usage in clinical trials spans participant identification, monitoring, and data integrity assurance~\cite{arneric2017biometric}. \gls{ML} assists in the efficient management and analysis of biometric data during clinical trials, informing researchers and ensuring study outcomes' validity~\cite{harrer2019artificial,weissler2021role}.

 \item \textbf{Voice and Speech Analysis for Diagnostics:} Voice pattern analysis serves diagnostic purposes, detecting potential markers for various medical conditions~\cite{lyakso2021voice,niebudek2006diagnostic}. \gls{ML} algorithms process voice and speech data to identify indications of conditions such as Parkinson's disease~\cite{moro2021advances,ngo2022computerized,shahbakhi2014speech} or respiratory disorders~\cite{lella2021automatic,idrisoglu2023applied}, contributing to diagnostic capabilities.

 \item \textbf{Facial Recognition for Patient Monitoring:} Facial recognition technology is employed for patient well-being monitoring and distress detection~\cite{jeon2019facial}. \gls{ML} algorithms analyze facial expressions, offering insights into patient comfort levels and potentially enhancing the quality of care in healthcare settings~\cite{onyema2021enhancement}. 
\end{itemize}

\subsection{Sensor Data}\label{SensorData}
Sensor data in medical contexts refers to the information collected from various devices and sensors used to monitor health, detect changes in patient conditions, and assist in diagnoses. These sensors can be wearable, implantable, or environmental, and they continuously collect data related to physiological parameters, activity levels, and environmental conditions. The components of sensor data in medical data can include:
\begin{itemize}
 \item \textbf{Wearable Sensors:} These data are extracted from wearable devices such as heart rate monitors, activity trackers, blood pressure monitors, blood glucose monitors, and Oxygen saturation sensors (pulse Oximeters).
 \item \textbf{Implantable Sensors:} These data are extracted from cardiac monitors, glucose monitors, and intraocular pressure sensors.
 \item \textbf{Environmental sensors:} consists of the data gathered from air quality monitors that measure pollutants and allergens in the environment, which can affect respiratory conditions like asthma, and temperature and humidity sensors.
 \item \textbf{Specialized Medical Sensors:} Consist the data gathered from \gls{ECG} sensors, \gls{EEG} sensors, \gls{EMG} sensors, gait sensors, and sleep monitors. 
 \item \textbf{Smart Health Homes:} These data are extracted from fall detectors (sensors that detect falls, especially important for elderly patients), and bed sensors that monitor sleep patterns, bed occupancy, and vital signs during sleep.
\end{itemize}

Sensor data used in healthcare \gls{ML} refers to information collected by various sensors, which can include wearable devices, medical equipment, and other monitoring tools~\cite{ghazal2021iot,balakrishna2020iot}. This data provides real-time insights into an individual's health and activities. Healthcare \gls{ML} algorithms analyze sensor data to make predictions, identify patterns, and offer personalized insights. The utilization of sensor data in healthcare \gls{ML} is detailed as follows:

\begin{itemize}
 \item \textbf{Vital Signs Monitoring:} sensors capture vital signs such as heart rate, blood pressure, respiratory rate, and temperature~\cite{khan2016monitoring,yilmaz2010detecting}. \gls{ML} algorithms analyze continuous vital sign data to detect anomalies, predict health deteriorations, and offer early warnings to healthcare professionals~\cite{wu2021internet,lv2022wearable,da2018internet}.
 
 \item \textbf{Wearable Devices for Activity Tracking:} Wearable sensors, including accelerometers and gyroscopes, monitor physical activities and movement patterns~\cite{attal2015physical,yang2010review,cornacchia2016survey}. \gls{ML} models analyze this data to assess overall physical health, detect abnormalities, and provide personalized insights for fitness and rehabilitation plans~\cite{zhang2022deep,nweke2018deep}.
 
 \item \textbf{Blood Glucose and Continuous Glucose Monitoring (CGM):} Sensors continuously measure blood glucose levels~\cite{rodriguez2021mobile}, offering a comprehensive glycemic profile. \gls{ML} algorithms analyze \gls{CGM} data to predict glucose level trends, recommend insulin dosages, and enhance diabetes management~\cite{zhu2022enhancing}.
 
 \item \textbf{Electrocardiogram (ECG) Monitoring:} \gls{ECG} sensors capture cardiac electrical activity. \gls{ML} interprets \gls{ECG} data to identify cardiac abnormalities, predict cardiovascular risk, and recommend early interventions~\cite{liu2021deep,somani2021deep,minchole2019machine,kwon2022flexible}.
 
 \item \textbf{Sleep Monitoring:} Wearable devices and bed-based sensors track sleep patterns, including duration and quality. \gls{ML} algorithms analyze sleep data to identify disorders, provide insights into sleep hygiene, and recommend personalized interventions for improved sleep health~\cite{sathyanarayana2016sleep,arora2020analysis}.
 
 \item \textbf{Environmental Sensors:} sensors measure environmental factors like air quality, temperature, and humidity. \gls{ML} correlates environmental data with health outcomes, aiding in identifying triggers for respiratory conditions, allergies, or other health issues~\cite{zhang2020comprehensive}.
 
 \item \textbf{Medication Adherence Monitoring:} Smart pill bottles and medication dispensers with sensors track adherence. \gls{ML} algorithms analyze adherence patterns, send reminders to patients, and provide healthcare providers with insights into patient compliance~\cite{bohlmann2021machine,roh2021deep}.
 
 \item \textbf{Fall Detection and Activity Recognition:} Motion sensors detect falls and recognize different activities. \gls{ML} models use this data for fall risk assessment, predicting accidents, and adapting care plans for individuals with mobility challenges~\cite{tunca2019deep,meyer2020wearables}.
 
 \item \textbf{Biometric Sensors for Stress and Emotion Monitoring:} Biometric sensors, measuring skin conductance or heart rate variability, assess stress levels and emotional states. \gls{ML} analyzes biometric signals to provide insights into mental health, stress management, and emotional well-being~\cite{al2019deep,kyamakya2021emotion,gedam2021review}.
 
\end{itemize}

\subsection{{Patient-Generated Data}}
\gls{PGD} within the realm of healthcare \gls{ML} pertains to health-related information actively contributed by patients~\cite{shapiro2012patient}. This data, distinct from traditional clinical records, is directly sourced from patients through various means, including wearables, mobile apps, and patient-reported outcomes. The incorporation of patient-generated data in healthcare \gls{ML} fosters a patient-centric and data-driven healthcare approach, enabling personalized interventions, early detection of health issues, and improved communication between patients and healthcare providers~\cite{hsueh2017making}. However, it is imperative to prioritize the privacy and security of patient-generated data and uphold ethical considerations throughout the development and implementation of \gls{ML} models in healthcare settings. Apart from the data gathered by sensors as mentioned in the previous subsection (i.e. Section~\ref{SensorData}), the utilization of patient-generated data in healthcare \gls{ML} is extended as follows:

\begin{itemize}
 
 
 \item \textbf{Mobile Health Apps and Surveys:} Patients leverage mobile apps to input health-related information, participate in surveys, and provide feedback on their health status~\cite{kao2017consumer,peng2016qualitative,sama2014evaluation}. \gls{ML} algorithms process patient-reported data to derive insights into treatment effectiveness, medication adherence, and overall patient satisfaction, thereby informing personalized care plans~\cite{istepanian2018m,mendo2021machine}.
 
 
 \item \textbf{Social Media and Online Communities:} Patients share health-related information, experiences, and concerns on social media platforms and online forums~\cite{van2013using}. \gls{ML} algorithms conduct analyses on social media data for health-related trends, sentiment, and public health monitoring, contributing valuable insights for population health research and understanding patient perspectives~\cite{gupta2020social,hasib2023depression}.
 
 \item \textbf{Genomic and Genetic Data Sharing:} Patients voluntarily contribute genetic information through platforms facilitating data sharing for research purposes. \gls{ML} analyzes aggregated genomic data to identify genetic factors associated with diseases, fostering advancements in precision medicine and contributing to genetic research~\cite{johnson2021precision}.
 
 
 \item \textbf{Telehealth and Virtual Visits:} Patients engage in virtual consultations, offering health-related updates and information during telehealth visits. \gls{ML} algorithms analyze patient-generated data from virtual visits to support clinical decision-making, monitor treatment progress, and enhance the overall quality of virtual healthcare interactions~\cite{schunke2022rapid,verma2022tele}.
\end{itemize}

\subsection{Clinical Trial Data}
Clinical trial data in the context of medical research refers to the information collected during the conduct of clinical trials. Clinical trials are systematic investigations designed to evaluate the safety, efficacy, and/or effectiveness of new drugs, medical devices, procedures, or interventions in human subjects. These trials follow a carefully designed protocol or plan and are conducted to generate data that can be used to inform medical decision-making, regulatory approvals, and advancements in medical knowledge~\cite{national2010prevention}. Integration of clinical trial data into \gls{ML} models enables the development of predictive algorithms, risk assessment tools, and decision support systems in healthcare. This approach contributes to evidence-based medicine, supports personalized treatment strategies, and enhances the overall efficiency of clinical research and practice~\cite{pettit2021artificial,liu2019advancing,liu2021deep,castiglioni2021ai}. key components of clinical trial data can be summarized as follows:

\begin{itemize}
 \item \textbf{Demographic Information:} This information consists of details about the study participants, including age, gender, race, ethnicity, and other relevant demographic characteristics. Demographic data can be incorporated into \gls{ML} models to assess how different demographic groups respond to interventions. It helps in developing personalized treatment plans based on age, gender, and other demographic factors.
 \item \textbf{Informed Consent:} Normally contains documentation confirming that participants have been informed about the trial, its risks, and benefits, and have voluntarily agreed to participate. \gls{ML} models can incorporate consent-related information to ensure that only data from participants who have provided informed consent are included in the analysis~\cite{cohen2019informed}. This is crucial for maintaining ethical standards and regulatory compliance~\cite{mckeown2021ethical}.
 \item \textbf{Medical History:} Information about participants' pre-existing medical conditions, relevant medical history, and details about any concurrent medications. Medical history data can be utilized to identify pre-existing conditions or comorbidities that may impact treatment outcomes. \gls{ML} models can predict how certain medical histories influence responses to interventions~\cite{miyoshi2021machine,chekroud2021promise}.
 \item \textbf{Intervention Details:} Specifics about the investigational product or procedure being tested, including dosage, administration methods, and treatment protocols. \gls{ML} algorithms can analyze intervention details to identify patterns associated with treatment success or failure~\cite{chien2020machine}. This information is valuable for predicting the efficacy of similar interventions in future cases.
 \item \textbf{Clinical Assessments:} Physical examinations, laboratory tests and results, imaging studies, and other clinical measurements were conducted to assess participants' health status and response to the intervention. \gls{ML} models can process clinical assessment data to identify trends, correlations, or anomalies that may be indicative of treatment responses or adverse events~\cite{benke2018artificial}. This assists in early detection and prediction of outcomes.
 \item \textbf{Adverse events:} Records of any adverse events or side effects experienced by participants during the trial, including their severity and relation to the intervention. \gls{ML} algorithms can learn from historical adverse event data to predict the likelihood of adverse events for new interventions. This supports risk assessment and enables proactive management~\cite{song2023using,yang2024machine}.
 \item \textbf{Efficacy Endpoints:} Measurements and assessments used to determine the effectiveness of the intervention, often focusing on outcomes such as improvements in disease markers or symptom relief. \gls{ML} models can analyze efficacy endpoint data to develop predictive models for treatment success or failure. This facilitates the identification of key factors contributing to positive outcomes.
 \item \textbf{Follow-Up Data:} Information collected during follow-up visits, which may include data on long-term outcomes, treatment adherence, and any sustained effects of the intervention. Follow-up data is essential for longitudinal analyses. \gls{ML} models can use this data to predict the long-term effects of interventions, allowing for the development of predictive models for sustained efficacy or potential relapses~\cite{badwan2023machine,gayvert2016data,ezzati2020machine}.
 \item \textbf{Protocol Deviations:} Documentation of any deviations from the original study protocol, including the reasons for these deviations. \gls{ML} models can analyze protocol deviation data to identify how deviations impact study outcomes. This information can be used to adjust for deviations during analysis or predict their potential effects~\cite{feijoo2020key}.
 \item \textbf{Statistical Analyses:} Methods and results of statistical analyses applied to the data, including primary and secondary endpoints, to draw conclusions about the efficacy and safety of the intervention. \gls{ML} models can incorporate statistical analysis results to understand the significance of various factors on study outcomes. This facilitates the development of \gls{ML} models that take into account the statistical nuances of the data~\cite{zame2020machine}.
\end{itemize}

\subsection{Prescription and Medication Data}
Prescription and medication data in medical records refer to information related to the medications prescribed to patients, including details about the prescribed drugs, dosage, frequency, and other relevant information. These data are crucial components of patient health records and play a significant role in managing and monitoring patient care. Prescription and medication data are important for ensuring patient safety, medication management, monitoring public health and treatment efficacy, research and analysis, regulatory compliance, and facilitating communication among healthcare providers. These data are integrated into \gls{EHRs} to provide a comprehensive view of a patient's medication history, allowing healthcare professionals to make informed decisions about treatment plans and avoid potential drug interactions or adverse events. Additionally, aggregated and anonymized prescription and medication data are used in research to assess the effectiveness and safety of medications on a broader scale. Key components of prescription and medication data can be summarized as follows:

\begin{itemize}
 \item \textbf{Patient Information:} These data consist of identification details of the patient, including name, date of birth, and other relevant demographic information. We already explained this kind of data and its use in ML. 
 \item \textbf{Prescriber Information:} These data consist of records of details about the healthcare provider who prescribed the medication, including their name, credentials, and contact information. Analyzing prescriber data can help identify patterns in prescribing behavior. \gls{ML} models can be trained to recognize prescribing practices associated with positive patient outcomes, contributing to more informed decision-making~\cite{maceachern2021machine}.
 \item \textbf{Prescription Date:} contains records of the date on which the prescription was issued, indicating when the patient should start taking the medication. Temporal analysis of prescription dates can assist in predicting medication adherence and treatment outcomes. \gls{ML} models may identify patterns related to the timing of prescription issuance and its impact on patient behavior~\cite{chalasani2023artificial}.
 \item \textbf{Medication Name:} Contains the name of the prescribed medication, including the generic and brand names. \gls{ML} models can categorize medications based on their therapeutic classes, aiding in the identification of commonalities and differences in treatment outcomes across different drug categories~\cite{askr2023deep}.
 \item \textbf{Dosage:} Consists of information about the prescribed amount or strength of the medication to be taken by the patient. Dosage and administration details are crucial for predicting medication adherence and potential adverse events. \gls{ML} models can identify optimal dosage regimens and administration routes based on historical data~\cite{nemati2016optimal,li2023machine}.
 \item \textbf{Frequency:} consists of instructions on how often the medication should be taken, such as daily, twice daily, or as needed. 
 \item \textbf{Route of Administration:} The method by which the medication should be administered (e.g., orally, intravenously, topically). 
 \item \textbf{Duration of Treatment:} Consists of the period for which the medication is prescribed, indicating the number of days or weeks the patient should take the medication. \gls{ML} models can analyze the duration of treatment to predict long-term outcomes, including the likelihood of treatment success and the potential development of drug resistance~\cite{lin2022can,del2023machine}.
 \item \textbf{Instructions for Use:} Additional guidance on how to take the medication, such as whether it should be taken with food, at specific times of the day, or with other medications. \gls{NLP} techniques can be applied to extract insights from free-text instructions, helping to identify specific nuances in how patients are instructed to take their medications~\cite{wang2018clinical,uzuner2010extracting}.
 \item \textbf{Refill Information:} Contains details regarding whether the prescription allows for refills and, if so, the number of authorized refills. Refill information is essential for predicting patient adherence and persistence with prescribed medications. \gls{ML} models can identify factors influencing refill behavior and predict the likelihood of medication discontinuation~\cite{galozy2020prediction,hasan2021machine}.
 \item \textbf{Allergies and contraindications:} contains records of information about any known allergies the patient has or contraindications that may affect the choice of medication. \gls{ML} models can identify associations between patient allergies, contraindications, and adverse drug reactions. This information aids in predicting the safety and suitability of specific medications for individual patients~\cite{kim2022analyzing,chandak2020using}.
 \item \textbf{Adverse Reactions:} contain documentation of any adverse reactions or side effects the patient experiences while taking the medication. Adverse reaction data can be analyzed to develop predictive models that identify patients at higher risk for experiencing specific side effects. This allows for proactive management and personalized interventions.
 \item \textbf{Medication Changes:} information about any changes made to the prescribed medication, including adjustments in dosage or switching to an alternative medication. \gls{ML} models can analyze historical medication changes to predict the likelihood of treatment modifications in the future, assisting in the development of personalized treatment plans~\cite{zhang2018learning,johnson2021precision}.
 \item \textbf{Medication Discontinuation:} If applicable, the reason for discontinuing the medication and the date of discontinuation can be considered in this class of data. \gls{ML} models can predict the factors contributing to medication discontinuation, helping healthcare providers identify patients at risk of stopping their medications and intervening to improve adherence~\cite{meng2023machine}.
 \item \textbf{Medication Administration Records (MAR):} contain records of actual medication administration, often documented in a healthcare setting, indicating when and by whom the medication was administered to the patient. \gls{MAR} data can be used to train \gls{ML} models for predicting medication administration patterns and identifying potential deviations from the prescribed regimen~\cite{visweswaran2010identifying,gu2021predicting}.
\end{itemize}

\subsection{Laboratory Data}
Laboratory data in medical records refer to the results of tests and analyses conducted on patient samples in a laboratory setting. These tests are essential for diagnosing, monitoring, and managing various medical conditions. The types of laboratory data collected can vary depending on the patient's symptoms, medical history, and the healthcare provider's assessment. \gls{ML} techniques, including supervised learning, unsupervised learning, and deep learning, can be employed depending on the nature of the data and the specific goals of the analysis. Common types of laboratory data are as follows:

\begin{figure*}\label{MedData}
\centering

\includegraphics[width=0.86\textwidth]{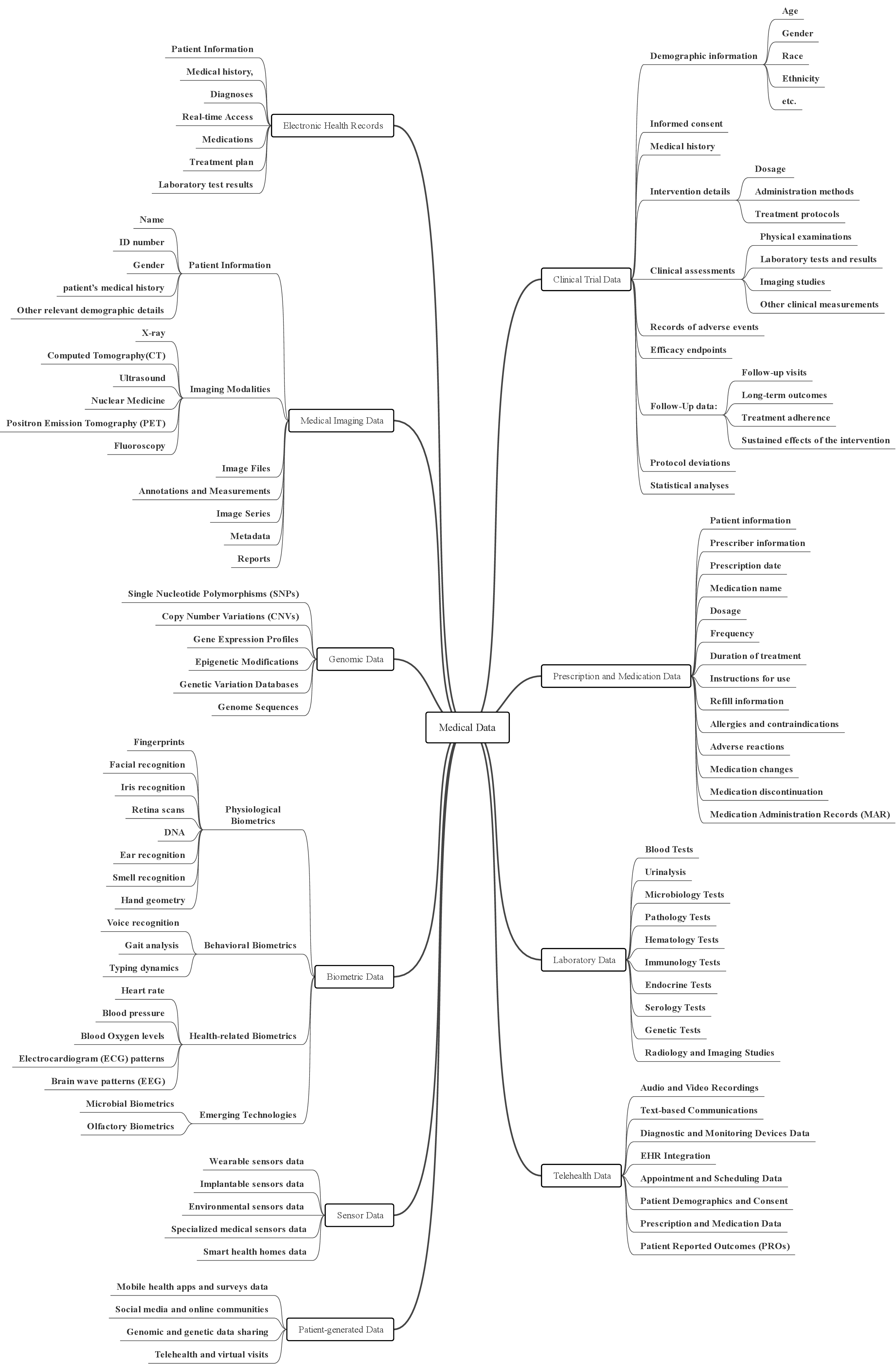}
\caption{Taxonomy of Medical Data Used for Machine Learning in Medical Applications}
\end{figure*}

\begin{itemize}
 \item \textbf{Blood Tests:} Consists of a \gls{CBC} that measures the number and types of blood cells, Blood Chemistry Panel (tests such as electrolytes, glucose, liver function tests, and kidney function tests), and Lipid Profile. In the context of ML, predictive models can be developed to identify patterns in \gls{CBC} results associated with specific diseases, such as anemia or infections~\cite{wu2020rapid,khan2020review,alam2019machine}. 
 \item \textbf{Urinalysis:} Examines the physical and chemical properties of urine to detect abnormalities. \gls{ML} algorithms can process urinalysis data to detect patterns indicative of kidney disorders, urinary tract infections, or diabetes~\cite{zeb2020towards,chittora2021prediction,de2023applications}. 
 \item \textbf{Microbiology Tests:} Results of this test identify and determine the sensitivity of microorganisms to specific antibiotics or identify the type of bacteria present. \gls{ML} can assist in the identification of microorganisms from culture data and predict antibiotic susceptibility, aiding in personalized treatment plans~\cite{peiffer2020machine,goodswen2021machine,qu2019application,ghannam2021machine}.
 \item \textbf{Pathology Tests:} These tests consist of tissue biopsy and Cytology where the former one examines tissue samples for abnormalities or diseases and the latter one examines cells for signs of disease, often used in cancer diagnosis. In the context of ML, image recognition models can be trained on pathology slides to assist pathologists in identifying abnormal tissue structures or cancerous cells~\cite{echle2021deep,madabhushi2016image,janowczyk2016deep}. 
 \item \textbf{Hematology Tests:} These tests can be classified as coagulation studies and \gls{ESR}. The first one assesses the blood clotting function while the second one measures inflammation in the body. In the context of ML, \gls{ML} models can analyze coagulation studies to predict the risk of bleeding or clotting disorders~\cite{obstfeld2023hematology,radakovich2020machine,fang2021using,guo2021predicting}.
 \item \textbf{Immunology Tests:} These tests consist of antibody tests that detect antibodies produced by the immune system and viral load that measures the amount of virus in the blood for certain infections. \gls{ML} algorithms can identify patterns in antibody levels to diagnose autoimmune diseases or infectious conditions~\cite{pertseva2021applications,danieli2023machine,usategui2023machine}. 
 \item \textbf{Endocrine Tests:} Assesses levels of hormones such as thyroid hormones, insulin, and others. \gls{ML} models can analyze hormone levels to predict and monitor endocrine disorders such as thyroid dysfunction or diabetes~\cite{thomasian2022machine,hong2020machine}.
 \item \textbf{Serology Tests:} Analyze components in the liquid portion of blood (serum), including proteins, enzymes, and electrolytes. \gls{ML} algorithms can be applied to serum test data to identify markers associated with specific diseases, aiding in early detection and monitoring~\cite{danieli2023machine,gunvcar2018application}.
 \item \textbf{Genetic Tests:} Genetic tests involve the identification of specific genetic markers associated with particular conditions. These markers represent distinct genetic variations and play a crucial role in understanding and diagnosing various health conditions through genetic analysis. \gls{ML} can be used to analyze genetic data for identifying disease risk, predicting treatment response, or diagnosing genetic disorders.
 \item \textbf{Radiology and Imaging Studies:} Although not conducted in a traditional laboratory, these tests are often considered part of the diagnostic process and contribute to the overall medical data. Image recognition and segmentation models can be trained on radiology images to assist in the diagnosis of conditions like tumors, fractures, or abnormalities. 
\end{itemize}

\subsection{Telehealth Data}
Telehealth~\cite{tuckson2017telehealth} data in medical records refers to the information collected during remote healthcare interactions between patients and healthcare providers. Telehealth data play a crucial role in modern healthcare, especially during times when remote interactions are necessary~\cite{gajarawala2021telehealth}. Telehealth involves the use of technology, such as video calls, phone calls, and online platforms, to deliver healthcare services remotely. The data generated during telehealth encounters contribute to the overall patient record and can include various types of information. Here are some key components of telehealth data:

\begin{itemize}
 \item \textbf{Audio and Video Recordings:} Recordings of virtual consultations between healthcare providers and patients can be stored for documentation, review, and quality assurance purposes. \gls{ML} applications may analyze these recordings to extract useful information, such as sentiment analysis or clinical insights. Speech recognition algorithms can transcribe audio recordings, enabling the analysis of verbal communication between healthcare providers and patients~\cite{schunke2022rapid}. Facial recognition and sentiment analysis can be applied to video recordings to assess patient emotions and engagement during telehealth consultations~\cite{torres2018patient}.
 \item \textbf{Text-Based Communications:} Consists of chat logs, text messages, or emails exchanged between patients and healthcare providers. These communications can contain important information about symptoms, treatment plans, and patient queries. \gls{NLP} models can extract and categorize information from text-based communications~\cite{farzindar2015natural}. This includes identifying symptoms, treatment discussions, and patient concerns. Sentiment analysis can help gauge patient satisfaction and emotional well-being.
 \item \textbf{Diagnostic and Monitoring Devices Data:} Includes all types of data from medical devices used by patients at home, such as remote monitoring devices for blood pressure, glucose meters, or wearable fitness trackers. These data enable continuous monitoring of patients' health remotely. \gls{ML} models can analyze trends and patterns in this data to provide insights into the patient's condition. Time series analysis and predictive modeling can be applied to data from remote monitoring devices to detect trends and anomalies in vital signs. For example, \gls{ML} algorithms can predict exacerbations of chronic conditions based on changes in physiological parameters~\cite{fernandez2015machine}.
 \item \textbf{EHR Integration:} Integration of telehealth data into \gls{EHRs} or \gls{EMRs} facilitates a comprehensive view of the patient's health history, including both in-person and remote interactions, for healthcare providers. Integrating telehealth data into \gls{EHRs} allows for a comprehensive patient profile. \gls{ML} models can analyze the combined data to identify correlations between in-person and virtual interactions, improving diagnostic accuracy and treatment planning.
 \item \textbf{Appointment and Scheduling Data:} Contains information related to telehealth appointments, including scheduling details, appointment duration, and attendance, and can support efficient management of healthcare services. Analysis of scheduling data can optimize appointment availability and improve patient access to care. As well, predictive analytics can optimize appointment scheduling by analyzing historical data to identify peak appointment times and anticipate patient demand. This can improve resource allocation and patient access to telehealth services.
 \item \textbf{Patient Demographics and Consent:} contains information about the patient, including demographic details and consent for telehealth services and ensures compliance with privacy regulations, and provides context for personalized care. \gls{ML} applications may analyze demographic data for population health management. Moreover, \gls{ML} models can analyze patient demographic data to identify population health trends, target interventions, and personalize healthcare services. Consent data can be used to ensure compliance with privacy regulations and customize communication preferences.
 \item \textbf{Prescription and Medication Data:} Contains information related to prescriptions, medication management, and adherence and Supports virtual prescription refills and medication management. Predictive modeling can be applied to medication adherence data to identify patterns and factors influencing adherence. This information can inform personalized interventions to improve patient compliance. \gls{ML} can also analyze prescription data to identify medication-related risks and interactions~\cite{segal2019reducing}.
 \item \textbf{\gls{PROs}:} Contains patient-reported data on symptoms, well-being, and treatment effectiveness and provides valuable insights into the patient's subjective experience. \gls{ML} can help analyze \gls{PROs} to predict treatment responses or identify trends in symptoms.
 Text mining and sentiment analysis can be applied to \gls{PROs} to extract valuable insights into patient experiences and treatment effectiveness. Predictive modeling can correlate \gls{PROs} with clinical outcomes to inform treatment decisions~\cite{verma2021application}. 
 
\end{itemize}

	\begin{figure*}[t]
		\centering
		\includegraphics[width=\linewidth]{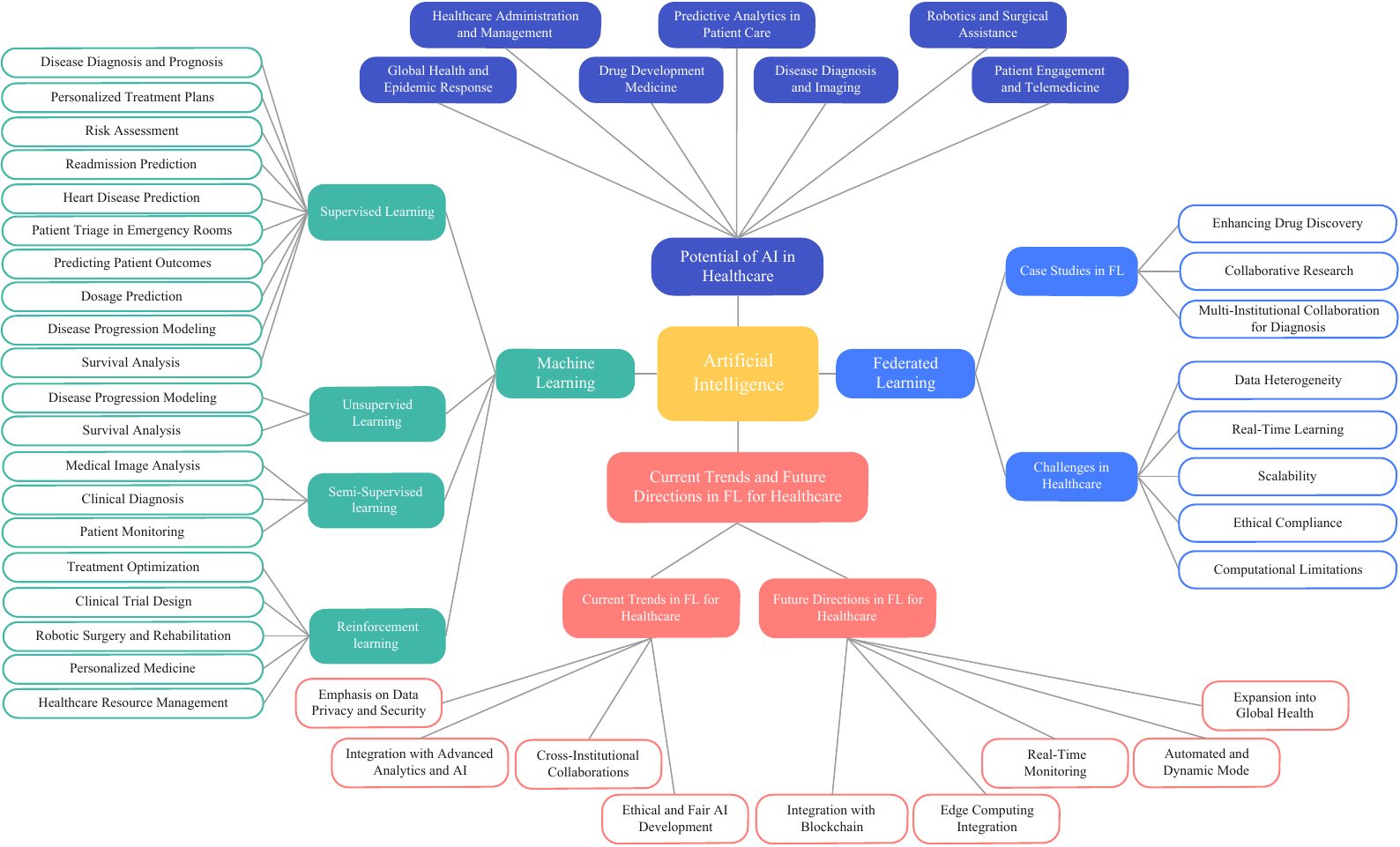}
		\caption{Exploring the Convergence of Artificial Intelligence with Healthcare: Trends, Applications, and Future Perspectives.}
		\label{fig:AI}
	\end{figure*}

\section{Artificial Intelligence}\label{sec:MedAI}

\gls{AI} is revolutionizing healthcare, impacting diagnosis, treatment, and patient care. This section explores the evolution of \gls{AI} in healthcare, from its historical roots to cutting-edge developments, showcasing its potential across various healthcare domains. We will further delve into \gls{FL} as a key technology for the future. \gls{FL}'s collaborative approach enables privacy-preserving data analysis, making it a powerful tool for healthcare research and delivery. We will discuss its applications, case studies, current challenges, and future directions.

\subsection{Evolution of \gls{AI} in Healthcare}

The journey of \gls{AI} in healthcare began in the late 1950s and early 1960s with foundational work in \gls{AI} and its potential applications \cite{russell2010artificial}. Early systems, such as ELIZA (1966) \cite{weizenbaum1966eliza} and MYCIN (1974) \cite{shortliffe1974mycin}, demonstrated \gls{AI}'s capabilities in mimicking human conversation and decision-making in medical diagnostics, respectively. These systems laid the groundwork for future \gls{AI} applications in healthcare. By delving into the key milestones marking the evolution of \gls{AI} in healthcare, we can categorize it into distinct periods:

 	\begin{itemize}
	 \item \textbf{1990s to Early 2000s}: The advancement of \gls{ML} algorithms, particularly in neural networks \cite{bishop1995neural}, led to increased interest in \gls{AI}'s applications in healthcare. The development of algorithms for pattern recognition in medical imaging \cite{lo1995artificial} and the emergence of \gls{EHRs} \cite{iakovidis1998towards} provided vast data sources for \gls{AI} analysis.
	 
	 \item \textbf{Mid-2000s to 2010s}: This period saw significant advancements in deep learning, dramatically improving the performance of \gls{AI} in image recognition \cite{duncan2000medical} and natural language processing \cite{demner2009can}. The launch of IBM's Watson in 2011 marked a significant milestone \cite{chen2016ibm}, showcasing \gls{AI}'s ability to analyze and interpret medical literature and patient data.
	 
	 \item \textbf{Recent Developments}: The last decade has witnessed exponential growth in \gls{AI} applications in healthcare \cite{younis2024systematic}. The advent of big data analytics \cite{nuseir2024role}, cloud computing \cite{gowda2024introduction}\, and improved algorithms have enabled more sophisticated and accurate \gls{AI} tools \cite{chaddad2023survey}. These developments have facilitated breakthroughs in precision medicine, predictive analytics, and patient care management.
	\end{itemize}

\subsection{Potential of \gls{AI} in Healthcare}
	\gls{AI} is transforming healthcare across a wide spectrum, from disease diagnosis to patient care management. Here are some key areas where \gls{AI} is making significant contributions:

 \begin{itemize}
\item{\textbf{Disease Diagnosis and Imaging:}} \gls{AI} has revolutionized medical imaging by providing tools for more accurate and faster diagnosis \cite{esteva2019guide}. Deep learning models, trained on large datasets of X-rays \cite{jaiswal2019identifying}, \gls{MRI}s \cite{havaei2017brain}, and \gls{CT} scans \cite{setio2017validation}, can identify patterns undetectable to the human eye. \gls{AI} aids in early detection of diseases like cancer, cardiovascular abnormalities, and neurological disorders.
	
\item{\textbf{Drug Development and Personalized Medicine:}} \gls{AI} algorithms have streamlined the drug development process by predicting molecular behavior \cite{zhavoronkov2019deep} and identifying potential drug candidates \cite{chan2019advancing}. In personalized medicine, \gls{AI} analyzes patient data, including genetic information, to tailor treatments to individual patients, improving efficacy and reducing side effects \cite{obermeyer2016predicting}.

\item{\textbf{Predictive Analytics in Patient Care:}} \gls{AI}'s predictive analytics are crucial to preventive medicine. By analyzing \gls{EHRs} \cite{rajkomar2018scalable}, \gls{AI} can predict patient risks for diseases \cite{gulshan2016development}, hospital readmission \cite{futoma2017learning}, and other adverse events, enabling proactive care and intervention \cite{henry2015targeted}.
	
\item{\textbf{Robotics and Surgical Assistance:}} Robotics integrated with \gls{AI} has improved surgical precision and outcomes \cite{kassahun2016surgical}. \gls{AI}-driven robots assist surgeons in complex procedures, reducing human error and patient recovery time. \gls{AI} also plays a role in training surgeons through virtual reality simulations \cite{biffi2017immersive}.
	
\item{\textbf{Patient Engagement and Telemedicine:}} \gls{AI}-powered chatbots and virtual health assistants provide 24/7 support and health monitoring \cite{miner2020chatbots}, enhancing patient engagement and adherence to treatment plans. In telemedicine, \gls{AI} tools assist in remote diagnosis and consultation \cite{dorsey2016state}, making healthcare more accessible.
	
\item{\textbf{Healthcare Administration and Management:}} \gls{AI} streamlines administrative tasks in healthcare, such as scheduling, billing, and claims processing \cite{mesko2018will}. It also optimizes hospital operations, resource allocation, and patient flow, improving overall efficiency and reducing costs \cite{harper2002framework}.
	
\item{\textbf{Global Health and Epidemic Response:}} \gls{AI} has been pivotal in global health, particularly in tracking and predicting the spread of infectious diseases \cite{wong2023leveraging}. During the COVID-19 pandemic, \gls{AI} models were instrumental in analyzing virus transmission \cite{li2020artificial}, vaccine development \cite{zhavoronkov2020potential}, and managing healthcare resources.
\end{itemize}	

\subsection{Fundamentals of Machine Learning} \gls{ML}, a pivotal branch of \gls{AI}, is fundamentally reshaping our approach to problem-solving across various domains \cite{chen2019artificial}, including healthcare \cite{pereira2009machine}. At its core, \gls{ML} involves the development and application of algorithms that enable computers to learn from and make decisions or predictions based on data. This capacity for self-improvement and adaptation without explicit programming is what sets \gls{ML} apart.

The cornerstone of \gls{ML} is data. Algorithms learn from data patterns, and the quality and quantity of this data significantly influence their performance as explained in the previous section. \gls{ML} algorithms are sets of rules or instructions given to computers to help them learn from data. These algorithms can be broadly categorized into supervised learning \cite{vercio2020supervised, kassahun2016surgical, erickson2017machine, kukreja2024review, bertsimas2020machine, asri2016using, talwar2023performance}, unsupervised learning \cite{colace2024unsupervised, eisen1998cluster, monti2003consensus, wu2017unsupervised, ringner2008principal, van2008visualizing, smith2002fast, menze2014multimodal}, semi-supervised \cite{chapelle2009semi,qiu2023federated,ren2020not,IBM_semi_supervised_learning,eckardt2022semi,jiao2023learning} and reinforcement learning \cite{gosavi2009reinforcement, yu2021reinforcement, komorowski2018artificial, zhao2009reinforcement, chi2019context, wang2022predicting, mclaverty2023unifying, almagrabi2022reinforcement}.\\

 \begin{enumerate}[wide, font=\itshape, labelwidth=!, labelindent=0pt, label*=\textit{C}.\arabic*.]
	\item \textit{Supervised Learning}
 
 Supervised learning, a dominant branch of \gls{ML}, plays a crucial role in healthcare by leveraging labeled datasets to train models that can make predictions or categorize data \cite{vercio2020supervised}. This approach is especially powerful in scenarios where the relationship between input data and the output is known and can be modeled \cite{kassahun2016surgical}. 
 		
 		\begin{itemize}
 		
		\item \textbf{Labeled Data:} In supervised learning, the training data is labeled, meaning that the outcome of each data point is known. This label guides the algorithm in learning the relationship between the input features and the output.
		
		\item \textbf{Classification and Regression:} The two primary tasks in supervised learning are classification (predicting discrete outcomes) and regression (predicting continuous outcomes). Classification might involve diagnosing whether a patient has a specific disease, while regression could involve predicting a patient's recovery time. We will explore these concepts in greater depth in the classification and regression subsection, providing more detailed insights and applications.
		
		\item \textbf{Model Training and Validation:} The process involves training the model on a portion of the data and then validating its accuracy on a separate, unseen dataset. This helps in ensuring that the model generalizes well to new data.
		
		\end{itemize}
		
		Supervised learning has a big impact in healthcare with use in applications such as disease diagnostics, treatment personalization, and patient management.
		
 		\begin{itemize}
		
		\item \textbf{Disease Diagnosis and Prognosis:} Perhaps the most significant application of supervised learning in healthcare is in disease diagnosis \cite{erickson2017machine}. \gls{ML} models are trained on clinical data, including patient symptoms, lab results, and medical imaging, to identify diseases. For instance, \gls{ML} models trained on imaging data can detect abnormalities such as tumors in radiographic images with high accuracy \cite{kukreja2024review}.
		
		\item \textbf{Personalized Treatment Plans:} Supervised learning algorithms analyze patient data to predict how individual patients might respond to different treatments. This personalized approach is particularly effective in oncology \cite{bertsimas2020machine}, where treatment plans are tailored based on the genetic makeup of a patient's tumor.
		
		\item \textbf{Risk Assessment:} Models trained on historical patient data can assess the risk of developing certain conditions, like diabetes or heart disease, based on a patient’s lifestyle, genetics, and other factors \cite{asri2016using}.
		
		\item \textbf{Readmission Prediction:} Supervised learning can predict a patient's likelihood of readmission to a hospital. This is vital for improving patient care and reducing healthcare costs \cite{talwar2023performance}.
		\end{itemize}

 \gls{ML} offers invaluable tools through its capabilities in classification and regression tasks, as previously mentioned. These foundational types of supervised learning tasks play a crucial role across a wide spectrum of applications, ranging from diagnosing diseases to forecasting patient outcomes. Below, we provide specific examples of both classification and regression tasks, highlighting their profound impact on improving healthcare delivery and enhancing patient care. Classification in the realm of \gls{ML} entails categorizing data into predetermined classes or categories. In healthcare, this capability is crucial for:
	 
		\begin{itemize}
		
		\item \textbf{Disease Diagnosis:} \gls{ML} models are trained to classify patient data into disease categories. An example includes the use of \gls{CNNs} for classifying dermatological images into benign or malignant skin lesions \cite{esteva2017dermatologist}.
		 
		 \item \textbf{Heart Disease Prediction:} Algorithms can analyze patient data such as age, blood pressure, cholesterol levels along with other clinical parameters to classify individuals into risk categories for heart disease. This classification aids in early intervention and preventive care \cite{shah2020heart}.
		 
		\item \textbf{Patient Triage in Emergency Rooms:} \gls{ML} models can classify patients based on the severity of their condition. By analyzing symptoms, vital signs, and medical history, algorithms can assist in determining the urgency of each case, optimizing patient flow and resource allocation in emergency departments \cite{levin2018machine}.
		
		\end{itemize}
		
		\begin{figure}
			\centering\includegraphics[width=0.7\linewidth]{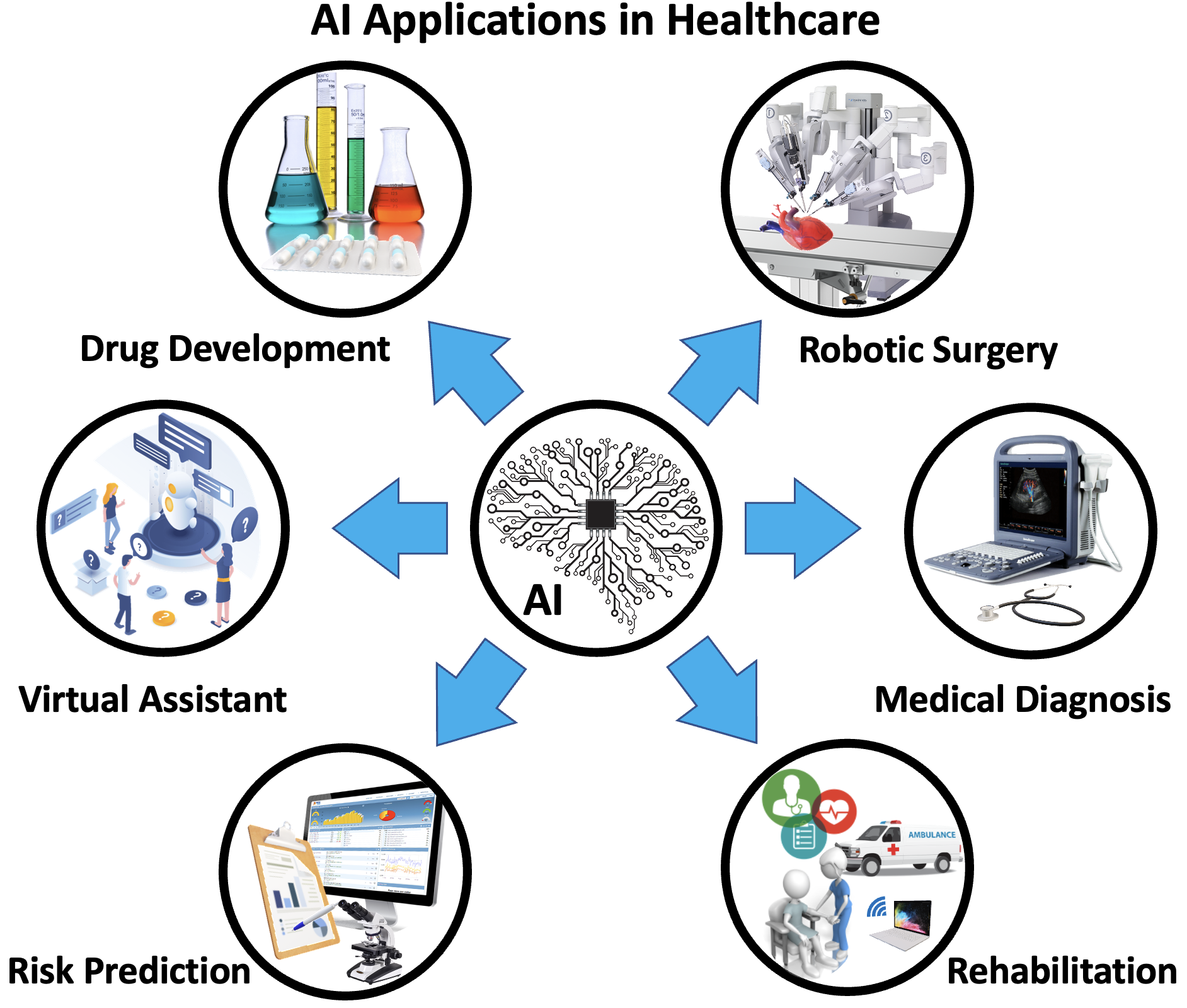}
			\caption{Schematic Diagram Showing Six Application Scenarios of Artificial Intelligence in Healthcare.}
			\label{fig:AI_Med}
		\end{figure}
		
		 Regression tasks in \gls{ML} deal with predicting continuous outcomes. In healthcare, regression models are applied to:
		
 	\begin{itemize}
		
		 \item \textbf{Predicting Patient Outcomes:} \gls{ML} models can predict quantitative outcomes, such as the length of hospital stay, recovery time after surgery, or progression of a disease. For example, regression models might be used to predict blood sugar levels in diabetic patients based on their diet, medication, and physical activity \cite{zou2018predicting}.
		 
		 \item \textbf{Dosage Prediction:} In pharmacotherapy, regression algorithms can predict the optimal drug dosage for individual patients \cite{tatonetti2012data}. This application is particularly important in treatments like chemotherapy, where the dosage needs to be carefully balanced to be effective yet not overly toxic \cite{mucaki2019predicting}.
		 
		\item \textbf{Disease Progression Modeling:} Regression models are used to understand and predict the progression of chronic diseases such as Alzheimer's, Parkinson's, or multiple sclerosis. By analyzing patient data over time, these models can forecast the rate of disease progression, aiding in treatment planning and patient counseling \cite{battineni2020applications}.
		 
		 \item \textbf{Survival Analysis:} In oncology, regression models are crucial for predicting patient survival times post-diagnosis or treatment \cite{dhiman2022methodological}. These predictions, based on patient characteristics and treatment variables, are vital for treatment planning and patient management.\\
		 
		\end{itemize}
		
		\item \textit{Unsupervised Learning} 
  
  Unsupervised learning, a fundamental category of \gls{ML}, involves analyzing and grouping unlabeled data based on similarities and differences, without any predefined labels \cite{colace2024unsupervised}. Two critical techniques in unsupervised learning are clustering \cite{eisen1998cluster, monti2003consensus, wu2017unsupervised} and dimensionality reduction \cite{ringner2008principal, van2008visualizing, smith2002fast, menze2014multimodal}, each playing a vital role in healthcare, particularly in genomics and medical imaging.
		
		Clustering is the process of grouping a set of objects in such a way that objects in the same group (or cluster) are more similar to each other than to those in other groups. Its applications in healthcare are significant:
		
		\begin{itemize}
		
		\item \textbf{Genomic Data Analysis:} In genomics, clustering helps in categorizing genes with similar expression patterns, which can be crucial for understanding gene functions, identifying disease markers, and revealing biological pathways. For instance, clustering can be used to identify groups of genes that are co-expressed in certain diseases, such as cancer and autoimmune disorders. This capability aids in discovering potential therapeutic targets by revealing genes that work in concert across these conditions, enhancing our understanding of disease mechanisms and treatment strategies \cite{eisen1998cluster}.
		
		\item \textbf{Patient Stratification:} Clustering algorithms can segment patients into groups based on similarities in their medical records or genetic information \cite{monti2003consensus}. This stratification aids in identifying subtypes of diseases with distinct clinical outcomes or responses to treatments, facilitating personalized medicine approaches.
		
		\item \textbf{Medical Imaging:} In medical imaging, clustering is used for image segmentation, which is the process of partitioning a digital image into multiple segments (sets of pixels). This technique plays a crucial role in identifying regions of interest, such as tumors in \gls{MRI} or \gls{CT} scans, and helps in the accurate diagnosis and treatment planning \cite{wu2017unsupervised}.
		
		\end{itemize}
		
		Dimensionality reduction is the process of reducing the number of random variables under consideration, by obtaining a set of principal variables. It's particularly important in dealing with high-dimensional data, as is often the case in healthcare:
		
		\begin{itemize}
		
		\item \textbf{Genomic Data Analysis:} Genomic data is inherently high-dimensional, with thousands of genes contributing to complex traits. Dimensionality reduction techniques like \gls{PCA} \cite{ringner2008principal} and \gls{t-SNE} \cite{van2008visualizing} are used to reduce the complexity of genomic data. This simplification helps in visualizing data, identifying genetic markers, and understanding the genetic architecture of diseases.
		
		\item \textbf{Medical Imaging:} High-resolution medical images contain a vast amount of data, making their analysis computationally intensive. Dimensionality reduction techniques can be applied to reduce the number of features in these images while retaining essential information \cite{smith2002fast}. This reduction is crucial for efficient storage, processing, and analysis of medical images, and facilitates the development of more efficient diagnostic algorithms \cite{menze2014multimodal}.\\
		
		\end{itemize}

 \item \textit{Semi-Supervised Learning} 
 
 Semi-supervised learning, a fundamental category of \gls{ML}, occupies a unique position in healthcare by leveraging both labeled and unlabeled data to improve model performance \cite{chapelle2009semi,qiu2023federated}. In scenarios where obtaining labeled data is expensive or time-consuming, semi-supervised learning offers a cost-effective solution by utilizing the abundance of unlabeled data available in healthcare settings.

\begin{itemize}

\item \textbf{Utilizing Unlabeled Data:} In semi-supervised learning, algorithms exploit the vast amounts of unlabeled data commonly present in healthcare databases~\cite{ren2020not}. This unlabeled data, although lacking explicit annotations, often contains valuable information that can complement the labeled data, enhancing the model's understanding of complex medical phenomena.

\item \textbf{Combining Labeled and Unlabeled Data:} By incorporating both labeled and unlabeled data during model training, semi-supervised learning algorithms can learn more robust representations of the underlying data distribution \cite{IBM_semi_supervised_learning}. This holistic approach improves the model's generalization capabilities, leading to more accurate predictions and classifications.

\item \textbf{Semi-Supervised Techniques:} Various techniques are employed in semi-supervised learning, including self-training, co-training, and semi-supervised support vector machines. These methods iteratively refine the model's predictions using the labeled data while leveraging the unlabeled data to enhance its overall performance \cite{IBM_semi_supervised_learning}.

\end{itemize}

In healthcare, semi-supervised learning finds applications in diverse areas such as medical image analysis, clinical diagnosis, and patient monitoring.

\begin{itemize}

\item \textbf{Medical Image Analysis:} Semi-supervised learning algorithms can analyze large volumes of unlabeled medical images to identify subtle patterns or anomalies that may not be apparent to human observers. By combining this unsupervised analysis with labeled data, these algorithms can improve the accuracy of tasks such as tumor detection, organ segmentation, and disease classification~\cite{jiao2023learning}.

\item \textbf{Clinical Diagnosis:} In clinical settings, semi-supervised learning can assist healthcare professionals in diagnosing diseases or predicting patient outcomes by leveraging both labeled patient data and unlabeled population health data~\cite{eckardt2022semi}. This integrated approach enhances the model's diagnostic accuracy and reliability, leading to more informed clinical decisions and improved patient care.

\item \textbf{Patient Monitoring:} Semi-supervised learning techniques can also be employed for continuous patient monitoring, where large streams of unlabeled patient data, such as electronic health records and physiological signals, are analyzed to detect deviations from normal health patterns \cite{sinha2023semi}. By incorporating this unlabeled data into predictive models, healthcare providers can proactively identify and intervene in adverse health events, minimizing patient risks and improving health outcomes.

\end{itemize}

Overall, semi-supervised learning offers a powerful framework for leveraging the wealth of unlabeled data in healthcare to enhance the performance of \gls{ML} models, ultimately advancing medical research, diagnosis, and treatment strategies~\cite{qiu2023review,christopoulou2024machine}. \\

\item \textit{Reinforcement Learning} 

\gls{RL} is a type of \gls{ML} that is particularly suited for situations where an agent must make a sequence of decisions to achieve a goal \cite{gosavi2009reinforcement}. Within \gls{RL}, there are several approaches to learning optimal policies, two of which are:
 \begin{itemize}
 \item \textbf{Dynamic Programming:} This approach to \gls{RL} involves breaking down a decision-making problem into simpler sub-problems and solving them recursively. It is particularly effective in environments with a perfect model, where all states and transitions are known beforehand.
 
 \item \textbf{Monte Carlo Methods:} These methods rely on repeated random sampling to approximate the optimal policy. They are model-free approaches, which means they do not require complete knowledge of the environment and are particularly useful for problems with stochastic dynamics and rewards.
 \end{itemize}
 
 In the context of healthcare, \gls{RL} offers innovative ways to approach complex, dynamic decision-making problems \cite{yu2021reinforcement}. It operates on the principle of reward and penalty, learning optimal actions through trial and error to maximize cumulative rewards.

		Some of the key concepts of reinforcement learning can be summarized as follows:
		
		\begin{itemize}
		
		\item \textbf{Agent and Environment:} In \gls{RL}, an 'agent' (e.g., a healthcare model) interacts with its 'environment' (e.g., patient data or medical scenarios). The agent makes decisions or actions, and the environment provides feedback in the form of rewards or penalties.
		
		\item \textbf{Policy:} A policy is a strategy used by the agent to determine the next action based on the current state of the environment. In healthcare, this might involve choosing a treatment plan based on a patient's current health status.
		
		\item \textbf{Reward Signal:} The agent's actions are guided by a reward signal. Positive rewards encourage the agent to continue making similar decisions, while negative rewards signal the agent to adjust its approach.
		
		\item \textbf{Value Function:} This function estimates the expected cumulative reward of taking a certain action in a given state, helping the agent predict long-term outcomes.
		
		\item \textbf{Exploration vs. Exploitation:} \gls{RL} involves balancing exploration (trying new actions) with exploitation (using known actions that yield high rewards). In healthcare, this might mean balancing between tried-and-tested treatments and experimental therapies.
		
		\end{itemize}
		
		Reinforcement learning can also be very useful in healthcare, with applications such as treatment optimization, clinical trial design, robotic surgery, and healthcare management.
		
		\begin{itemize}
		
		\item \textbf{Treatment Optimization:} \gls{RL} can optimize treatment strategies, adjusting them over time based on patient response \cite{komorowski2018artificial}. For chronic diseases like diabetes, \gls{RL} models can suggest insulin dosages, dietary recommendations, and exercise plans that adapt to changing health indicators.
		
		\item \textbf{Clinical Trial Design:} In designing clinical trials, \gls{RL} can help in determining the most effective trial structures, treatment regimens, and patient selection criteria, enhancing the efficiency and success rates of trials \cite{zhao2009reinforcement}.
		
		\item \textbf{Robotic Surgery and Rehabilitation:} \gls{RL} is used in training robotic systems for surgery and rehabilitation, allowing them to adapt to patient-specific conditions and improve over time based on feedback from surgical outcomes or patient recovery rates \cite{chi2019context}.
		
		\item \textbf{Personalized Medicine:} \gls{RL} models can analyze patient data over time to predict the most effective treatment plans, considering the unique health trajectory and response patterns of each patient \cite{wang2022predicting, mclaverty2023unifying}.
		
		\item \textbf{Healthcare Resource Management:} \gls{RL} algorithms can assist in managing healthcare resources, such as hospital bed allocation, staff scheduling, and equipment usage, by learning optimal allocation strategies based on demand patterns and resource availability \cite{almagrabi2022reinforcement}.\\
		\end{itemize}

 \item \textit{Case Studies of Machine Learning in Healthcare}
 
 \gls{ML} has made significant inroads into the healthcare sector, offering groundbreaking applications that are transforming patient care, diagnostics, treatment planning, and disease management. Below are some notable real-world examples of \gls{ML} applications in healthcare, highlighting both their successes and the challenges they face.
	
 \begin{itemize}

	\item \textbf{Diagnostic Imaging and Radiology:} \gls{ML} algorithms, particularly deep learning models, have achieved remarkable success in interpreting medical images. For instance, Google Health developed an \gls{ML} model for breast cancer screening that outperformed human radiologists in detecting cancer in mammograms \cite{mckinney2020international}. However, challenges remain in integrating these systems into clinical workflows, dealing with diverse data quality, and ensuring consistent performance across different populations and equipment.

	\item \textbf{Drug Discovery and Development:} \gls{ML} has accelerated the drug discovery process, reducing costs and time. Atomwise uses \gls{AI} to predict which molecules could lead to effective drugs, and in 2020, they used their platform to identify promising compounds for treating COVID-19~\cite{zhavoronkov2019deep}. Nevertheless, the primary challenge lies in validating \gls{AI}-discovered drugs in clinical trials, a process that is time-consuming and requires substantial investment.

	\item \textbf{Predictive Analytics in Patient Care:} \gls{ML} models are increasingly used for predictive analytics in patient care. An example is the use of \gls{ML} by a team at Johns Hopkins University, to predict sepsis in hospitalized patients, enabling early intervention~\cite{henry2015targeted}. However, the challenge here involves ensuring data privacy, overcoming data silos in healthcare settings, and dealing with the potential biases in the data used to train these models.

	\item \textbf{Personalized Medicine:} In personalized medicine, \gls{ML} aids in tailoring treatments to individual patients' genetic profiles. A team in Earle A. Chiles Research Institute uses \gls{ML} to analyze genetic data from cancer patients to identify the most effective treatment plans~\cite{piening2023improved}. Challenges include managing the vast amounts of genetic data, ensuring accurate interpretations of this data, and integrating these insights into routine clinical practice.

	\item \textbf{ Mental Health Applications:} \gls{ML} models are used for monitoring and diagnosing mental health conditions. For example, apps like Ginger.io use \gls{ML} algorithms to analyze user interaction and provide personalized mental health support~\cite{iyortsuun2023review}. Challenges include addressing privacy concerns, ensuring the sensitivity and specificity of the algorithms in diverse populations, and integrating these tools with traditional mental health services.

	\end{itemize}
\end{enumerate}	

\subsection{Federated Learning}\label{sec:federated-learning}
Within the field of \gls{ML}, data security and privacy are paramount concerns. Traditional \gls{ML} approaches often require centralized data storage, which can raise privacy issues and limit participation due to data ownership restrictions. \gls{FL} emerges as a groundbreaking solution, offering a decentralized paradigm for collaborative \gls{ML} \cite{konevcny2016federated}. In \gls{FL}, multiple entities, such as healthcare institutions or research centers, collaborate to train a model without sharing their raw data. Each entity trains a local model on its own data and shares only model updates, such as gradients or parameters, with a central server. This collaborative approach allows for distributed learning while preserving data privacy and security.\\

 \begin{enumerate}[wide, font=\itshape, labelwidth=!, labelindent=0pt, label*=\textit{D}.\arabic*.]
 
 \item \textit{Federated Learning Types} 
 
 \gls{FL} encompasses various implementations tailored to diverse scenarios and constraints. In the following, we delve deeper into the three prominent types of \gls{FL} implementations: horizontal \gls{FL}, vertical \gls{FL}, and federated transfer learning. Each type addresses specific challenges and offers unique advantages in \gls{FL} environments.

	\begin{itemize}
			\item \textbf{\gls{HFL}:} \gls{HFL} involves training models across multiple devices or clients that possess similar data distributions but are unable to share raw data due to privacy or regulatory concerns. In this approach, each client trains a local model using its data and shares only model updates, such as gradients or model parameters, with a central server \cite{yang2019federated}. The server aggregates these updates to refine a global model, which is then redistributed to clients for further refinement. This iterative process continues until convergence is achieved. Horizontal federated learning is particularly suitable for scenarios where data is distributed across devices or clients with similar characteristics, such as mobile phones in a collaborative learning setting, edge devices in \gls{IoT} networks \cite{zhang2023federated} or \gls{EEG} data for electroencephalography \cite{gao2019hhhfl}. Challenges in horizontal federated learning often revolve around privacy-preserving techniques, communication efficiency, and model aggregation strategies to ensure convergence and accuracy while preserving data privacy and security \cite{huang2022fairness}.
	
	\item \textbf{\gls{VFL}:} Vertical federated learning addresses scenarios where data is distributed across multiple parties with complementary features but cannot be directly shared due to privacy or proprietary concerns. Unlike horizontal federated learning, which focuses on data with similar distributions, vertical federated learning operates on data with different but complementary features \cite{liu2024vertical}.
	
	In vertical federated learning, each party holds a subset of features relevant to the learning task, and the goal is to collaboratively train a model without sharing raw data. \gls{SMPC} and homomorphic encryption techniques are commonly employed to enable computations on encrypted data while preserving privacy \cite{feng2024mmvfl}. These techniques allow parties to jointly compute model updates or predictions without revealing sensitive information about their data.
	
	Vertical federated learning finds applications in scenarios such as healthcare, where different institutions may hold complementary patient data (e.g., medical records, lab results) that are crucial for training accurate models while preserving patient privacy and data ownership \cite{gupta2024federated}.
	
	\item \textbf{\gls{FTL}:} Federated transfer learning extends traditional transfer learning to federated settings, where models are trained across multiple decentralized datasets to leverage knowledge from related tasks or domains \cite{liu2020secure}. Unlike traditional transfer learning, where a pre-trained model is fine-tuned on a target dataset, federated transfer learning involves aggregating knowledge from multiple decentralized datasets to improve model performance.
	
	In federated transfer learning, a base model is initialized either with a pre-trained model or from scratch, and model updates from decentralized clients are aggregated to refine the base model. This approach allows for leveraging knowledge from diverse data sources while accommodating variations in data distributions and characteristics across clients \cite{chen2020fedhealth}.
	
	Federated transfer learning is beneficial in scenarios where labeled data is scarce or unevenly distributed across clients such as electrocardiogram signal analysis \cite{chorney2024towards}, enabling collaborative model training while leveraging knowledge from related tasks or domains to improve model performance.\\

\begin{figure}[!h]
\begin{subfigure}[t]{\linewidth}
 \centering
 \resizebox{\linewidth}{!}{
 \includegraphics[width=\linewidth, trim=0 118 0 0,clip]{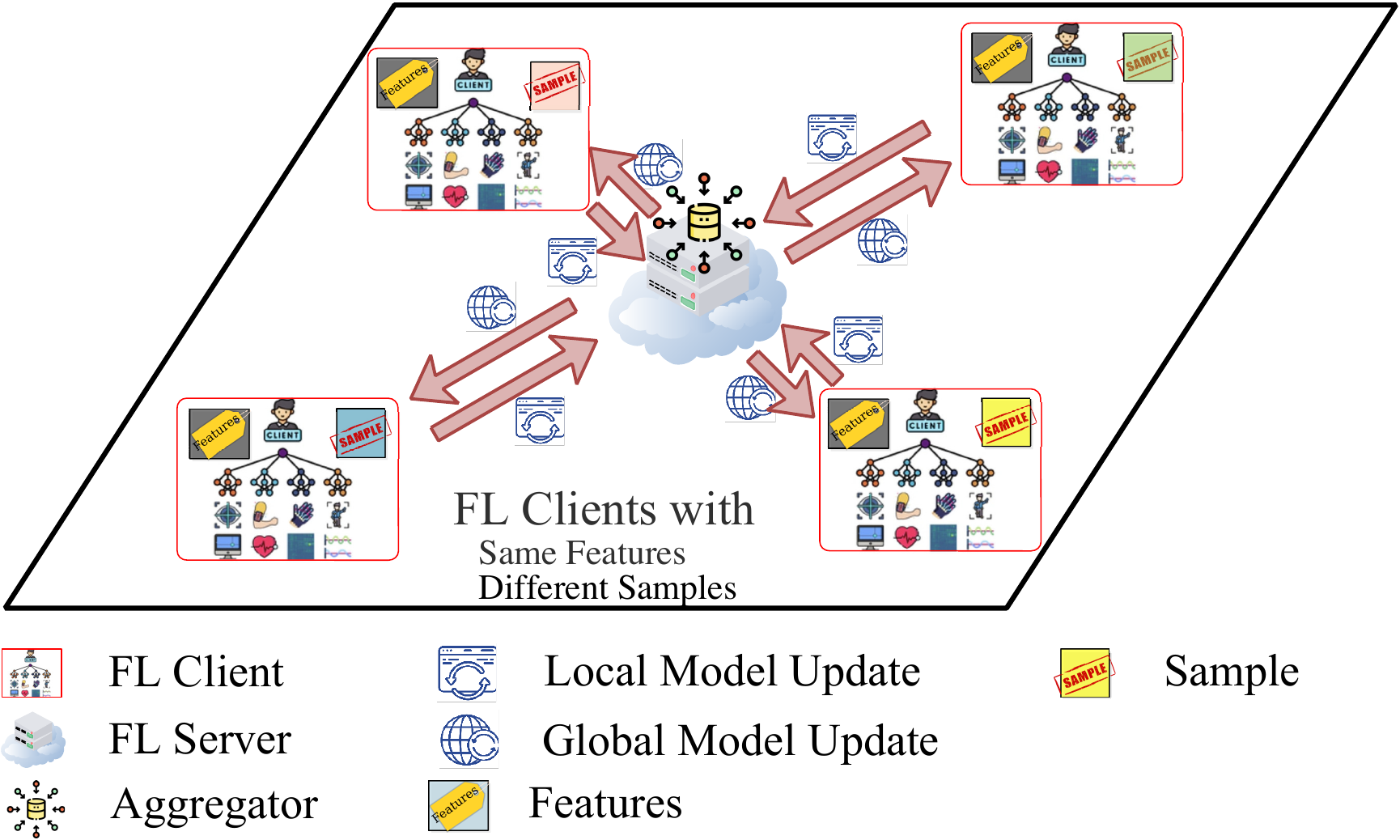}
 }
 \caption{HFL}
 \end{subfigure}%
 \hfill
 \begin{subfigure}[t]{\linewidth}
 \centering
 \resizebox{\linewidth}{!}{
 \includegraphics[width=\linewidth, trim=0 118 0 0,clip]{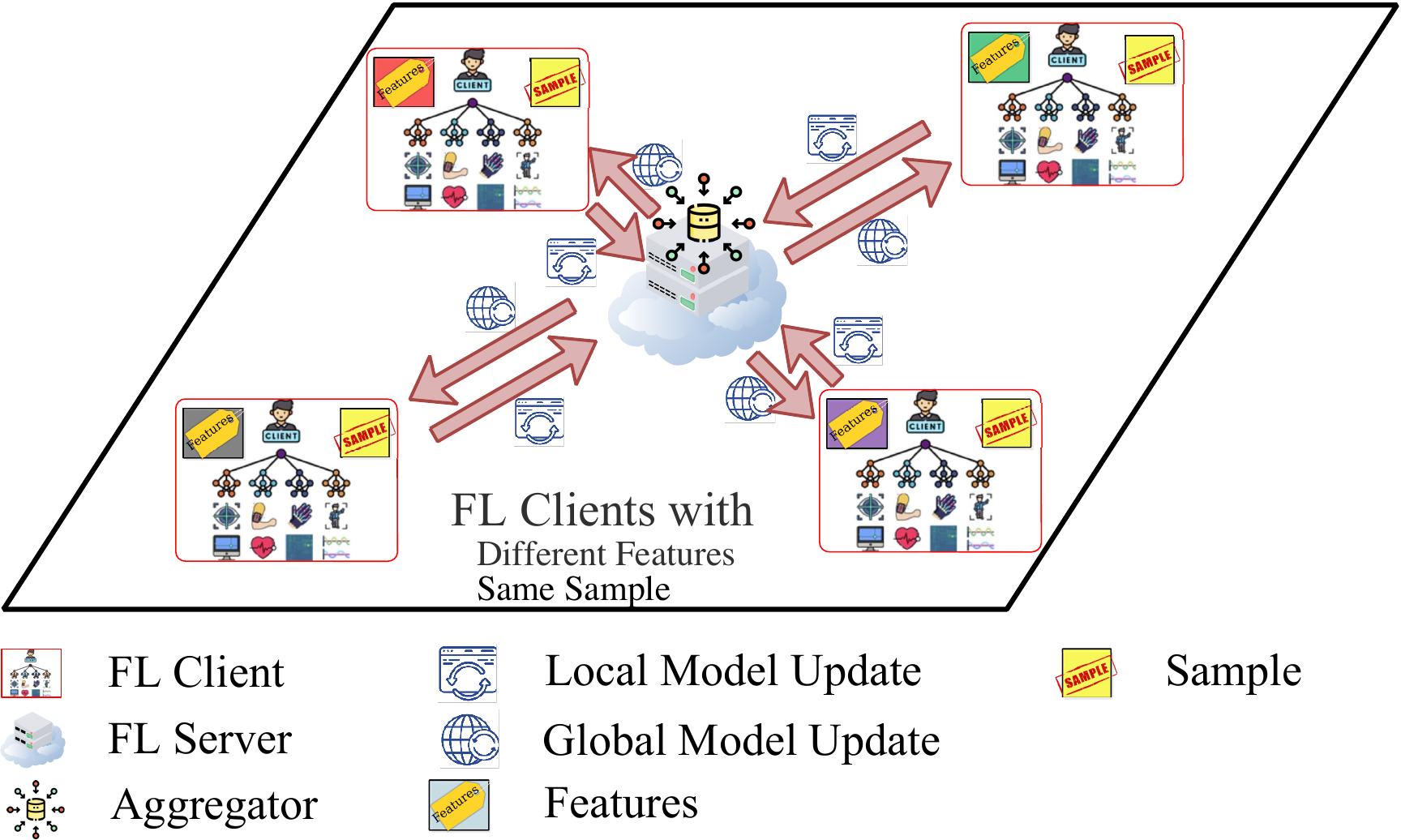}
 }
 \caption{VFL}
 \end{subfigure}%
 \hfill
 \begin{subfigure}[t]{\linewidth}
 \centering
 \resizebox{\linewidth}{!}{
 \includegraphics[width=\linewidth, trim=0 118 0 0,clip]{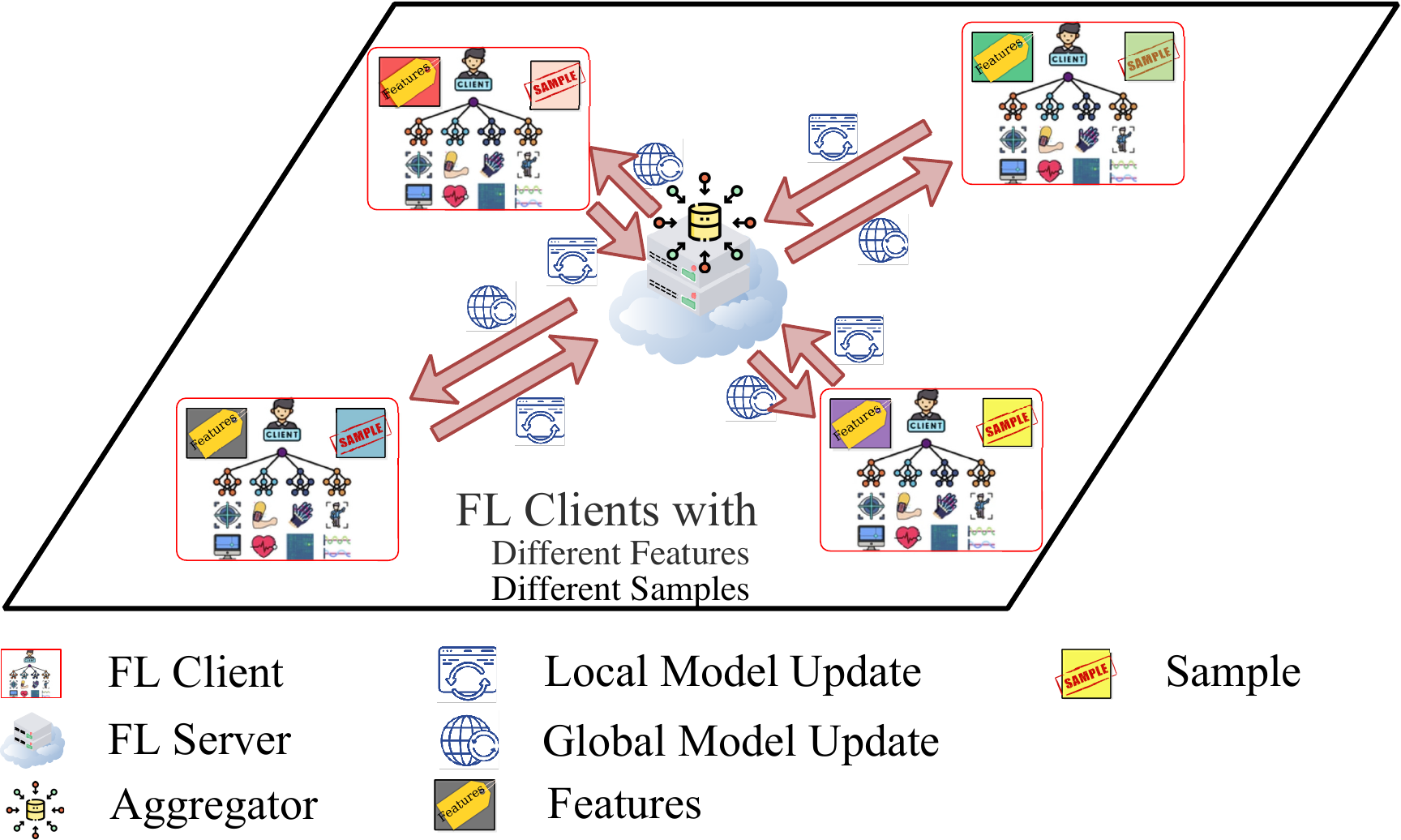}
 }
 \caption{FTL}
 \end{subfigure}%
 \hfill
 \begin{subfigure}[t]{\linewidth}
 \centering
 \resizebox{\linewidth}{!}{
 \includegraphics[width=\linewidth, trim=0 0 0 370,clip]{Figures/FTL.pdf}
 }
 \end{subfigure}%

\caption{Types of Federated Learning}

\label{fig:domain}
\end{figure}%
	\end{itemize}
	
\item \textit{Case Studies of Federated Learning in Healthcare}

\gls{FL} holds promise in healthcare by facilitating the collaborative development of robust \gls{ML} models across various institutions, such as hospitals and research centers, while safeguarding patient data privacy~\cite{rieke2020future, nguyen2022federated}. This approach entails training \gls{ML} models on local datasets, with only model updates shared, rather than raw data, to a central server for aggregation, thus mitigating privacy concerns and reducing data transfer costs~\cite{li2020federated}. \gls{FL} in healthcare offers several advantages over traditional centralized learning, including enhanced privacy and the utilization of diverse datasets without necessitating centralization. Below, we delve into real-world examples of \gls{FL} applications in the healthcare sector:

		\begin{itemize}
 \item \textbf{Multi-Institutional Collaboration for Disease Diagnosis:} One of the most significant successes of \gls{FL} in healthcare is observed in collaborative projects for disease diagnosis. For example, a consortium of international medical centers used \gls{FL} to develop models for predicting patient outcomes in critical care. By leveraging data from diverse populations while maintaining data privacy, these models achieved high accuracy in predicting outcomes such as mortality and length of hospital stay. One successful implementation is seen in the Rhino Health consortium collaboration, which involves prestigious institutions and experts from around the world. Participants in this collaboration include Massachusetts General Hospital (U.S.), University of Cambridge School of Medicine (UK), Lahey Hospital and Medical Center (U.S.), Assuta Medical Centers (Israel), Dasa S.A. (Brazil), National Taiwan University (Taiwan), and Seoul National University Hospital (South Korea). These collaborations underscore the global effort to harness the power of \gls{FL} in enhancing medical research and healthcare delivery while upholding data privacy and security.
		\item \textbf{Enhancing Drug Discovery and Development:} \gls{FL} has been used in pharmaceutical research to create predictive models for drug response and toxicity. A notable instance is a project where multiple pharmaceutical companies shared algorithmic models, not data, to predict the success of drug compounds, expediting the drug discovery process while maintaining the confidentiality of proprietary data \cite{heyndrickx2023melloddy}.
		
		\item \textbf{Collaborative Research:} In oncology, \gls{FL} enables institutions to collaborate on cancer research without sharing sensitive patient data. A notable example is the collaboration facilitated by Intel and Penn Medicine, utilizing \gls{FL} to identify brain tumors \cite{foley2022openfl}. These case studies demonstrate how \gls{FL} enables institutions to collaborate in advancing medical research while maintaining the confidentiality of their data.\\
		\end{itemize}	
\item \textit{Challenges of Federated Learning in Healthcare} 

\gls{FL} presents a groundbreaking opportunity in healthcare, facilitating the collaborative development of potent \gls{ML} models while safeguarding patient data privacy and minimizing data transfer costs. However, realizing the full potential of \gls{FL} requires overcoming significant challenges. Below, we delineate these challenges to underscore the complexities \gls{FL} encounters in healthcare:
		\begin{itemize}
		
		\item \textbf{Data Heterogeneity and Model Generalizability:} One of the primary challenges in \gls{FL} is the heterogeneity of medical data across different institutions having data in various formats and from diverse populations. This variability can lead to challenges in model convergence and performance, as models must be generalizable to diverse patient populations and data types.
		
		\item \textbf{Technical and Computational Limitations:} Implementing \gls{FL} requires significant computational resources and technical expertise, which may not be uniformly available across all participating institutions. Balancing these computational disparities is a challenge that needs to be addressed for the successful implementation of \gls{FL}.
		
		\item \textbf{Regulatory and Ethical Compliance:} \gls{FL} must navigate complex regulatory and ethical landscapes. This includes ensuring compliance with healthcare regulations such as \gls{HIPAA} in the US, and the \gls{GDPR} in Europe, which govern patient data privacy and security. Ethical considerations include ensuring that patient consent is obtained for the use of their data in such models and that the benefits of such research are equitably distributed. Addressing these considerations requires clear data governance policies and ethical frameworks for \gls{FL} deployment in healthcare.
		
		\item \textbf{Scalability and Real-Time Learning:} While \gls{FL} allows for collaborative model building, scaling these models to accommodate real-time learning and large datasets poses technical challenges. Efficiently managing and updating models with new data in real time remains a hurdle.
		
		\item \textbf{Future Prospects:} Looking forward, the integration of \gls{FL} with emerging technologies such as \gls{IoT} devices and real-time health monitoring systems holds great potential. This could enable the continuous improvement of \gls{ML} models with real-time data from diverse and distributed sources, leading to more dynamic and responsive healthcare solutions. Additionally, advancements in edge computing could further enhance the efficiency and scalability of \gls{FL} in healthcare.
		
		\item \textbf{Possible Solutions:} Addressing these challenges can be achieved through several approaches: firstly, by establishing standardized data formats across participating entities, ensuring consistency and compatibility in the data used for \gls{FL}. Secondly, by implementing comprehensive data preprocessing pipelines, which enhance the quality and usability of the data before it is fed into the learning models. Additionally, designing \gls{FL} algorithms that are adept at handling diverse data types and can operate effectively across varying computational resources is crucial.
		
		\end{itemize}
\end{enumerate}
\subsection{Current Trends and Future Directions in \gls{FL} for Healthcare}\label{sec:FL-threat}

The landscape of \gls{FL} in healthcare is rapidly evolving, driven by technological advancements and the growing need for collaborative and privacy-preserving data analysis. This section outlines the current trends shaping \gls{FL} in healthcare and forecasts its future trajectory.\\
	
 \begin{enumerate}[wide, font=\itshape, labelwidth=!, labelindent=0pt, label*=\textit{E}.\arabic*.]
 \item \textit{Current Trends in \gls{FL} for Healthcare} 
 
 The prevailing trends shaping \gls{FL} in healthcare reflect a dynamic evolution towards enhanced analytics, privacy, collaboration, and ethical considerations. These trends include:
\begin{itemize}
\item \textbf{Integration with Advanced Analytics and \gls{AI}:} \gls{FL} is increasingly being integrated with sophisticated \gls{AI} techniques, such as deep learning, to enhance its analytical capabilities \cite{malik2024federated, dasaradharami2023comprehensive}. This integration allows for more complex and accurate models, capable of addressing intricate healthcare challenges like personalized medicine and predictive analytics \cite{rieke2020future}.
\item \textbf{Emphasis on Data Privacy and Security:} In the wake of heightened concerns about data privacy, \gls{FL} is gaining traction as a preferred method for collaborative healthcare research \cite{kaissis2020secure}. Its inherent design, which allows for model training without sharing raw data, aligns well with stringent data privacy regulations like those of \gls{HIPAA} and \gls{GDPR}.
\item \textbf{Cross-Institutional Collaborations:} There is a growing trend of cross-institutional collaborations facilitated by \gls{FL} as is explained above. These collaborations can unite hospitals, research centers, and academic institutions, enabling them to pool their knowledge and data resources for collective model improvement while maintaining data sovereignty.
	
\item \textbf{Ethical and Fair \gls{AI} Development:} As \gls{FL} continues to evolve, there will be an increased focus on ethical \gls{AI} development. This includes ensuring that \gls{FL} models are fair, unbiased, and representative of all patient demographics, thereby addressing concerns around algorithmic bias \cite{rafi2024fairness, djebrouni2024bias, chen2024credible}.\\
	
\end{itemize}
	
\item \textit{Future Directions in \gls{FL} for Healthcare} 

\gls{FL} in healthcare stands on the brink of significant expansion, propelled by emerging technological advancements and evolving healthcare landscapes. This forthcoming evolution encompasses:
	
\begin{itemize}
	
\item \textbf{Expansion into Global Health Initiatives:} \gls{FL} has the potential to significantly impact global health research, particularly in areas with stringent privacy laws or limited data-sharing capabilities. It could facilitate the analysis of global health trends and the development of models that are representative of diverse populations \cite{rieke2020future}.
	
	\item \textbf{Automated and Dynamic Model Updating:} The future of \gls{FL} in healthcare might see more automated and dynamic updating of models \cite{you2024slmfed}. This would enable healthcare systems to respond quickly to new data or changing health trends, making \gls{FL} models more adaptive and responsive.
	
	\item \textbf{Use in Remote and Real-Time Monitoring:} With the proliferation of wearable devices and \gls{IoT} in healthcare, \gls{FL} is poised to play a significant role in real-time patient monitoring and remote healthcare, providing personalized insights and treatments based on data collected from diverse patient populations \cite{mazzocca2024enabling}.
	
	\item \textbf{Edge Computing Integration:} Integrating \gls{FL} with edge computing could decentralize the computational workload, allowing for faster and more efficient model training and updates, especially in real-time applications \cite{djebrouni2024bias, ji2024edge, elhattab2024pastel}.
		
	\item \textbf{Integration with Blockchain for Enhanced Security:} The integration of \gls{FL} with blockchain technology is a promising development, which forms the central theme of this tutorial. This combination not only bolsters data security but also adds a layer of transparency and traceability to the \gls{FL} process, ensuring immutable record-keeping and verifiable model updates in \gls{FL} networks \cite{myrzashova2023blockchain}.\\
	
	\end{itemize}
	
	In summary, \gls{FL} in healthcare is at a dynamic juncture, with its trajectory poised to reshape healthcare research and delivery. Its alignment with current needs for privacy, collaboration, and advanced analytics, coupled with its adaptability for future technological trends, positions \gls{FL} as a key player in the future landscape of healthcare technology. The ongoing advancements in \gls{FL} are not just technological but also pave the way for more equitable, secure, and efficient use of healthcare data globally.
	
\end{enumerate}

\section{Blockchain} \label{sec:MedBC}
\gls{BC} technology has undergone a remarkable evolution since its inception in 2009 with the creation of Bitcoin by an individual or group using the pseudonym Satoshi Nakamoto~\cite{bitcoin2008bitcoin}. The primary purpose of Bitcoin was to establish a decentralized digital currency, and the innovation that made this possible was the blockchain—a distributed ledger that records transactions across a network of computers securely and transparently. 

In the following years, the potential applications of blockchain technology expanded beyond cryptocurrency. Vitalik Buterin introduced Ethereum~\cite{wood2014ethereum} in 2015, introducing the concept of smart contracts—self-executing contracts with the terms of the agreement directly written into code. This development opened up a broader spectrum of \gls{DApps} and laid the foundation for blockchain's role in facilitating not only \gls{P2P} transactions but also complex programmable interactions.

The years that followed witnessed a surge in blockchain projects and platforms, each aiming to address specific challenges across various industries. The technology gained recognition for its potential to enhance transparency, security, and efficiency. Consortia and collaborations emerged, with enterprises exploring how blockchain could optimize supply chains~\cite{azzi2019power,dutta2020blockchain,casado2018blockchain,korpela2017digital,queiroz2020blockchain}, streamline financial transactions, and enhance data integrity.

To the best of our knowledge in January 2024, blockchain continues to evolve, with ongoing efforts to address scalability issues, energy consumption concerns, and regulatory considerations. From its humble beginnings as the underlying technology for Bitcoin, blockchain has grown into a versatile tool with the potential to reshape how industries manage and verify data. The technology's journey reflects an ongoing quest for innovative solutions to long-standing challenges in the digital realm.

In this section, we provide a concise overview of fundamental concepts, features, structure, and taxonomy within the realm of blockchain technology. 


\subsection{Blockchain Technology: An Overview}

\gls{BC} technology is a decentralized and distributed ledger system designed to facilitate secure and transparent transactions without the need for a central authority. At its core, a blockchain consists of a chain of blocks, each containing a list of transactions. These blocks are linked together in a chronological and immutable manner, forming a continuous chain. One of the key features of blockchain is its decentralization, meaning that the ledger is maintained by a network of nodes rather than a single central entity. This distributed nature enhances security, reduces the risk of fraud, and ensures transparency in the transaction process.


\gls{BC} technology possesses several distinctive features that contribute to its popularity and versatility across various industries. We summarize the features of blockchain technology as follows: 

\begin{itemize}
 \item \textbf{Decentralization:} blockchain operates on a \gls{P2P} network. This fact eliminates the need for a central authority or intermediary. Decentralization enhances security, reduces the risk of a single point of failure, and promotes trust among mutually untrusted participants~\cite{raval2016decentralized}.
 \item \textbf{Immutability:} Immutability is the capability of a blockchain ledger to remain unchanged. once a block is added to the blockchain, it becomes virtually impossible to alter or delete the information within it. Immutability ensures the integrity of the transaction history and builds trust in the accuracy of recorded data.
 \item \textbf{Transparency:} means the entire transaction history is visible to all participants in the network. Transparency fosters trust and accountability as participants can independently verify transactions and the state of the blockchain.
 \item \textbf{Security:} blockchain employs cryptographic techniques to secure transactions and control access to the network. Consensus mechanisms, such as \gls{PoW} or \gls{PoS}, enhance security by preventing unauthorized changes to the blockchain~\cite{joshi2018survey,mohanta2019blockchain,hassan2022anomaly}. 
 \item \textbf{Distributed Ledger:} The ledger is distributed among the nodes over the network and each node in the network holds a copy of the blockchain. This distribution ensures redundancy, resilience, and a shared source of trust among the participants. 
 \item \textbf{Consensus Mechanisms:} Consensus is a mechanism that gives the ability to the network to agree upon the validity of transactions (and blocks) and the order in which they can be added to the blockchain. 
 \item \textbf{Anonymity and Privacy:} While all the transactions in the blockchain network are transparent, participants will remain pseudonymous due to the use of public/private key pairs. 
 \item \textbf{Efficiency and Speed:} \gls{BC} reduces the need for intermediaries and manual processes, leading to faster and more efficient transactions. In some cases, however, the speed of transactions may depend on the specific consensus mechanism employed. For instance, \gls{PoW} blockchains, like Bitcoin, tend to be slower (i.e. with less throughput) compared to traditional payment systems like Visa and Mastercard. The primary reason for this is the inefficiency of the underlying consensus mechanisms and the way transactions are processed~\cite{shahsavari2020theoretical}.
 \item \textbf{Interoperability:} \gls{BC} interoperability refers to the capacity of various blockchain networks to interact seamlessly, facilitating the exchange of messages, data, and tokens among them.~\cite{belchior2021survey,lafourcade2020blockchain,schulte2019towards,hardjono2019toward,zhang2017applying}. Standards and protocols are evolving to enable such a communication and data exchange between disparate blockchain platforms. 
The \gls{IBC} protocol~\cite{qasse2019inter} is designed to facilitate this interoperability by providing a standardized way for independent blockchains to transfer and communicate with each other.
 \item \textbf{Ability to Support Smart Contracts:} Smart contracts are self-executing contracts with the terms directly written into code. These contracts automate and enforce predefined rules and agreements, reducing the need for intermediaries and streamlining processes~\cite{zou2019smart,delmolino2016step}. 
 
\end{itemize}

\subsection{Consensus in Blockchain} 
Consensus in the context of blockchain refers to the mechanism by which a distributed network of nodes agrees on the state of the system or the validity of transactions~\cite{xu2023survey,nguyen2018survey,monrat2019survey}. Since blockchain operates in a decentralized and trustless environment, consensus is crucial to ensure that all participants have a consistent view of the blockchain's history and current state. The consensus mechanism is responsible for preventing double-spending (where the same digital asset is spent more than once) and maintaining the integrity of the blockchain. Different blockchain networks employ various consensus algorithms, each with its own set of rules and processes. The most prominent and currently existing consensus protocols are as follows:\\

 \begin{enumerate}[wide, font=\itshape, labelwidth=!, labelindent=0pt, label*=B.\arabic*.]
 \item \textit{Proof of Work} 
 
 This is the original consensus algorithm used by Bitcoin~\cite{bitcoin2008bitcoin} and many more cryptocurrencies. In \gls{PoW}, participants (miners) solve complex mathematical puzzles to validate transactions and create new blocks. The first miner to solve the puzzle gets the right to add a new block to the blockchain. \gls{PoW} is resource-intensive and requires a significant amount of computational power~\cite{gervais2016security}. \gls{BC}s that rely on \gls{PoW} are more prone to forks. A fork in blockchain technology refers to a split in the blockchain's transaction history, resulting in two or more separate paths. This can occur for various reasons, such as changes in the protocol rules, disagreements among participants, or software upgrades~\cite{shahsavari2019theoretical}. \\

 \item \textit{Proof of Stake} 
 
 In \gls{PoS}~\cite{nguyen2019proof}, validators (i.e. block proposer participants) are chosen to create new blocks based on the amount of cryptocurrency they hold and are willing to "stake" as collateral. This eliminates the need for energy-intensive mining and aims to provide a more energy-efficient alternative to \gls{PoW}. Participants are chosen to create new blocks and validate transactions based on the amount of cryptocurrency they hold and are willing to stake as collateral. Examples include Ethereum's~\cite{buterin2014next} transition to Ethereum 2.0, Cardano~\cite{cardanodocs}, and Algorand~\cite{gilad2017algorand}. \gls{DPoS} is an improvement over traditional \gls{PoS} in terms of scalability and efficiency~\cite{wang2020revisiting}.\\

 


 \item \textit{Byzantine Fault Tolerance}
 
 \gls{BFT} consensus algorithms are a class of protocols designed to achieve consensus in distributed systems, even in the presence of faulty or malicious nodes. In a \gls{BFT-based consensus} algorithm, a network of nodes collaborates to agree on the state of the system or the validity of transactions.
The term "Byzantine Fault" originates from the Byzantine Generals' Problem~\cite{lamport2019byzantine}, a theoretical scenario where a group of generals must come to a unanimous agreement on a coordinated action, despite the possibility of some generals being traitors and sending conflicting messages. \gls{pBFT} is the most prominent variant of \gls{BFT-based consensus} protocols~\cite{castro1999practical}. This algorithm is designed to tolerate up to one-third of the total number of nodes being faulty or malicious. This means that as long as no more than one-third of the nodes in the network exhibit Byzantine behavior (i.e., they may fail arbitrarily or behave maliciously), \gls{pBFT} can still reach consensus and continue to operate correctly. Some variants of \gls{BFT} (e.g. \gls{pBFT} and \gls{IBFT}) are supported by Hyperledger Fabric~\cite{androulaki2018hyperledger}. \\

 

\item \textit{Proof of Capacity} 

\gls{PoC} is a consensus mechanism used in blockchain networks as an alternative to \gls{PoW} and \gls{PoS}. In \gls{PoC}, the ability to mine or validate transactions is determined by the amount of storage space (capacity) a participant allocates rather than computational power or stake in the network's native cryptocurrency. Participants, also known as miners or validators, allocate a certain amount of their available storage space to participate in the consensus process, often precomputing and dedicating this storage space solely for mining purposes. They generate plots, which are essentially pre-computed sets of data representing potential solutions to cryptographic puzzles, and store these plots in the allocated storage space. When a new block needs to be added to the blockchain, miners search their precomputed plots for solutions to a cryptographic puzzle, and the miner who finds a valid solution first is eligible to add the new block to the blockchain and is rewarded accordingly. To verify the validity of a block generated through \gls{PoC}, other nodes in the network check that the solution provided by the miner corresponds to the pre-computed data in their plots, ensuring consensus. \gls{PoC} offers advantages such as energy efficiency, decentralization, and fair reward distribution, as it consumes less electricity, allows for a more decentralized network without the need for expensive specialized hardware, and distributes block rewards more evenly among participants based on the amount of storage space allocated. However, challenges include the initial investment of time and resources required for generating plots and allocating storage space, as well as the significant storage space requirements, which may limit participation for some individuals or entities with limited resources. Overall, \gls{PoC} presents an alternative approach to achieving consensus in blockchain networks, emphasizing energy efficiency, decentralization, and fair reward distribution. A more energy-efficient variant of \gls{PoC} called \gls{PoST} is being used by Chia~\cite{cohen2019chia}.\\ 

\item \textit{Direct Acyclic Graph Tangle} 

\gls{DAG} consensus protocols are a class of distributed consensus algorithms that use a data structure called a directed acyclic graph to achieve agreement on the order of transactions or events in a decentralized network. Unlike traditional blockchain-based consensus protocols where transactions are organized into linear blocks, \gls{DAG}-based protocols organize transactions in a more flexible graph structure~\cite{li2020direct}.

One of the most well-known implementations of \gls{DAG} consensus is the Tangle~\cite{popov2018tangle}, which is used in the IOTA cryptocurrency network~\cite{saa2023iota}. In the Tangle, each transaction directly references and approves two previous transactions, forming a directed acyclic graph structure. The most prominent blockchains that run on proof of capacity include Signum, Chia, and SpaceMint. \\

\item \textit{Proof of Burn} 

\gls{PoB} is a consensus mechanism utilized in which the participants, also known as burners, demonstrate their commitment to the network by sending cryptocurrency tokens to a verifiably unspendable address, termed the "burn address," effectively destroying or burning the tokens~\cite{karantias2020proof}. This process is integral to \gls{PoB}, as it signifies participants' investment in the network. Following the token burning, participants provide evidence, or proof of the burned tokens, typically recorded transparently and verifiably on the blockchain. This proof serves to validate the burn event and participants' commitment to the network. In some \gls{PoB} systems, participants demonstrating proof of burn may be eligible for rewards or incentives, which can include newly minted tokens, voting rights, or other benefits within the blockchain ecosystem. \gls{PoB} offers several key advantages, including the allocation of resources by participants as a form of investment or commitment, leading to a fairer distribution of tokens and resistance against Sybil attacks~\cite{douceur2002sybil} due to the deterrent effect of burning tokens. However, \gls{PoB} also presents challenges such as potential token scarcity resulting from the reduction in the overall token supply, as well as economic considerations for participants regarding the permanent loss of token value. Overall, Proof of Burn represents an innovative consensus mechanism that emphasizes commitment and resource allocation, offering advantages such as fair distribution and Sybil resistance, albeit accompanied by challenges related to token scarcity and economic implications.\\

\item \textit{Hybrid Consensus Protocols} 

Hybrid consensus models in blockchain combine elements of multiple traditional consensus mechanisms to leverage their respective strengths and mitigate their weaknesses. These models aim to achieve a balance between decentralization, security, scalability, and energy efficiency~\cite{pass2016hybrid}. Here are some examples of hybrid consensus models in blockchain:\\
\item \textit{Proof of Stake and Proof of Work Hybrid} 

Some blockchain networks combine \gls{PoS} and \gls{PoW} mechanisms to achieve consensus (e.g. TwinsCoin~\cite{chepurnoy2017twinscoin}). For example, a \gls{PoW} component may be used for initial block creation, while \gls{PoS} is utilized for subsequent block validation or as a way to elect validators. This hybrid approach aims to maintain security through \gls{PoW} while improving scalability and energy efficiency with \gls{PoS}.\\

\item \textit{Proof of Authority and Proof of Work Hybrid} 

In this hybrid model, a network may utilize \gls{PoW} for initial block creation and \gls{PoA} for block validation \gls{PoW} ensures the initial distribution of tokens and secures the network against Sybil attacks, while \gls{PoA} provides fast finality and scalability by relying on known and trusted validators.\\

\item \textit{Delegated Proof of Stake  and Proof of Authority Hybrid}

\gls{DPoS} allows token holders to vote for a limited number of delegates who are responsible for block validation. In a hybrid approach, \gls{DPoS} can be combined with \gls{PoA}, where the initial set of validators is determined through \gls{PoA}, and then token holders can vote for additional delegates using \gls{DPoS}. This hybrid model aims to achieve both decentralization and scalability.\\

\item \textit{Proof of Work and Byzantine Fault Tolerance Hybrid} 

This hybrid model combines the energy-intensive \gls{PoW} with a \gls{BFT-based consensus} algorithm such as \gls{pBFT} or Tendermint~\cite{kwon2014tendermint}. \gls{PoW} is used for block creation, while \gls{BFT} consensus ensures finality and Byzantine fault tolerance. This approach aims to achieve both security and efficiency in blockchain networks~\cite{cheng2018new}.\\

\item \textit{Hybrid Voting Systems}

Some blockchain networks combine different voting mechanisms, such as direct voting by token holders and voting by elected delegates. This hybrid voting system aims to balance the influence of token holders with the expertise and accountability of elected representatives.
\end{enumerate}

\subsection{Blockchain Data Structure}
In blockchain technology, the data structure plays a pivotal role in ensuring the integrity, security, and immutability of the distributed ledger. At its core, a blockchain is composed of a series of blocks, each containing a bundle of transactions. These blocks are cryptographically linked together sequentially, forming a continuous chain. The data structure of a block typically includes several key components: a header, a list of transactions, and a cryptographic hash. The header contains metadata such as the block's unique identifier (block hash), a timestamp, and a reference to the previous block's hash, thus establishing the chronological order of blocks. The list of transactions records the details of all transactions included in the block, such as sender and receiver addresses, transaction amounts, and cryptographic signatures for verification. Additionally, each block is assigned a cryptographic hash, computed based on its contents using a hashing algorithm like SHA-256. This hash serves as a unique identifier for the block and is crucial for maintaining the integrity of the blockchain. Any alteration to the data within a block would result in a change in its hash, thereby breaking the chain's continuity and signaling tampering. This inherent immutability and tamper-resistance of the data structure in blockchain ensure that once recorded, transactions cannot be altered or deleted without consensus from the network participants, establishing a reliable and transparent system for recording and verifying transactions~\cite{dinh2018untangling}.

\subsection{Blockchain Network and Architecture}
The network architecture in blockchain is a distributed and decentralized system that enables the secure and transparent exchange of data and value across a network of interconnected nodes. At its core, blockchain operates as a  \gls{P2P} network where each participant, or node, maintains a copy of the entire blockchain ledger. This distributed architecture ensures that there is no single point of failure, as the data is replicated and synchronized across multiple nodes. Nodes communicate with each other through a consensus mechanism that we already explained. Depending on the consensus algorithm employed, nodes may take on different roles, such as miners in \gls{PoW} or validators in \gls{PoS} systems. Transactions are broadcasted to the network and validated by consensus, typically requiring confirmation from a majority of nodes before being added to the blockchain. This network architecture provides several benefits, including resilience against censorship and tampering, increased transparency and accountability, and enhanced security through cryptographic techniques. Additionally, the decentralized nature of blockchain networks promotes trust among participants by eliminating the need for intermediaries and central authorities, thereby fostering a more inclusive and democratic ecosystem for conducting transactions and exchanging value~\cite{decker2013information}.

\gls{BC} technology typically consists of several layers, each serving a specific purpose in the functioning and security of the blockchain network. The most common layers include:

\begin{enumerate}
 \item \textit{Network Layer:} The network layer serves as the foundation of the blockchain system, facilitating communication between nodes (individual computers) in the network. It is responsible for transmitting data, such as transactions and blocks, across the network using protocols like TCP/IP, HTTP, and \gls{P2P} protocols.
 \item \textit{Data Layer:} The data layer stores the actual blockchain data, including blocks, transactions, smart contracts, and other relevant information. It includes data structures, databases, and storage mechanisms optimized for storing and retrieving blockchain data in a secure and efficient manner.
 \item \textit{Consensus Layer:} The consensus layer ensures that all nodes in the network agree on the validity of transactions and the order in which they are added to the blockchain. Different blockchain networks employ various consensus mechanisms, such as \gls{PoW}, \gls{PoS}, \gls{DPoS}, and Practical \gls{pBFT}, to achieve agreement among participants.
 \item \textit{Smart Contract Layer:} Smart contracts are self-executing contracts with terms directly written into code. This layer enables the creation and execution of programmable contracts, powering various \gls{DApps}. Smart contract platforms like Ethereum provide the infrastructure for developers to build applications ranging from \gls{DeFi} to supply chain management.
 \item \textit{Incentive Layer:} The incentive layer provides mechanisms to incentivize participants, such as miners or validators, to contribute resources and maintain the security and integrity of the blockchain network. Incentives typically include block rewards and transaction fees, which compensate participants for their contributions. Examples include Bitcoin's block rewards and Ethereum's gas fees.
 \item \textit{Application Layer:} This layer encompasses the user-facing applications and interfaces that interact with the blockchain network. It includes \gls{DApps}, wallets, smart contracts, and other software built on top of the blockchain protocol.

\end{enumerate}
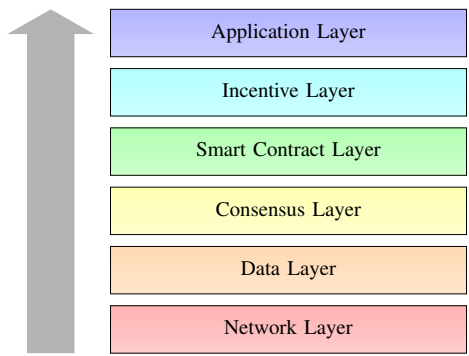
\begin{figure}[!ht]
\centering
\resizebox{0.7\linewidth}{!}{
\begin{tikzpicture}[
 body1/.style={draw, minimum width=6cm, minimum height=0.8cm, font={\normalsize}},
 typetag/.style={rectangle, draw=black!100, anchor=west}
 ]
 \draw [thick arrow]
 (-1,-5.4) -- (-1, 0.4) [set mark={\scriptsize{ }},color=gray!60];
 \node (d1) [body1, typetag, top color=blue!30, bottom color=blue!20] {\centering Application Layer};
 \node (d2) [body1, below=of d1.west, top color=cyan!30, bottom color=cyan!20 , typetag] {\begin{minipage}[c]{5.00cm}
\centering Incentive Layer\end{minipage}};
 \node (d3) [body1, below=of d2.west, typetag, top color=green!30, bottom color=green!20] {\begin{minipage}[c]{5.00cm}\centering Smart Contract Layer\end{minipage}};
 \node (d4) [body1, below=of d3.west, typetag, top color=yellow!30, bottom color=yellow!20] {\begin{minipage}[c]{5.00cm}\centering Consensus Layer\end{minipage}};
 \node (d5) [body1, below=of d4.west, typetag, top color=orange!30, bottom color=orange!20] {\begin{minipage}[c]{5.00cm}\centering Data Layer\end{minipage}};
 \node (d6) [body1, below=of d5.west, typetag, top color=red!30, bottom color=red!20] {\begin{minipage}[c]{5.00cm}\centering Network Layer\end{minipage}};
\end{tikzpicture}
}
\caption{Blockchain Layers}
\label{fig:Blockchain Layers}
\end{figure}

\subsection{Inter-Blockchain Communication (IBC) Protocol}
Interoperability is one of the most important features of the next-generation blockchain networks and refers to the ability of different blockchain platforms to communicate, share data, and transact with each other seamlessly. It enables interoperability between disparate blockchain networks, allowing them to interact and exchange information or assets without the need for intermediaries or centralized exchanges. Interoperability is essential for realizing the full potential of blockchain technology by facilitating cross-chain transactions, asset transfers, and data sharing between different blockchain ecosystems~\cite{qasse2019inter}.

The \gls{IBC} protocol is a set of standards and protocols designed to enable communication and interoperability between independent blockchain networks. \gls{IBC} facilitates the secure and trustless transfer of assets and data across different blockchains, allowing them to interact and transact with each other directly. The protocol defines a standardized messaging format and a set of rules for validating and verifying transactions between participating blockchains. By implementing the \gls{IBC} protocol, blockchain networks can establish interconnectivity, enabling cross-chain transactions, decentralized exchanges, and interoperable \gls{DApps}~\cite{kan2018multiple,chen2017inter,schulte2019towards}. 

\gls{IBC} is one of the pillars of the \gls{IoB}~\cite{vo2018internet,zarrin2021blockchain}. The concept of \gls{IoB} refers to a vision where blockchain networks are interconnected in a similar way to how the Internet connects various computer networks globally. \gls{IoB} aims to create a decentralized and interoperable network of blockchains, where different blockchain platforms can seamlessly communicate and transact with each other, just as different devices and systems are connected over the internet. The \gls{IoB} enables a decentralized and open ecosystem where data, assets, and services can flow freely between different blockchains, unlocking new possibilities for innovation and collaboration. Examples of interoperability solutions and projects in the blockchain space include:
\begin{enumerate}
 \item \textit{Cosmos:} Cosmos is a decentralized network of interconnected blockchains that utilize the \gls{IBC} protocol to enable communication and interoperability between different blockchain platforms~\cite{kwon2019cosmos}. Cosmos Hub serves as the primary hub for connecting various blockchains within the Cosmos ecosystem, allowing them to transfer assets and data securely and efficiently.

 \item \textit{Polkadot:} Polkadot is a multi-chain blockchain platform that enables interoperability between different parachains (parallel blockchains) within its network. Polkadot's relay chain facilitates communication and interoperability between parachains, allowing them to share data and assets and interact with each other seamlessly~\cite{wood2016polkadot}.

 \item \textit{Wanchain:} Wanchain is a cross-chain blockchain platform that focuses on interoperability and connecting different blockchain networks. Wanchain's interoperability protocol enables the secure and decentralized exchange of assets between different blockchains, including Bitcoin, Ethereum, and others~\cite{wanchain}.
\end{enumerate}
\subsection{Blockchain Taxonomy} At a high level, blockchain networks are classified into three main categories: private, public, and consortium blockchains. We briefly explain each of them as follows.\\

 \begin{enumerate}[wide, font=\itshape, labelwidth=!, labelindent=0pt, label*=F.\arabic*.]
 \item \textit{Private Blockchain} 
 
 A private blockchain is a permissioned blockchain network where access and participation are restricted to authorized entities only. These entities typically have known identities and are granted permission to join the network by a central authority or administrator. Private blockchains are often used by enterprises and organizations to build internal blockchain solutions for specific use cases such as supply chain management, document verification, or intercompany transactions. They offer enhanced privacy, control, and scalability compared to public blockchains~\cite{dinh2017blockbench,pahlajani2019survey,li2017towards,gramoli2016danger}. For instance, Hyperledger Fabric is a private blockchain framework developed by the Linux Foundation's Hyperledger project~\cite{androulaki2018hyperledger,cachin2016architecture}. It is designed for enterprise use cases and enables organizations to create permissioned blockchain networks with customizable features and governance models.\\

 \item \textit{Public Blockchain}
 
 A public blockchain is a permissionless blockchain network that is open to anyone to join, participate, and transact without requiring permission or identification. Public blockchains are decentralized networks where transactions are transparent, immutable, and verifiable by anyone. They offer high levels of transparency, censorship resistance, and security but may sacrifice scalability and privacy due to their open nature~\cite{irresberger2021public,ferdous2021survey,benhamouda2020can,rebello2024survey}. Public blockchains are often used for cryptocurrencies, \gls{DApps}, and tokenized assets. For instance, Bitcoin is the first and most well-known public blockchain, created by an anonymous entity or group of individuals using the pseudonym Satoshi Nakamoto. It operates as a decentralized peer-to-peer network for sending and receiving the Bitcoin cryptocurrency.\\

 \item \textit{Consortium Blockchain} 
 
 A consortium blockchain is a semi-decentralized blockchain network governed by a consortium or group of organizations rather than a single centralized entity. Consortium blockchains are permissioned networks where the consensus process and governance are shared among a predefined set of participants. Consortium blockchains are commonly used in industries or sectors where multiple organizations collaborate on shared processes or infrastructure while still maintaining some level of control and privacy~\cite{dib2018consortium,li2017consortium,yao2021survey,zhang2018towards}. They offer a balance between the decentralization of public blockchains and the control of private blockchains. for instance, R3 Corda is a consortium blockchain platform developed by the enterprise blockchain consortium R3. Corda is designed for use cases that require privacy, scalability, and interoperability in multiple organizations in sectors such as finance, healthcare, and supply chain.
\end{enumerate}

\section{Integration of Federated Learning with Blockchain}\label{sec:MedIngr}
Integrating blockchain technology with \gls{FL} has emerged as a novel approach to address inherent data privacy, security, and trust challenges within distributed \gls{ML} systems~\cite{qammar2023securing,li2022blockchain,qu2022blockchain}. \gls{FL}, characterized by training \gls{ML} models across decentralized devices without centrally aggregating raw data, offers significant advantages in preserving user privacy and data confidentiality. In other words, in federated \gls{ML}, the exchange occurs at the parameters level instead of transmitting raw data. This approach mitigates the risk associated with centralized computing architectures, which are susceptible to targeted attacks and potential denial of service due to their single-point-of-failure vulnerability. A detailed discussion on \gls{FL} is presented in Section~\ref{sec:federated-learning}, including its different types (horizontal, vertical, federated transfer learning), applications (particularly in healthcare), and potential future directions (integration with advanced analytics, cross-institutional collaborations), highlighting \gls{FL}'s pivotal role in reshaping healthcare technology.

However, concerns persist regarding the integrity of \gls{FL} systems, particularly regarding data tampering or manipulation by malicious or compromised nodes. \gls{BC}, renowned for its immutable and transparent ledger capabilities, presents a compelling solution to these challenges. By leveraging blockchain's decentralized consensus mechanisms and cryptographic primitives, \gls{FL} systems can ensure data integrity, traceability, and transparency throughout the \gls{ML} model training process ~\cite{issa2023blockchain,passerat2020blockchain,aich2022protecting}. Moreover, blockchain's smart contract functionality enables the establishment of auditable and self-executing agreements among participants, further enhancing the trustworthiness of \gls{FL} collaborations~\cite{qammar2023securing}. This integration not only addresses privacy and security concerns but also fosters a more collaborative and inclusive environment for distributed \gls{ML} research and applications. As such, the motivation behind the integration of blockchain and \gls{FL} lies in the pursuit of enhancing data privacy, security, and trust in decentralized \gls{ML} ecosystems, ultimately advancing the adoption and efficacy of \gls{FL} methodologies in various domains such as healthcare~\cite{myrzashova2023blockchain,el2020blockchain,aich2022protecting}.

\gls{BC} technology holds immense potential to significantly enhance the security, transparency, and trustworthiness of \gls{FL} systems. By addressing key challenges like verifying local model updates, aggregating the global model, and incentivizing participants, blockchain can pave the way for a more secure and collaborative future in \gls{AI} development. In the following, we explore how blockchain technology can significantly enhance these aspects of \gls{FL}.

\begin{itemize}
 \item \textbf{Verifying Local Model Updates:} \gls{BC} offers an immutable and tamper-proof record of transactions, acting as a secure ledger for local model updates from participants. These updates can be recorded as transactions, ensuring their validity and preventing any unauthorized modifications. Additionally, smart contracts can be employed to define specific rules for model updates, functioning as verification mechanisms. These smart contracts can even execute algorithms to validate the integrity and accuracy of the updates~\cite{kim2019blockchain}.

 \item \textbf{Global Model Aggregation:} \gls{BC} empowers \gls{FL} with decentralized consensus mechanisms, such as \gls{PoW} or \gls{PoS}. These algorithms enable participants to collectively agree on the process of aggregating the global model. Furthermore, the entire aggregation process can be transparently recorded on the blockchain. This transparency allows participants to verify the fairness and accuracy of the final model~\cite{lu2020communication}.

 \item \textbf{Incentivizing Participants:} One of the key strengths of blockchain lies in its ability to create tokenized incentives. These tokens or cryptocurrencies can be used to reward participants who contribute data or computational resources to the \gls{FL} process \cite{pandey2209fedtoken}. Smart contracts further enhance this system by automating the distribution of incentives based on predefined criteria. For instance, the quality of contributions or the amount of computational resources provided by a participant could determine the reward~\cite{behera2021federated}. Importantly, the entire distribution process remains transparently recorded on the blockchain, ensuring accountability and fairness~\cite{ma2021transparent}.
 
\end{itemize}

By leveraging the unique strengths of blockchain technology, \gls{FL} systems can achieve a new level of security, transparency, and trust, ultimately fostering a more collaborative and efficient environment for \gls{AI} development.

\subsection{Integration Architecture}
Several research efforts have identified distinct architectures for integrating blockchain and \gls{FL}. This paper proposes a similar framework to~\cite{wang2021blockchain} with slight variations in terminology. Based on the level of interaction between blockchain and \gls{FL} entities, we categorize these architectures as fully coupled, semi-coupled, and loosely coupled.\\

\begin{figure}
			\includegraphics[width=\linewidth]{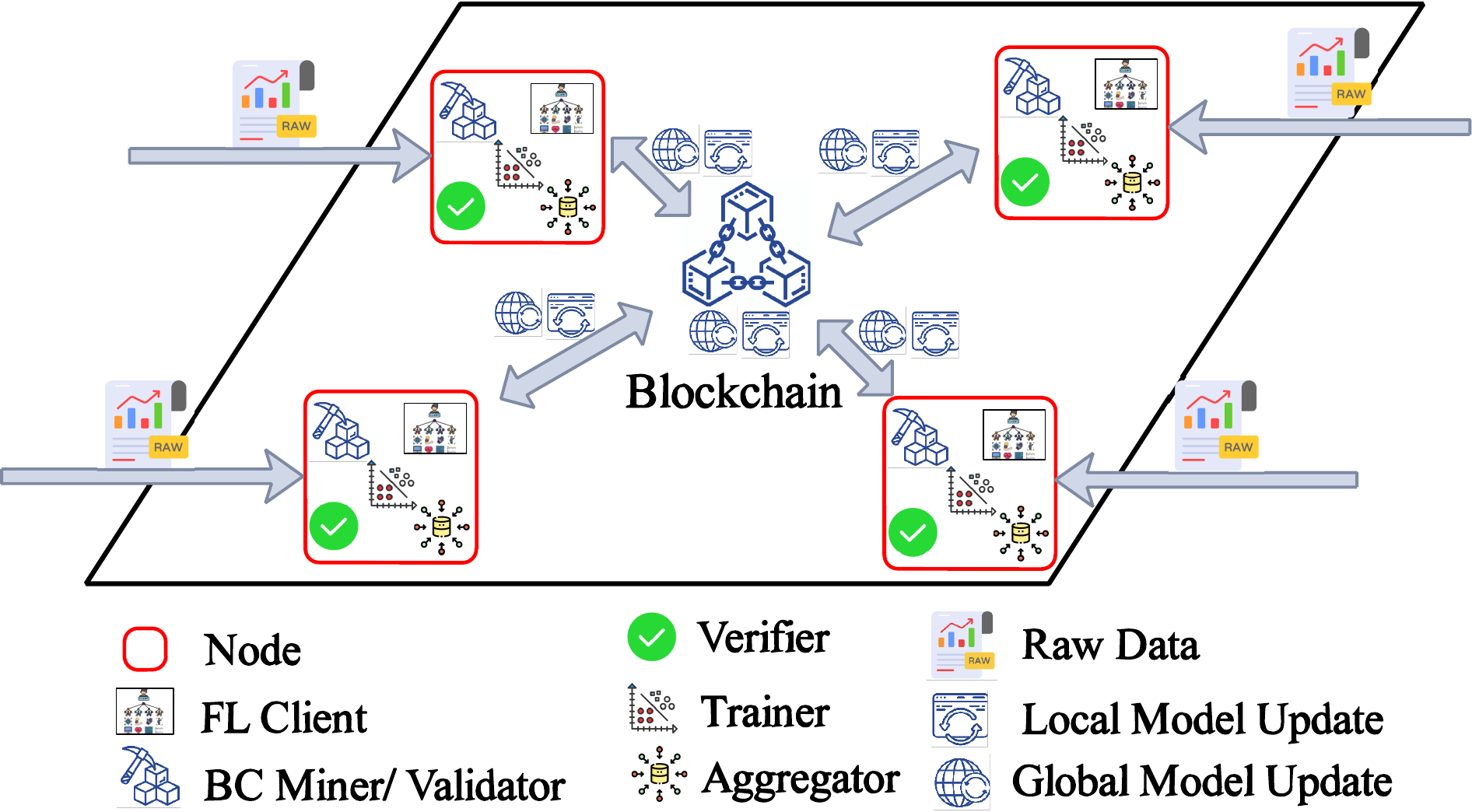}
			\caption{Fully Coupled Architecture}
			\label{fig:FC}
		\end{figure}

 \begin{enumerate}[wide, font=\itshape, labelwidth=!, labelindent=0pt, label*=A.\arabic*.]
\item \textit{Fully Coupled Architecture}

In the fully coupled architecture, blockchain nodes (i.e. miners or validators), perform dual roles as \gls{FL} clients. Within this architecture, \gls{FL} clients engage in computing local model updates as well as validating these updates as blockchain nodes. Notably, blockchain nodes not only partake in training local models but also participate as the global model aggregator. The aggregator, which may be a selected node, a designated leader, or a collection of nodes based on the predefined protocol, is responsible for gathering local model updates. Every node in this model has the opportunity to function as a blockchain validator, a local model trainer, and a global aggregator concurrently. Consequently, both local model updates and global model updates are contained within the blockchain. Importantly, the absence of a necessity to transmit the global model to a central server mitigates the risk of a single point of failure within this architecture. A schematic of this architecture is depicted in Figure~\ref{fig:FC}.\\ 

\begin{figure}
			\includegraphics[width=\linewidth]{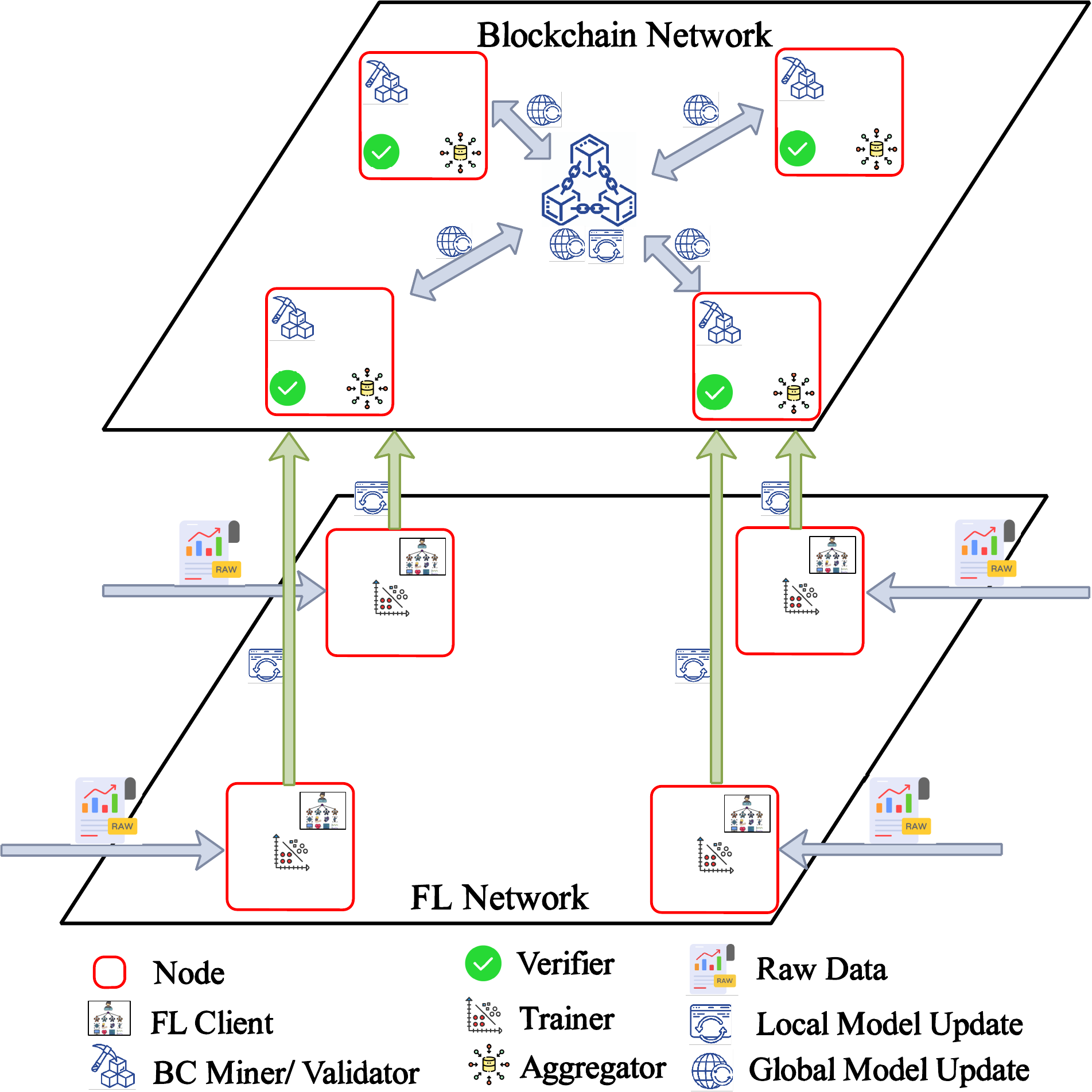}
			\caption{Semi-Coupled Architecture}
			\label{fig:SC}
		\end{figure}

 \item \textit{Semi-Coupled Architecture}

In the semi-coupled architecture, blockchain and \gls{FL} clients inhabit separate networks, although \gls{FL} clients retain the capability to interact with the blockchain and manipulate the distributed ledger. \gls{FL} clients gather data from diverse sources, train local models, and subsequently upload local model updates to the blockchain. \gls{BC} nodes (i.e. miners or validators) are tasked to validate the uploaded local model updates that will be used for training the global model. Upon the preparation of the global model, blockchain nodes will store it within the blockchain. Participant rewards are allocated based on a predefined incentive mechanism. This architecture also circumvents the potential for a single point of failure. A schematic of this architecture is depicted in Figure~\ref{fig:SC}. \\

\begin{figure}
			\includegraphics[width=\linewidth]{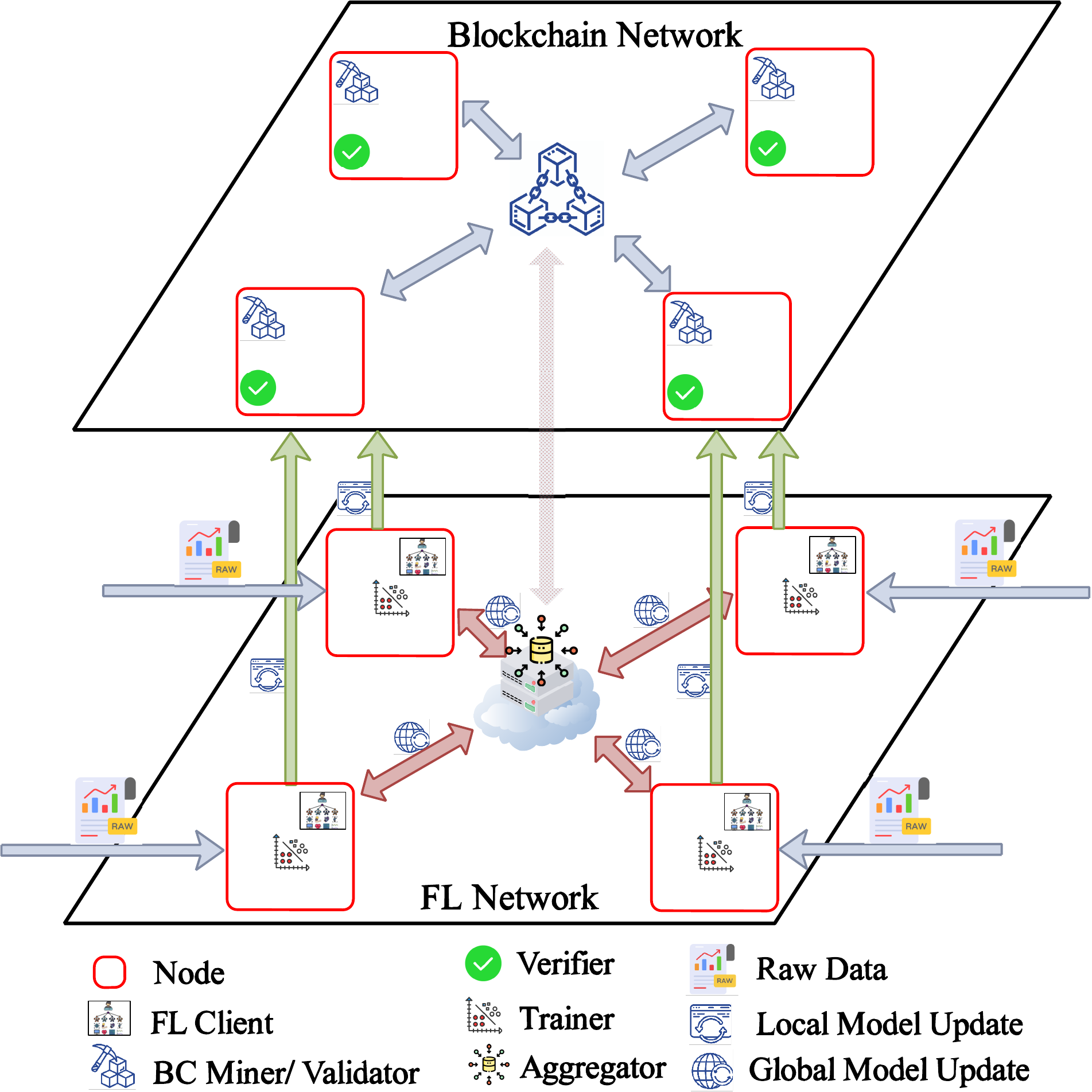}
			\caption{Loosely Coupled architecture}
			\label{fig:LC}
		\end{figure}
 
 \item \textit{Loosely Coupled Architecture}

In the loosely coupled architecture, blockchain nodes and \gls{FL} clients are in two distinct networks. This architecture introduces reputation as a criterion for measuring the reliability of the clients. In this architecture, the primary function of the blockchain is to furnish a coordination mechanism for clients, manage their reputation, authentication, validation of local model updates, and incentives management (i.e. contributions are managed to ascertain reputation and incentivize participation). While the blockchain validates local model updates, it refrains from storing them. In contrast, it stores the data related to the reputation of the participants. The responsibility of the \gls{FL} clients is to train local models and upload the updates to the blockchain for validation. After validations, these updates will be sent to an aggregator, which can be a distinct server or a cloud space. A schematic of this architecture is depicted in Figure~\ref{fig:LC}. 
\end{enumerate}

\begin{table*}[htbp]
\renewcommand{\arraystretch}{1.2}
 \centering
 \caption{Related works on Blockchain-assisted Federated Learning in Healthcare}
 \resizebox{\textwidth}{!}{%
 \begin{tabular}{|p{0.7cm}|p{5.5cm}|p{1.5cm}|p{1.85cm}|p{1cm}|p{2cm}|p{4.5cm}|}
 \hline
 \textbf{Ref.} & \textbf{Main Contribution} & \textbf{Architecture} & \textbf{\gls{BC} platform} & \textbf{FL type} & \textbf{Data type}& \textbf{Limitations} \\
 \hline
 \cite{liang2023architectural} & Software architecture integrating \gls{FL} and blockchain to tackle bias and fairness issues in healthcare predictive modeling while safeguarding patient privacy & Semi-coupled & Rahasak & Unknown & Unknown & Reliance on simulated validations rather than experimental implementations\\
 \hline
 
 ~\cite{om2023securing} & Introduces a mechanism to reward organizations participating in the \gls{FL} process, ensuring privacy-preserving model transfer between users and organizations using \gls{BC} & Loosely coupled & Ethereum & \gls{HFL} &Medical Imaging Data (Covid-19 \gls{CT}) & Not explicitly stated \\
 \hline
 
 ~\cite{moulahi2023blockchain} & Integrating \gls{FL} and \gls{BC} technology to develop a trusted system for predicting diabetes risk while ensuring data privacy and model integrity & Semi-coupled & Ethereum & \gls{HFL} & Sensor Data (\gls{IoT}) & Not explicitly stated \\
 \hline

 \cite{ali2023empowering} & Integration of \gls{BC} with \gls{FL} to enable secure and decentralized analysis of \gls{EMRs} in precision medicine & Loosely coupled & Ethereum, Hyperledger Fabric (simulation) & \gls{HFL} & EMR & Reliance on simulated validations rather than experimental implementations \\
 \hline

 \cite{chang2021blockchain} & Integration of adaptive differential privacy and gradient verification-based consensus protocols & Fully coupled. & Ethereum & \gls{HFL}/\gls{VFL} & \gls{IoMT} Sensor Data & scalability challenges, increased complexity, potential computational overhead, and the need for further validation across diverse medical conditions and datasets\\
 \hline

 \cite{lian2023blockchain} & \gls{BC}-based personalized \gls{FL} system to address security and privacy concerns in the \gls{IoMT} & Fully coupled & Unknown (Consortium \gls{PoS} \gls{BC}) & \gls{VFL} & \gls{IoMT} Sensor Data (Fashion-MNIST) & Reliance on simulated validations rather than experimental implementations \\
 \hline
 
 \cite{farooq2022blockchain} & Developing an automated system for analyzing patients' live data & Fully coupled & Ethereum & \gls{HFL} & \gls{IoMT} Sensor Data & Limited empirical validation or real-world testing of the proposed framework \\
 \hline

 \cite{zhang2021blockchain} & Propose a \gls{BCFL} framework for healthcare data privacy protection & Fully coupled & Unknown (conceptual) & \gls{HFL} & EHR (MNIST) & Theoretical model only, not supported via experimental implementations \\
 \hline
 
 \cite{aich2022protecting} & Proposes a \gls{BC}-assisted \gls{FL} framework for personal data preserving & Loosely coupled & Unknown (conceptual) & - & No data & Reliance on conceptual assumptions without real-world application \\
 \hline
 
 \cite{passerat2019blockchain} & Propose a novel architecture for \gls{FL} integrated with blockchain within healthcare system & Loosely Coupled & Ethereum & - & No data & Not explicitly stated \\
 \hline

 \cite{rahman2020secure} & Proposes a \gls{BCFL} framework for COVID-19 applications to classify IoHT data & Loosely coupled & Ethereum & \gls{FL} type & Sensor Data (\gls{IoMT}) & Not explicitly stated\\
 \hline
 
 \cite{singh2022framework} & Integrating \gls{BC} technology with \gls{FL} to enhance privacy preservation and scalability in healthcare data management & Fully coupled & Unknown & Unknown & Sensor Data (\gls{IoT}) & Theoretical model only, not supported via experimental implementations \\
 \hline

 \cite{nguyen2021federated} & Propose a new \gls{BC}-enabled Fed-based \gls{GANs} framework for secure COVID-19 data analytics & Semi-coupled & Unknown (conceptual) & HLF & Medical Imaging Data (Covid- 19 \gls{CT}) & \gls{BC} model is conceptual and real-world implementation was not reported \\
 \hline
 
 \cite{liu2022blockchain} & proposes A framework of \gls{BC}-empowered 
 \gls{FL} in healthcare-based \gls{CPS} & Fully coupled & Ethereum & \gls{HFL} & EHR (MINIST HAM10000) & Not explicitly stated \\
 \hline

 \cite{otoum2021preventing} & Proposes a novel solution for revolutionizing healthcare systems by considering concepts like distributivity, self-learnability, and autonomy. & - & Unknown (conceptual) & Unknown & No Data & Only introduces a theoretical framework without an empirical validation\\
 \hline
 
 \cite{kumar2021blockchain} & blockchain empowered method to detect patterns of COVID-19 from the lung \gls{CT} scans & Fully coupled & Unknown & \gls{HFL} & Medical Imaging Data (Covid- 19 \gls{CT}) & The paper does not extensively discuss the generalization capabilities of the proposed model to handle variations in COVID-19 manifestations across different patients. \\
 \hline

 \cite{durga2021federated} & A concise review of  \gls{BCFL} in healthcare, concepts and taxonomy & - & Unknown & Unknown & Unknown & article is very brief and does not cover all concepts \\
 \hline
 
 \cite{lakhan2022federated} & Propose a \gls{FL-BETS} framework with different dynamic heuristics for healthcare applications & Loosely coupled & Unknown & Unknown & Sensor Data (\gls{IoMT}) & Dynamic and run-time unknown attacks are against \gls{IoMT} were not considered in this work \\
 \hline

 \cite{samuel2022iomt} & Introduces FedMedChain as a \gls{BCFL} framework for medical data privacy-preserving & Loosely coupled & Unknown & Unknown & Sensor Data (Unknown) & Not explicitly stated \\
 \hline

 \cite{yang2024federated} & Introduces a privacy protection framework for medical data using \gls{BC} and \gls{FL} for secure and auditable data sharing among medical institutions using a secure aggregation scheme based on homomorphic encryption. & Loosely coupled & Ethereum & \gls{HFL} & EHR (MNIST different datasets) & Not explicitly stated \\
 \hline

 \cite{lo2022toward} & Proposes a \gls{BCFL} architecture to enable accountability in \gls{FL} systems. & Loosely coupled & Ethereum & \gls{HFL} & Medical Imaging Data (Covid-19 \gls{CT}) & Lack of real-world deployment analysis \\
 \hline
 \end{tabular}}
 \label{tab:sample}
\end{table*}

\section{Related Works}\label{sec:relatedWork}

In this section, we comprehensively explore existing research endeavors focusing on integrating \gls{BCFL} frameworks within healthcare contexts. Initially, we delve into the corpus of literature dedicated to elucidating the practical applications, methodologies, and outcomes of utilizing blockchain technology in conjunction with \gls{FL} for healthcare use cases. In addition, we aim to provide a taxonomy and categorize the existing works based on the insights garnered throughout this paper. To achieve this, we classify the works according to their integration architecture (i.e., fully coupled, semi-coupled, and loosely coupled), blockchain platform, \gls{FL} type, and data type. Furthermore, we highlight their primary contributions as well as their limitations. In the sequel, we will examine existing surveys and reviews focusing on blockchain-assisted \gls{FL} in healthcare. This will give us insights into the overall research landscape and help identify any gaps or areas for further exploration. 
\subsection{Literature review}


Liang et al.\cite{liang2023architectural} propose a software architecture integrating \gls{FL} and blockchain to mitigate bias and fairness issues in healthcare predictive modeling while safeguarding patient privacy. Om et al. \cite{om2023securing} introduce a mechanism to reward organizations participating in the \gls{FL} process, ensuring privacy-preserving model transfer using loosely coupled integration with the Ethereum blockchain. Moulahi et al. \cite{moulahi2023blockchain} integrate \gls{FL} and blockchain to develop a trusted system for predicting diabetes risk while ensuring data privacy and model integrity.

Ali et al. \cite{ali2023empowering} focus on integrating blockchain with \gls{FL} for secure and decentralized analysis of \gls{EMRs} in precision medicine. Chang et al. \cite{chang2021blockchain} propose an integration of adaptive differential privacy and gradient verification-based consensus protocols in a fully coupled architecture for healthcare analytics. Lian et al. \cite{lian2023blockchain} present a blockchain-based personalized \gls{FL} system for ensuring security and privacy in the \gls{IoMT}.

Farooq et al. \cite{farooq2022blockchain} develop an automated system for analyzing patients' live data within a fully coupled architecture, while Zhang et al. \cite{zhang2021blockchain} propose a blockchain-enabled \gls{FL} framework for healthcare data privacy protection. Aich et al. \cite{aich2022protecting} introduce a blockchain-assisted \gls{FL} framework for personal data preservation, and Passerat et al. \cite{passerat2019blockchain} propose a novel architecture for \gls{FL} integrated with blockchain within healthcare systems. A taxonomy of the existing research studies, along with additional relevant works, has been compiled in Table~\ref{tab:sample}.

\subsection{Existing Surveys}

To the best of our knowledge, comprehensive surveys or reviews focusing on the integration of blockchain and \gls{FL} for healthcare use cases are scarce. While individual studies have explored the potential of each technology independently within healthcare settings, there is a notable lack of resources delving into the synergistic benefits and challenges of combining blockchain's immutable ledger capabilities with \gls{FL}'s decentralized model training approach.

Noteworthy initial explorations, such as those conducted by Myrzashova et al. (2023)~\cite{myrzashova2023blockchain} and Nguyen et al. (2021)~\cite{nguyen2021blockchain}, offer valuable insights but often lack a broader perspective. Myrzashova et al.~\cite{myrzashova2023blockchain}, for instance, analyze the advantages and disadvantages of \gls{BC}-\gls{FL} integration in healthcare, but they overlook the importance of a data type taxonomy. Understanding the diverse medical data types (e.g., genomics, imaging) used in healthcare \gls{ML} is crucial as different data may have varying security and privacy requirements.
Similarly, Nguyen et al.~\cite{nguyen2021blockchain} introduce a new conceptual architecture that integrates blockchain and \gls{AI} for combating the COVID-19 pandemic. While offering valuable insights into addressing specific challenges posed by the pandemic, its scope is limited to COVID-19 and related data, lacking a broader analysis of the integration's potential for various healthcare use cases beyond this specific context.

In contrast to these existing works, this tutorial offers a more comprehensive perspective. We present a taxonomy of medical data used for \gls{ML}, providing a foundational understanding of the diverse data types relevant to this integration. Furthermore, we unveil an innovative architecture meticulously designed for the seamless integration of blockchain and \gls{FL} within healthcare systems, addressing the need for secure and privacy-preserving healthcare analytics.

\section{Conclusion}\label{sec:conclusion}

This paper explored the transformative potential of integrating blockchain technology and \gls{FL} for secure and privacy-preserving healthcare analytics. We highlighted how this synergy leverages the strengths of both blockchain's immutable ledger and ensures data integrity and transparency, while \gls{FL} facilitates collaborative model development without compromising patient privacy. The three proposed architectural models (fully coupled, semi-coupled, and loosely coupled) offer flexibility in tailoring \gls{FL} to specific healthcare requirements, balancing decentralization, scalability, and reliability.

We explored how \gls{FL} empowers various healthcare applications, including disease prediction, medical image analysis, patient monitoring, and drug discovery. By enabling model training on local devices at healthcare institutions, \gls{FL} eliminates the need for centralized data storage, addressing privacy concerns associated with traditional \gls{ML} approaches.

While \gls{FL} offers privacy benefits, it introduces vulnerabilities. \gls{BC}'s tamper-proof nature and smart contracts address these challenges by ensuring data integrity and mitigating malicious activities. This tutorial provided a foundational understanding of medical data, \gls{FL}, and their integration for healthcare applications.

Looking forward, continued research, collaboration, and innovation across the research community hold immense promise to unlock the full potential of \gls{FL}. This collaborative effort will pave the way for a more secure, transparent, and inclusive healthcare ecosystem, ultimately leading to improved patient outcomes.

\section*{Acknowledgment}
This work was supported by the Natural Sciences and Engineering Research Council of Canada (NSERC) and FLEX Group Company. 


\ifCLASSOPTIONcaptionsoff
 \newpage
\fi
\bibliographystyle{IEEEtran}
\bibliography{bibliographies/bib/References.bib} 

\begin{thebibliography}{100}
\providecommand{\url}[1]{#1}
\csname url@samestyle\endcsname
\providecommand{\newblock}{\relax}
\providecommand{\bibinfo}[2]{#2}
\providecommand{\BIBentrySTDinterwordspacing}{\spaceskip=0pt\relax}
\providecommand{\BIBentryALTinterwordstretchfactor}{4}
\providecommand{\BIBentryALTinterwordspacing}{\spaceskip=\fontdimen2\font plus
\BIBentryALTinterwordstretchfactor\fontdimen3\font minus \fontdimen4\font\relax}
\providecommand{\BIBforeignlanguage}[2]{{%
\expandafter\ifx\csname l@#1\endcsname\relax
\typeout{** WARNING: IEEEtran.bst: No hyphenation pattern has been}%
\typeout{** loaded for the language `#1'. Using the pattern for}%
\typeout{** the default language instead.}%
\else
\language=\csname l@#1\endcsname
\fi
#2}}
\providecommand{\BIBdecl}{\relax}
\BIBdecl

\bibitem{seneviratne2017survey}
S.~Seneviratne, Y.~Hu, T.~Nguyen, G.~Lan, S.~Khalifa, K.~Thilakarathna, M.~Hassan, and A.~Seneviratne, ``A survey of wearable devices and challenges,'' \emph{IEEE Communications Surveys \& Tutorials}, vol.~19, no.~4, pp. 2573--2620, 2017.

\bibitem{ometov2021survey}
A.~Ometov, V.~Shubina, L.~Klus, J.~Skibi{\'n}ska, S.~Saafi, P.~Pascacio, L.~Flueratoru, D.~Q. Gaibor, N.~Chukhno, O.~Chukhno \emph{et~al.}, ``A survey on wearable technology: History, state-of-the-art and current challenges,'' \emph{Computer Networks}, vol. 193, p. 108074, 2021.

\bibitem{kang2022wearing}
H.~S. Kang and M.~Exworthy, ``Wearing the future—wearables to empower users to take greater responsibility for their health and care: Scoping review,'' \emph{JMIR mHealth and uHealth}, vol.~10, no.~7, p. e35684, 2022.

\bibitem{haghi2017wearable}
M.~Haghi, K.~Thurow, and R.~Stoll, ``Wearable devices in medical internet of things: scientific research and commercially available devices,'' \emph{Healthcare informatics research}, vol.~23, no.~1, pp. 4--15, 2017.

\bibitem{vijayan2021review}
V.~Vijayan, J.~P. Connolly, J.~Condell, N.~McKelvey, and P.~Gardiner, ``Review of wearable devices and data collection considerations for connected health,'' \emph{Sensors}, vol.~21, no.~16, p. 5589, 2021.

\bibitem{wu2019wearable}
M.~Wu and J.~Luo, ``Wearable technology applications in healthcare: a literature review,'' \emph{Online J. Nurs. Inform}, vol.~23, no.~3, 2019.

\bibitem{datta2018survey}
P.~Datta, A.~S. Namin, and M.~Chatterjee, ``A survey of privacy concerns in wearable devices,'' in \emph{2018 IEEE International Conference on Big Data (Big Data)}.\hskip 1em plus 0.5em minus 0.4em\relax IEEE, 2018, pp. 4549--4553.

\bibitem{cilliers2020wearable}
L.~Cilliers, ``Wearable devices in healthcare: Privacy and information security issues,'' \emph{Health information management journal}, vol.~49, no. 2-3, pp. 150--156, 2020.

\bibitem{arias2015privacy}
O.~Arias, J.~Wurm, K.~Hoang, and Y.~Jin, ``Privacy and security in internet of things and wearable devices,'' \emph{IEEE transactions on multi-scale computing systems}, vol.~1, no.~2, pp. 99--109, 2015.

\bibitem{qayyum2020secure}
A.~Qayyum, J.~Qadir, M.~Bilal, and A.~Al-Fuqaha, ``Secure and robust machine learning for healthcare: A survey,'' \emph{IEEE Reviews in Biomedical Engineering}, vol.~14, pp. 156--180, 2020.

\bibitem{kim2015reliability}
Y.~Kim, W.~Lee, A.~Raghunathan, V.~Raghunathan, and N.~K. Jha, ``Reliability and security of implantable and wearable medical devices,'' in \emph{implantable biomedical microsystems}.\hskip 1em plus 0.5em minus 0.4em\relax Elsevier, 2015, pp. 167--199.

\bibitem{rieke2020future}
N.~Rieke, J.~Hancox, W.~Li, F.~Milletari, H.~R. Roth, S.~Albarqouni, S.~Bakas, M.~N. Galtier, B.~A. Landman, K.~Maier-Hein \emph{et~al.}, ``The future of digital health with federated learning,'' \emph{NPJ digital medicine}, vol.~3, no.~1, p. 119, 2020.

\bibitem{zhang2021survey}
C.~Zhang, Y.~Xie, H.~Bai, B.~Yu, W.~Li, and Y.~Gao, ``A survey on federated learning,'' \emph{Knowledge-Based Systems}, vol. 216, p. 106775, 2021.

\bibitem{li2020review}
L.~Li, Y.~Fan, M.~Tse, and K.-Y. Lin, ``A review of applications in federated learning,'' \emph{Computers \& Industrial Engineering}, vol. 149, p. 106854, 2020.

\bibitem{pfitzner2021federated}
B.~Pfitzner, N.~Steckhan, and B.~Arnrich, ``Federated learning in a medical context: a systematic literature review,'' \emph{ACM Transactions on Internet Technology (TOIT)}, vol.~21, no.~2, pp. 1--31, 2021.

\bibitem{antunes2022federated}
R.~S. Antunes, C.~Andr{\'e}~da Costa, A.~K{\"u}derle, I.~A. Yari, and B.~Eskofier, ``Federated learning for healthcare: Systematic review and architecture proposal,'' \emph{ACM Transactions on Intelligent Systems and Technology (TIST)}, vol.~13, no.~4, pp. 1--23, 2022.

\bibitem{xu2021federated}
J.~Xu, B.~S. Glicksberg, C.~Su, P.~Walker, J.~Bian, and F.~Wang, ``Federated learning for healthcare informatics,'' \emph{Journal of Healthcare Informatics Research}, vol.~5, pp. 1--19, 2021.

\bibitem{fang2020local}
M.~Fang, X.~Cao, J.~Jia, and N.~Gong, ``Local model poisoning attacks to $\{$Byzantine-Robust$\}$ federated learning,'' in \emph{29th USENIX security symposium (USENIX Security 20)}, 2020, pp. 1605--1622.

\bibitem{cao2019understanding}
D.~Cao, S.~Chang, Z.~Lin, G.~Liu, and D.~Sun, ``Understanding distributed poisoning attack in federated learning,'' in \emph{2019 IEEE 25th International Conference on Parallel and Distributed Systems (ICPADS)}.\hskip 1em plus 0.5em minus 0.4em\relax IEEE, 2019, pp. 233--239.

\bibitem{lyu2020threats}
L.~Lyu, H.~Yu, and Q.~Yang, ``Threats to federated learning: A survey,'' \emph{arXiv preprint arXiv:2003.02133}, 2020.

\bibitem{bouacida2021vulnerabilities}
N.~Bouacida and P.~Mohapatra, ``Vulnerabilities in federated learning,'' \emph{IEEE Access}, vol.~9, pp. 63\,229--63\,249, 2021.

\bibitem{lyu2022privacy}
L.~Lyu, H.~Yu, X.~Ma, C.~Chen, L.~Sun, J.~Zhao, Q.~Yang, and S.~Y. Philip, ``Privacy and robustness in federated learning: Attacks and defenses,'' \emph{IEEE transactions on neural networks and learning systems}, 2022.

\bibitem{ye2022decentralized}
H.~Ye, L.~Liang, and G.~Y. Li, ``Decentralized federated learning with unreliable communications,'' \emph{IEEE journal of selected topics in signal processing}, vol.~16, no.~3, pp. 487--500, 2022.

\bibitem{nguyen2021federated}
D.~C. Nguyen, M.~Ding, Q.-V. Pham, P.~N. Pathirana, L.~B. Le, A.~Seneviratne, J.~Li, D.~Niyato, and H.~V. Poor, ``Federated learning meets blockchain in edge computing: Opportunities and challenges,'' \emph{IEEE Internet of Things Journal}, vol.~8, no.~16, pp. 12\,806--12\,825, 2021.

\bibitem{rahman2015electronic}
R.~Rahman and C.~K. Reddy, ``Electronic health records: A survey.'' \emph{Healthcare Data Analytics}, vol.~36, p.~21, 2015.

\bibitem{yadav2018mining}
P.~Yadav, M.~Steinbach, V.~Kumar, and G.~Simon, ``Mining electronic health records (ehrs) a survey,'' \emph{ACM Computing Surveys (CSUR)}, vol.~50, no.~6, pp. 1--40, 2018.

\bibitem{hoerbst2010electronic}
A.~Hoerbst and E.~Ammenwerth, ``Electronic health records,'' \emph{Methods of information in medicine}, vol.~49, no.~04, pp. 320--336, 2010.

\bibitem{simon2007physicians}
S.~R. Simon, R.~Kaushal, P.~D. Cleary, C.~A. Jenter, L.~A. Volk, E.~J. Orav, E.~Burdick, E.~G. Poon, and D.~W. Bates, ``Physicians and electronic health records: a statewide survey,'' \emph{Archives of internal medicine}, vol. 167, no.~5, pp. 507--512, 2007.

\bibitem{jin2009patient}
J.~Jin, G.-J. Ahn, H.~Hu, M.~J. Covington, and X.~Zhang, ``Patient-centric authorization framework for sharing electronic health records,'' in \emph{Proceedings of the 14th ACM symposium on Access control models and technologies}, 2009, pp. 125--134.

\bibitem{blobel2004authorisation}
B.~Blobel, ``Authorisation and access control for electronic health record systems,'' \emph{International journal of medical informatics}, vol.~73, no.~3, pp. 251--257, 2004.

\bibitem{fernandez2013security}
J.~L. Fern{\'a}ndez-Alem{\'a}n, I.~C. Se{\~n}or, P.~{\'A}.~O. Lozoya, and A.~Toval, ``Security and privacy in electronic health records: A systematic literature review,'' \emph{Journal of biomedical informatics}, vol.~46, no.~3, pp. 541--562, 2013.

\bibitem{tang2019efficient}
F.~Tang, S.~Ma, Y.~Xiang, and C.~Lin, ``An efficient authentication scheme for blockchain-based electronic health records,'' \emph{IEEE access}, vol.~7, pp. 41\,678--41\,689, 2019.

\bibitem{arbet2021lessons}
J.~Arbet, C.~Brokamp, J.~Meinzen-Derr, K.~E. Trinkley, and H.~M. Spratt, ``Lessons and tips for designing a machine learning study using ehr data,'' \emph{Journal of Clinical and Translational Science}, vol.~5, no.~1, p. e21, 2021.

\bibitem{wu2020statistics}
H.~Wu, J.~M. Yamal, A.~Yaseen, and V.~Maroufy, \emph{Statistics and machine learning methods for EHR data: from data extraction to data analytics}.\hskip 1em plus 0.5em minus 0.4em\relax CRC Press, 2020.

\bibitem{deo2016learning}
R.~C. Deo and B.~K. Nallamothu, ``Learning about machine learning: the promise and pitfalls of big data and the electronic health record,'' pp. 618--620, 2016.

\bibitem{wu2010prediction}
J.~Wu, J.~Roy, and W.~F. Stewart, ``Prediction modeling using ehr data: challenges, strategies, and a comparison of machine learning approaches,'' \emph{Medical care}, pp. S106--S113, 2010.

\bibitem{johnston2019using}
S.~S. Johnston, J.~M. Morton, I.~Kalsekar, E.~M. Ammann, C.-W. Hsiao, and J.~Reps, ``Using machine learning applied to real-world healthcare data for predictive analytics: an applied example in bariatric surgery,'' \emph{Value in health}, vol.~22, no.~5, pp. 580--586, 2019.

\bibitem{dev2022predictive}
S.~Dev, H.~Wang, C.~S. Nwosu, N.~Jain, B.~Veeravalli, and D.~John, ``A predictive analytics approach for stroke prediction using machine learning and neural networks,'' \emph{Healthcare Analytics}, vol.~2, p. 100032, 2022.

\bibitem{muniasamy2020deep}
A.~Muniasamy, S.~Tabassam, M.~A. Hussain, H.~Sultana, V.~Muniasamy, and R.~Bhatnagar, ``Deep learning for predictive analytics in healthcare,'' in \emph{The International Conference on Advanced Machine Learning Technologies and Applications (AMLTA2019) 4}.\hskip 1em plus 0.5em minus 0.4em\relax Springer, 2020, pp. 32--42.

\bibitem{sajda2006machine}
P.~Sajda, ``Machine learning for detection and diagnosis of disease,'' \emph{Annu. Rev. Biomed. Eng.}, vol.~8, pp. 537--565, 2006.

\bibitem{singh2021diagnosing}
P.~Singh, N.~Singh, K.~K. Singh, and A.~Singh, ``Diagnosing of disease using machine learning,'' in \emph{Machine learning and the internet of medical things in healthcare}.\hskip 1em plus 0.5em minus 0.4em\relax Elsevier, 2021, pp. 89--111.

\bibitem{li2020heart}
J.~P. Li, A.~U. Haq, S.~U. Din, J.~Khan, A.~Khan, and A.~Saboor, ``Heart disease identification method using machine learning classification in e-healthcare,'' \emph{IEEE access}, vol.~8, pp. 107\,562--107\,582, 2020.

\bibitem{mall2022implementation}
S.~Mall, A.~Srivastava, B.~D. Mazumdar, M.~Mishra, S.~L. Bangare, and A.~Deepak, ``Implementation of machine learning techniques for disease diagnosis,'' \emph{Materials Today: Proceedings}, vol.~51, pp. 2198--2201, 2022.

\bibitem{ahsan2022machine}
M.~M. Ahsan, S.~A. Luna, and Z.~Siddique, ``Machine-learning-based disease diagnosis: A comprehensive review,'' in \emph{Healthcare}, vol.~10, no.~3.\hskip 1em plus 0.5em minus 0.4em\relax MDPI, 2022, p. 541.

\bibitem{alanazi2022identification}
R.~Alanazi \emph{et~al.}, ``Identification and prediction of chronic diseases using machine learning approach,'' \emph{Journal of Healthcare Engineering}, vol. 2022, 2022.

\bibitem{battineni2020applications}
G.~Battineni, G.~G. Sagaro, N.~Chinatalapudi, and F.~Amenta, ``Applications of machine learning predictive models in the chronic disease diagnosis,'' \emph{Journal of personalized medicine}, vol.~10, no.~2, p.~21, 2020.

\bibitem{chekroud2021promise}
A.~M. Chekroud, J.~Bondar, J.~Delgadillo, G.~Doherty, A.~Wasil, M.~Fokkema, Z.~Cohen, D.~Belgrave, R.~DeRubeis, R.~Iniesta \emph{et~al.}, ``The promise of machine learning in predicting treatment outcomes in psychiatry,'' \emph{World Psychiatry}, vol.~20, no.~2, pp. 154--170, 2021.

\bibitem{bica2021real}
I.~Bica, A.~M. Alaa, C.~Lambert, and M.~Van Der~Schaar, ``From real-world patient data to individualized treatment effects using machine learning: current and future methods to address underlying challenges,'' \emph{Clinical Pharmacology \& Therapeutics}, vol. 109, no.~1, pp. 87--100, 2021.

\bibitem{wong2018using}
J.~Wong, M.~Murray~Horwitz, L.~Zhou, and S.~Toh, ``Using machine learning to identify health outcomes from electronic health record data,'' \emph{Current epidemiology reports}, vol.~5, pp. 331--342, 2018.

\bibitem{moehring2021development}
R.~W. Moehring, M.~Phelan, E.~Lofgren, A.~Nelson, E.~D. Ashley, D.~J. Anderson, and B.~A. Goldstein, ``Development of a machine learning model using electronic health record data to identify antibiotic use among hospitalized patients,'' \emph{JAMA Network Open}, vol.~4, no.~3, pp. e213\,460--e213\,460, 2021.

\bibitem{chu2020treatment}
J.~Chu, W.~Dong, J.~Wang, K.~He, and Z.~Huang, ``Treatment effect prediction with adversarial deep learning using electronic health records,'' \emph{BMC Medical Informatics and Decision Making}, vol.~20, pp. 1--14, 2020.

\bibitem{komal2019drug}
N.~Komal~Kumar and D.~Vigneswari, ``A drug recommendation system for multi-disease in health care using machine learning,'' in \emph{International Conference on Advanced Communication and Computational Technology}.\hskip 1em plus 0.5em minus 0.4em\relax Springer, 2019, pp. 1--12.

\bibitem{lalitha2022medical}
S.~Lalitha, T.~Sanjana, H.~Bhavana, I.~Bhan, and G.~Harshith, ``Medical imaging modalities and different image processing techniques: State of the art review,'' \emph{Disruptive Developments in Biomedical Applications}, pp. 17--36, 2022.

\bibitem{abhisheka2023recent}
B.~Abhisheka, S.~K. Biswas, B.~Purkayastha, D.~Das, and A.~Escargueil, ``Recent trend in medical imaging modalities and their applications in disease diagnosis: a review,'' \emph{Multimedia Tools and Applications}, pp. 1--36, 2023.

\bibitem{elangovan2016medical}
A.~Elangovan and T.~Jeyaseelan, ``Medical imaging modalities: a survey,'' in \emph{2016 International Conference on emerging trends in engineering, technology and science (ICETETS)}.\hskip 1em plus 0.5em minus 0.4em\relax ieee, 2016, pp. 1--4.

\bibitem{doi2006diagnostic}
K.~Doi, ``Diagnostic imaging over the last 50 years: research and development in medical imaging science and technology,'' \emph{Physics in Medicine \& Biology}, vol.~51, no.~13, p.~R5, 2006.

\bibitem{khoon2016survey}
L.~L. Khoon and L.~S. Chuin, ``A survey of medical image processing tools,'' \emph{International Journal of Software Engineering and Computer Systems (IJSECS)}, vol.~2, no.~1, pp. 10--27, 2016.

\bibitem{santhi2022survey}
K.~Santhi, ``A survey on medical imaging techniques and applications,'' \emph{Journal of Innovative Image Processing}, vol.~4, no.~3, pp. 173--182, 2022.

\bibitem{huda2015x}
W.~Huda and R.~B. Abrahams, ``X-ray-based medical imaging and resolution,'' \emph{American Journal of Roentgenology}, vol. 204, no.~4, pp. W393--W397, 2015.

\bibitem{mustapha2021comparative}
M.~T. Mustapha, B.~Uzun, D.~U. Ozsahin, and I.~Ozsahin, ``A comparative study of x-ray based medical imaging devices,'' in \emph{Applications of Multi-Criteria Decision-Making Theories in Healthcare and Biomedical Engineering}.\hskip 1em plus 0.5em minus 0.4em\relax Elsevier, 2021, pp. 163--180.

\bibitem{aaslund2010detectors}
M.~{\AA}slund, E.~Fredenberg, M.~Telman, and M.~Danielsson, ``Detectors for the future of x-ray imaging,'' \emph{Radiation protection dosimetry}, vol. 139, no. 1-3, pp. 327--333, 2010.

\bibitem{brooks1993computed}
S.~L. Brooks, ``Computed tomography,'' \emph{Dental Clinics of North America}, vol.~37, no.~4, pp. 575--590, 1993.

\bibitem{sluimer2006computer}
I.~Sluimer, A.~Schilham, M.~Prokop, and B.~Van~Ginneken, ``Computer analysis of computed tomography scans of the lung: a survey,'' \emph{IEEE transactions on medical imaging}, vol.~25, no.~4, pp. 385--405, 2006.

\bibitem{withers2021x}
P.~J. Withers, C.~Bouman, S.~Carmignato, V.~Cnudde, D.~Grimaldi, C.~K. Hagen, E.~Maire, M.~Manley, A.~Du~Plessis, and S.~R. Stock, ``X-ray computed tomography,'' \emph{Nature Reviews Methods Primers}, vol.~1, no.~1, p.~18, 2021.

\bibitem{kalender2006x}
W.~A. Kalender, ``X-ray computed tomography,'' \emph{Physics in Medicine \& Biology}, vol.~51, no.~13, p. R29, 2006.

\bibitem{shrimpton2005doses}
P.~Shrimpton, M.~Hillier, M.~Lewis, and M.~Dunn, \emph{Doses from computed tomography (CT) examinations in the UK-2003 review}.\hskip 1em plus 0.5em minus 0.4em\relax NRPB Chilton, 2005, vol.~67.

\bibitem{katti2011magnetic}
G.~Katti, S.~A. Ara, and A.~Shireen, ``Magnetic resonance imaging (mri)--a review,'' \emph{International journal of dental clinics}, vol.~3, no.~1, pp. 65--70, 2011.

\bibitem{khoo1997magnetic}
V.~S. Khoo, D.~P. Dearnaley, D.~J. Finnigan, A.~Padhani, S.~F. Tanner, and M.~O. Leach, ``Magnetic resonance imaging (mri): considerations and applications in radiotherapy treatment planning,'' \emph{Radiotherapy and Oncology}, vol.~42, no.~1, pp. 1--15, 1997.

\bibitem{fatahi2015magnetic}
M.~Fatahi, O.~Speck \emph{et~al.}, ``Magnetic resonance imaging (mri): A review of genetic damage investigations,'' \emph{Mutation Research/Reviews in Mutation Research}, vol. 764, pp. 51--63, 2015.

\bibitem{tirotta201519f}
I.~Tirotta, V.~Dichiarante, C.~Pigliacelli, G.~Cavallo, G.~Terraneo, F.~B. Bombelli, P.~Metrangolo, and G.~Resnati, ``19f magnetic resonance imaging (mri): from design of materials to clinical applications,'' \emph{Chemical reviews}, vol. 115, no.~2, pp. 1106--1129, 2015.

\bibitem{prasad2005making}
A.~Prasad, ``Making images/making bodies: Visibilizing and disciplining through magnetic resonance imaging (mri),'' \emph{Science, Technology, \& Human Values}, vol.~30, no.~2, pp. 291--316, 2005.

\bibitem{mori1999diffusion}
S.~Mori and P.~B. Barker, ``Diffusion magnetic resonance imaging: its principle and applications,'' \emph{The Anatomical Record: An Official Publication of the American Association of Anatomists}, vol. 257, no.~3, pp. 102--109, 1999.

\bibitem{avola2021ultrasound}
D.~Avola, L.~Cinque, A.~Fagioli, G.~Foresti, and A.~Mecca, ``Ultrasound medical imaging techniques: a survey,'' \emph{ACM Computing Surveys (CSUR)}, vol.~54, no.~3, pp. 1--38, 2021.

\bibitem{huang2017review}
Q.~Huang, Z.~Zeng \emph{et~al.}, ``A review on real-time 3d ultrasound imaging technology,'' \emph{BioMed research international}, vol. 2017, 2017.

\bibitem{whittaker2011ultrasound}
J.~L. Whittaker and M.~Stokes, ``Ultrasound imaging and muscle function,'' \emph{Journal of Orthopaedic \& Sports Physical Therapy}, vol.~41, no.~8, pp. 572--580, 2011.

\bibitem{ellis2020exploring}
R.~Ellis, J.~Helsby, J.~Naus, S.~Bassett, C.~Fern{\'a}ndez-de Las-Pe{\~n}as, S.~F. Carnero, J.~Hides, C.~O'sullivan, D.~Teyhen, M.~Stokes \emph{et~al.}, ``Exploring the use of ultrasound imaging by physiotherapists: An international survey,'' \emph{Musculoskeletal Science and Practice}, vol.~49, p. 102213, 2020.

\bibitem{ortiz2012ultrasound}
S.~H.~C. Ortiz, T.~Chiu, and M.~D. Fox, ``Ultrasound image enhancement: A review,'' \emph{Biomedical Signal Processing and Control}, vol.~7, no.~5, pp. 419--428, 2012.

\bibitem{fenster2001three}
A.~Fenster, D.~B. Downey, and H.~N. Cardinal, ``Three-dimensional ultrasound imaging,'' \emph{Physics in medicine \& biology}, vol.~46, no.~5, p. R67, 2001.

\bibitem{lees2001ultrasound}
W.~Lees, ``Ultrasound imaging in three and four dimensions,'' in \emph{Seminars in Ultrasound, CT and MRI}, vol.~22, no.~1.\hskip 1em plus 0.5em minus 0.4em\relax Elsevier, 2001, pp. 85--105.

\bibitem{schellpfeffer2013ultrasound}
M.~A. Schellpfeffer, ``Ultrasound imaging in research and clinical medicine,'' \emph{Birth Defects Research Part C: Embryo Today: Reviews}, vol.~99, no.~2, pp. 83--92, 2013.

\bibitem{hricak2021medical}
H.~Hricak, M.~Abdel-Wahab, R.~Atun, M.~M. Lette, D.~Paez, J.~A. Brink, L.~Donoso-Bach, G.~Frija, M.~Hierath, O.~Holmberg \emph{et~al.}, ``Medical imaging and nuclear medicine: a lancet oncology commission,'' \emph{The Lancet Oncology}, vol.~22, no.~4, pp. e136--e172, 2021.

\bibitem{schoder2003pet}
H.~Sch{\"o}der, Y.~E. Erdi, S.~M. Larson, and H.~W. Yeung, ``Pet/ct: a new imaging technology in nuclear medicine,'' \emph{European journal of nuclear medicine and molecular imaging}, vol.~30, pp. 1419--1437, 2003.

\bibitem{blankenberg2002nuclear}
F.~G. Blankenberg and H.~W. Strauss, ``Nuclear medicine applications in molecular imaging,'' \emph{Journal of Magnetic Resonance Imaging: An Official Journal of the International Society for Magnetic Resonance in Medicine}, vol.~16, no.~4, pp. 352--361, 2002.

\bibitem{vaz2020nuclear}
S.~C. Vaz, F.~Oliveira, K.~Herrmann, and P.~Veit-Haibach, ``Nuclear medicine and molecular imaging advances in the 21st century,'' \emph{The British journal of radiology}, vol.~93, no. 1110, p. 20200095, 2020.

\bibitem{van2003nuclear}
C.~Van~de Wiele, C.~Lahorte, W.~Oyen, O.~Boerman, I.~Goethals, G.~Slegers, and R.~A. Dierckx, ``Nuclear medicine imaging to predict response to radiotherapy: a review,'' \emph{International Journal of Radiation Oncology* Biology* Physics}, vol.~55, no.~1, pp. 5--15, 2003.

\bibitem{prandini2006nuclear}
N.~Prandini, E.~Lazzeri, B.~Rossi, P.~Erba, M.~G. Parisella, and A.~Signore, ``Nuclear medicine imaging of bone infections,'' \emph{Nuclear medicine communications}, vol.~27, no.~8, pp. 633--644, 2006.

\bibitem{bailey2005positron}
D.~L. Bailey, M.~N. Maisey, D.~W. Townsend, and P.~E. Valk, \emph{Positron emission tomography}.\hskip 1em plus 0.5em minus 0.4em\relax Springer, 2005, vol.~2.

\bibitem{tai2004applications}
Y.~Tai and P.~Piccini, ``Applications of positron emission tomography (pet) in neurology,'' \emph{Journal of neurology, neurosurgery, and psychiatry}, vol.~75, no.~5, p. 669, 2004.

\bibitem{gallamini2014positron}
A.~Gallamini, C.~Zwarthoed, and A.~Borra, ``Positron emission tomography (pet) in oncology,'' \emph{Cancers}, vol.~6, no.~4, pp. 1821--1889, 2014.

\bibitem{wagner1998brief}
H.~N. Wagner~Jr, ``A brief history of positron emission tomography (pet),'' in \emph{Seminars in nuclear medicine}, vol.~28, no.~3.\hskip 1em plus 0.5em minus 0.4em\relax Elsevier, 1998, pp. 213--220.

\bibitem{townsend2008positron}
D.~W. Townsend, ``Positron emission tomography/computed tomography,'' in \emph{Seminars in nuclear medicine}, vol.~38, no.~3.\hskip 1em plus 0.5em minus 0.4em\relax Elsevier, 2008, pp. 152--166.

\bibitem{jones2014medical}
A.~K. Jones, S.~Balter, P.~Rauch, and L.~K. Wagner, ``Medical imaging using ionizing radiation: optimization of dose and image quality in fluoroscopy,'' \emph{Medical physics}, vol.~41, no.~1, p. 014301, 2014.

\bibitem{maher1986digital}
K.~Maher and J.~Malone, ``Digital fluoroscopy: a new development in medical imaging,'' \emph{Contemporary physics}, vol.~27, no.~6, pp. 533--552, 1986.

\bibitem{krohmer1989radiography}
J.~S. Krohmer, ``Radiography and fluoroscopy, 1920 to the present.'' \emph{Radiographics}, vol.~9, no.~6, pp. 1129--1153, 1989.

\bibitem{rudin1992region}
S.~Rudin and D.~R. Bednarek, ``Region of interest fluoroscopy,'' \emph{Medical physics}, vol.~19, no.~5, pp. 1183--1189, 1992.

\bibitem{mahesh2001fluoroscopy}
M.~Mahesh, ``Fluoroscopy: patient radiation exposure issues,'' \emph{Radiographics}, vol.~21, no.~4, pp. 1033--1045, 2001.

\bibitem{hoheisel2006review}
M.~Hoheisel, ``Review of medical imaging with emphasis on x-ray detectors,'' \emph{Nuclear Instruments and Methods in Physics Research Section A: Accelerators, Spectrometers, Detectors and Associated Equipment}, vol. 563, no.~1, pp. 215--224, 2006.

\bibitem{ulrich2022understanding}
H.~Ulrich, A.-K. Kock-Schoppenhauer, N.~Deppenwiese, R.~G{\"o}tt, J.~Kern, M.~Lablans, R.~W. Majeed, M.~R. St{\"o}hr, J.~Stausberg, J.~Varghese \emph{et~al.}, ``Understanding the nature of metadata: systematic review,'' \emph{Journal of medical Internet research}, vol.~24, no.~1, p. e25440, 2022.

\bibitem{sweet2013electronic}
L.~E. Sweet and H.~L. Moulaison, ``Electronic health records data and metadata: challenges for big data in the united states,'' \emph{Big data}, vol.~1, no.~4, pp. 245--251, 2013.

\bibitem{mclean2008electronic}
T.~R. McLean, L.~Burton, C.~C. Haller, and P.~B. McLean, ``Electronic medical record metadata: uses and liability,'' \emph{Journal of the American College of Surgeons}, vol. 206, no.~3, pp. 405--411, 2008.

\bibitem{vardaki2009statistical}
M.~Vardaki, H.~Papageorgiou, and F.~Pentaris, ``A statistical metadata model for clinical trials’ data management,'' \emph{Computer methods and programs in biomedicine}, vol.~95, no.~2, pp. 129--145, 2009.

\bibitem{de2015data}
D.~D. de~Macedo, A.~Von~Wangenheim, and M.~A. Dantas, ``A data storage approach for large-scale distributed medical systems,'' in \emph{2015 Ninth International Conference on Complex, Intelligent, and Software Intensive Systems}.\hskip 1em plus 0.5em minus 0.4em\relax IEEE, 2015, pp. 486--490.

\bibitem{Lim2006}
J.~Lim and R.~Zein, ``The digital imaging and communications in medicine (dicom): Description, structure and applications,'' in \emph{Rapid Prototyping: Theory and Practice}.\hskip 1em plus 0.5em minus 0.4em\relax Springer US, 2006, pp. 63--86.

\bibitem{gur2017towards}
Y.~Gur, M.~Moradi, H.~Bulu, Y.~Guo, C.~Compas, and T.~Syeda-Mahmood, ``Towards an efficient way of building annotated medical image collections for big data studies,'' in \emph{Intravascular Imaging and Computer Assisted Stenting, and Large-Scale Annotation of Biomedical Data and Expert Label Synthesis: 6th Joint International Workshops, CVII-STENT 2017 and Second International Workshop, LABELS 2017, Held in Conjunction with MICCAI 2017, Qu{\'e}bec City, QC, Canada, September 10--14, 2017, Proceedings 2}.\hskip 1em plus 0.5em minus 0.4em\relax Springer, 2017, pp. 87--95.

\bibitem{patel2018annotation}
P.~Patel, D.~Davey, V.~Panchal, and P.~Pathak, ``Annotation of a large clinical entity corpus,'' in \emph{Proceedings of the 2018 Conference on Empirical Methods in Natural Language Processing}, 2018, pp. 2033--2042.

\bibitem{seifert2010semantic}
S.~Seifert, M.~Kelm, M.~Moeller, S.~Mukherjee, A.~Cavallaro, M.~Huber, and D.~Comaniciu, ``Semantic annotation of medical images,'' in \emph{Medical Imaging 2010: Advanced PACS-based Imaging Informatics and Therapeutic Applications}, vol. 7628.\hskip 1em plus 0.5em minus 0.4em\relax SPIE, 2010, pp. 43--50.

\bibitem{aljabri2022towards}
M.~Aljabri, M.~AlAmir, M.~AlGhamdi, M.~Abdel-Mottaleb, and F.~Collado-Mesa, ``Towards a better understanding of annotation tools for medical imaging: a survey,'' \emph{Multimedia tools and applications}, vol.~81, no.~18, pp. 25\,877--25\,911, 2022.

\bibitem{hajnal2001medical}
J.~V. Hajnal and D.~L. Hill, \emph{Medical image registration}.\hskip 1em plus 0.5em minus 0.4em\relax CRC press, 2001.

\bibitem{chileshe2023large}
E.~Chileshe and L.~Phiri, ``Large-scale analysis of medical image metadata,'' in \emph{Proceedings of International Conference for ICT (ICICT)-Zambia}, vol.~5, no.~1, 2023, pp. 44--48.

\bibitem{razzak2018deep}
M.~I. Razzak, S.~Naz, and A.~Zaib, ``Deep learning for medical image processing: Overview, challenges and the future,'' \emph{Classification in BioApps: Automation of Decision Making}, pp. 323--350, 2018.

\bibitem{ker2017deep}
J.~Ker, L.~Wang, J.~Rao, and T.~Lim, ``Deep learning applications in medical image analysis,'' \emph{Ieee Access}, vol.~6, pp. 9375--9389, 2017.

\bibitem{yadav2019deep}
S.~S. Yadav and S.~M. Jadhav, ``Deep convolutional neural network based medical image classification for disease diagnosis,'' \emph{Journal of Big data}, vol.~6, no.~1, pp. 1--18, 2019.

\bibitem{rana2023machine}
M.~Rana and M.~Bhushan, ``Machine learning and deep learning approach for medical image analysis: diagnosis to detection,'' \emph{Multimedia Tools and Applications}, vol.~82, no.~17, pp. 26\,731--26\,769, 2023.

\bibitem{latif2019medical}
J.~Latif, C.~Xiao, A.~Imran, and S.~Tu, ``Medical imaging using machine learning and deep learning algorithms: a review,'' in \emph{2019 2nd International conference on computing, mathematics and engineering technologies (iCoMET)}.\hskip 1em plus 0.5em minus 0.4em\relax IEEE, 2019, pp. 1--5.

\bibitem{chan2020deep}
H.-P. Chan, R.~K. Samala, L.~M. Hadjiiski, and C.~Zhou, ``Deep learning in medical image analysis,'' \emph{Deep Learning in Medical Image Analysis: Challenges and Applications}, pp. 3--21, 2020.

\bibitem{castiglioni2021ai}
I.~Castiglioni, L.~Rundo, M.~Codari, G.~Di~Leo, C.~Salvatore, M.~Interlenghi, F.~Gallivanone, A.~Cozzi, N.~C. D'Amico, and F.~Sardanelli, ``Ai applications to medical images: From machine learning to deep learning,'' \emph{Physica Medica}, vol.~83, pp. 9--24, 2021.

\bibitem{jeyaraj2019computer}
P.~R. Jeyaraj and E.~R. Samuel~Nadar, ``Computer-assisted medical image classification for early diagnosis of oral cancer employing deep learning algorithm,'' \emph{Journal of cancer research and clinical oncology}, vol. 145, pp. 829--837, 2019.

\bibitem{chan2020computer}
H.-P. Chan, L.~M. Hadjiiski, and R.~K. Samala, ``Computer-aided diagnosis in the era of deep learning,'' \emph{Medical physics}, vol.~47, no.~5, pp. e218--e227, 2020.

\bibitem{wang2022medical}
R.~Wang, T.~Lei, R.~Cui, B.~Zhang, H.~Meng, and A.~K. Nandi, ``Medical image segmentation using deep learning: A survey,'' \emph{IET Image Processing}, vol.~16, no.~5, pp. 1243--1267, 2022.

\bibitem{lai2015deep}
M.~Lai, ``Deep learning for medical image segmentation,'' \emph{arXiv preprint arXiv:1505.02000}, 2015.

\bibitem{wang2018interactive}
G.~Wang, W.~Li, M.~A. Zuluaga, R.~Pratt, P.~A. Patel, M.~Aertsen, T.~Doel, A.~L. David, J.~Deprest, S.~Ourselin \emph{et~al.}, ``Interactive medical image segmentation using deep learning with image-specific fine tuning,'' \emph{IEEE transactions on medical imaging}, vol.~37, no.~7, pp. 1562--1573, 2018.

\bibitem{echle2021deep}
A.~Echle, N.~T. Rindtorff, T.~J. Brinker, T.~Luedde, A.~T. Pearson, and J.~N. Kather, ``Deep learning in cancer pathology: a new generation of clinical biomarkers,'' \emph{British journal of cancer}, vol. 124, no.~4, pp. 686--696, 2021.

\bibitem{savadjiev2019image}
P.~Savadjiev, J.~Chong, A.~Dohan, V.~Agnus, R.~Forghani, C.~Reinhold, and B.~Gallix, ``Image-based biomarkers for solid tumor quantification,'' \emph{European radiology}, vol.~29, pp. 5431--5440, 2019.

\bibitem{xu2019deep}
Y.~Xu, A.~Hosny, R.~Zeleznik, C.~Parmar, T.~Coroller, I.~Franco, R.~H. Mak, and H.~J. Aerts, ``Deep learning predicts lung cancer treatment response from serial medical imaging,'' \emph{Clinical Cancer Research}, vol.~25, no.~11, pp. 3266--3275, 2019.

\bibitem{wang2018image}
G.~Wang, J.~C. Ye, K.~Mueller, and J.~A. Fessler, ``Image reconstruction is a new frontier of machine learning,'' \emph{IEEE transactions on medical imaging}, vol.~37, no.~6, pp. 1289--1296, 2018.

\bibitem{zhang2020review}
H.-M. Zhang and B.~Dong, ``A review on deep learning in medical image reconstruction,'' \emph{Journal of the Operations Research Society of China}, vol.~8, pp. 311--340, 2020.

\bibitem{ahishakiye2021survey}
E.~Ahishakiye, M.~Bastiaan Van~Gijzen, J.~Tumwiine, R.~Wario, and J.~Obungoloch, ``A survey on deep learning in medical image reconstruction,'' \emph{Intelligent Medicine}, vol.~1, no.~03, pp. 118--127, 2021.

\bibitem{vamathevan2019applications}
J.~Vamathevan, D.~Clark, P.~Czodrowski, I.~Dunham, E.~Ferran, G.~Lee, B.~Li, A.~Madabhushi, P.~Shah, M.~Spitzer \emph{et~al.}, ``Applications of machine learning in drug discovery and development,'' \emph{Nature reviews Drug discovery}, vol.~18, no.~6, pp. 463--477, 2019.

\bibitem{louie2007data}
B.~Louie, P.~Mork, F.~Martin-Sanchez, A.~Halevy, and P.~Tarczy-Hornoch, ``Data integration and genomic medicine,'' \emph{Journal of biomedical informatics}, vol.~40, no.~1, pp. 5--16, 2007.

\bibitem{he2017big}
K.~Y. He, D.~Ge, and M.~M. He, ``Big data analytics for genomic medicine,'' \emph{International journal of molecular sciences}, vol.~18, no.~2, p. 412, 2017.

\bibitem{libbrecht2015machine}
M.~W. Libbrecht and W.~S. Noble, ``Machine learning applications in genetics and genomics,'' \emph{Nature Reviews Genetics}, vol.~16, no.~6, pp. 321--332, 2015.

\bibitem{chafai2023emerging}
N.~Chafai, L.~Bonizzi, S.~Botti, and B.~Badaoui, ``Emerging applications of machine learning in genomic medicine and healthcare,'' \emph{Critical Reviews in Clinical Laboratory Sciences}, pp. 1--24, 2023.

\bibitem{diao2018biomedical}
J.~A. Diao, I.~S. Kohane, and A.~K. Manrai, ``Biomedical informatics and machine learning for clinical genomics,'' \emph{Human molecular genetics}, vol.~27, no.~R1, pp. R29--R34, 2018.

\bibitem{wu2018deep}
Q.~Wu, A.~Boueiz, A.~Bozkurt, A.~Masoomi, A.~Wang, D.~L. DeMeo, S.~T. Weiss, and W.~Qiu, ``Deep learning methods for predicting disease status using genomic data,'' \emph{Journal of biometrics \& biostatistics}, vol.~9, no.~5, 2018.

\bibitem{lin2017machine}
E.~Lin and H.-Y. Lane, ``Machine learning and systems genomics approaches for multi-omics data,'' \emph{Biomarker research}, vol.~5, pp. 1--6, 2017.

\bibitem{khan2023bioinformatics}
M.~Khan, ``Bioinformatics and machine learning: Analyzing genomic data for personalized medicine,'' 2023.

\bibitem{whalen2022navigating}
S.~Whalen, J.~Schreiber, W.~S. Noble, and K.~S. Pollard, ``Navigating the pitfalls of applying machine learning in genomics,'' \emph{Nature Reviews Genetics}, vol.~23, no.~3, pp. 169--181, 2022.

\bibitem{akgun2015privacy}
M.~Akg{\"u}n, A.~O. Bayrak, B.~Ozer, and M.~{\c{S}}. Sa{\u{g}}{\i}ro{\u{g}}lu, ``Privacy preserving processing of genomic data: A survey,'' \emph{Journal of biomedical informatics}, vol.~56, pp. 103--111, 2015.

\bibitem{leache2017utility}
A.~D. Leach{\'e} and J.~R. Oaks, ``The utility of single nucleotide polymorphism (snp) data in phylogenetics,'' \emph{Annual Review of Ecology, Evolution, and Systematics}, vol.~48, pp. 69--84, 2017.

\bibitem{zarrei2015copy}
M.~Zarrei, J.~R. MacDonald, D.~Merico, and S.~W. Scherer, ``A copy number variation map of the human genome,'' \emph{Nature reviews genetics}, vol.~16, no.~3, pp. 172--183, 2015.

\bibitem{audic1997significance}
S.~Audic and J.-M. Claverie, ``The significance of digital gene expression profiles,'' \emph{Genome research}, vol.~7, no.~10, pp. 986--995, 1997.

\bibitem{portela2010epigenetic}
A.~Portela and M.~Esteller, ``Epigenetic modifications and human disease,'' \emph{Nature biotechnology}, vol.~28, no.~10, pp. 1057--1068, 2010.

\bibitem{sherry2001dbsnp}
S.~T. Sherry, M.-H. Ward, M.~Kholodov, J.~Baker, L.~Phan, E.~M. Smigielski, and K.~Sirotkin, ``dbsnp: the ncbi database of genetic variation,'' \emph{Nucleic acids research}, vol.~29, no.~1, pp. 308--311, 2001.

\bibitem{gulamali2022machine}
F.~F. Gulamali, A.~S. Sawant, and G.~N. Nadkarni, ``Machine learning for risk stratification in kidney disease,'' \emph{Current Opinion in Nephrology and Hypertension}, vol.~31, no.~6, pp. 548--552, 2022.

\bibitem{okser2013genetic}
S.~Okser, T.~Pahikkala, and T.~Aittokallio, ``Genetic variants and their interactions in disease risk prediction--machine learning and network perspectives,'' \emph{BioData mining}, vol.~6, no.~1, pp. 1--16, 2013.

\bibitem{tseng2020development}
Y.-J. Tseng, H.-Y. Wang, T.-W. Lin, J.-J. Lu, C.-H. Hsieh, and C.-T. Liao, ``Development of a machine learning model for survival risk stratification of patients with advanced oral cancer,'' \emph{JAMA network open}, vol.~3, no.~8, pp. e2\,011\,768--e2\,011\,768, 2020.

\bibitem{kruppa2012risk}
J.~Kruppa, A.~Ziegler, and I.~R. K{\"o}nig, ``Risk estimation and risk prediction using machine-learning methods,'' \emph{Human genetics}, vol. 131, pp. 1639--1654, 2012.

\bibitem{quazi2022artificial}
S.~Quazi, ``Artificial intelligence and machine learning in precision and genomic medicine,'' \emph{Medical Oncology}, vol.~39, no.~8, p. 120, 2022.

\bibitem{west2006embracing}
M.~West, G.~S. Ginsburg, A.~T. Huang, and J.~R. Nevins, ``Embracing the complexity of genomic data for personalized medicine,'' \emph{Genome research}, vol.~16, no.~5, pp. 559--566, 2006.

\bibitem{ginsburg2009genomic}
G.~S. Ginsburg and H.~F. Willard, ``Genomic and personalized medicine: foundations and applications,'' \emph{Translational research}, vol. 154, no.~6, pp. 277--287, 2009.

\bibitem{vadapalli2022artificial}
S.~Vadapalli, H.~Abdelhalim, S.~Zeeshan, and Z.~Ahmed, ``Artificial intelligence and machine learning approaches using gene expression and variant data for personalized medicine,'' \emph{Briefings in bioinformatics}, vol.~23, no.~5, p. bbac191, 2022.

\bibitem{maceachern2021machine}
S.~J. MacEachern and N.~D. Forkert, ``Machine learning for precision medicine,'' \emph{Genome}, vol.~64, no.~4, pp. 416--425, 2021.

\bibitem{gupta2021artificial}
R.~Gupta, D.~Srivastava, M.~Sahu, S.~Tiwari, R.~K. Ambasta, and P.~Kumar, ``Artificial intelligence to deep learning: machine intelligence approach for drug discovery,'' \emph{Molecular diversity}, vol.~25, pp. 1315--1360, 2021.

\bibitem{carracedo2021review}
P.~Carracedo-Reboredo, J.~Li{\~n}ares-Blanco, N.~Rodr{\'\i}guez-Fern{\'a}ndez, F.~Cedr{\'o}n, F.~J. Novoa, A.~Carballal, V.~Maojo, A.~Pazos, and C.~Fernandez-Lozano, ``A review on machine learning approaches and trends in drug discovery,'' \emph{Computational and structural biotechnology journal}, vol.~19, pp. 4538--4558, 2021.

\bibitem{ramsundar2019deep}
B.~Ramsundar, P.~Eastman, P.~Walters, and V.~Pande, \emph{Deep learning for the life sciences: applying deep learning to genomics, microscopy, drug discovery, and more}.\hskip 1em plus 0.5em minus 0.4em\relax " O'Reilly Media, Inc.", 2019.

\bibitem{dana2018deep}
D.~Dana, S.~V. Gadhiya, L.~G. St.~Surin, D.~Li, F.~Naaz, Q.~Ali, L.~Paka, M.~A. Yamin, M.~Narayan, I.~D. Goldberg \emph{et~al.}, ``Deep learning in drug discovery and medicine; scratching the surface,'' \emph{Molecules}, vol.~23, no.~9, p. 2384, 2018.

\bibitem{gordon2018future}
E.~S. Gordon, D.~Babu, and D.~A. Laney, ``The future is now: Technology's impact on the practice of genetic counseling,'' in \emph{American Journal of Medical Genetics Part C: Seminars in Medical Genetics}, vol. 178, no.~1.\hskip 1em plus 0.5em minus 0.4em\relax Wiley Online Library, 2018, pp. 15--23.

\bibitem{kearney2020artificial}
E.~Kearney, A.~Wojcik, and D.~Babu, ``Artificial intelligence in genetic services delivery: utopia or apocalypse?'' \emph{Journal of genetic counseling}, vol.~29, no.~1, pp. 8--17, 2020.

\bibitem{dias2019artificial}
R.~Dias and A.~Torkamani, ``Artificial intelligence in clinical and genomic diagnostics,'' \emph{Genome medicine}, vol.~11, no.~1, pp. 1--12, 2019.

\bibitem{benning2022advances}
L.~Benning, A.~Peintner, and L.~Peintner, ``Advances in and the applicability of machine learning-based screening and early detection approaches for cancer: A primer,'' \emph{Cancers}, vol.~14, no.~3, p. 623, 2022.

\bibitem{venugopalan2021multimodal}
J.~Venugopalan, L.~Tong, H.~R. Hassanzadeh, and M.~D. Wang, ``Multimodal deep learning models for early detection of alzheimer’s disease stage,'' \emph{Scientific reports}, vol.~11, no.~1, p. 3254, 2021.

\bibitem{tao2020machine}
K.~Tao, Z.~Bian, Q.~Zhang, X.~Guo, C.~Yin, Y.~Wang, K.~Zhou, S.~Wan, M.~Shi, D.~Bao \emph{et~al.}, ``Machine learning-based genome-wide interrogation of somatic copy number aberrations in circulating tumor dna for early detection of hepatocellular carcinoma,'' \emph{EBioMedicine}, vol.~56, 2020.

\bibitem{mullen2021race}
M.~Mullen, A.~Zhang, G.~K. Lui, A.~W. Romfh, J.-W. Rhee, and J.~C. Wu, ``Race and genetics in congenital heart disease: application of ipscs, omics, and machine learning technologies,'' \emph{Frontiers in Cardiovascular Medicine}, vol.~8, p. 635280, 2021.

\bibitem{shameer2018machine}
K.~Shameer, K.~W. Johnson, B.~S. Glicksberg, J.~T. Dudley, and P.~P. Sengupta, ``Machine learning in cardiovascular medicine: are we there yet?'' \emph{Heart}, 2018.

\bibitem{leung2015machine}
M.~K. Leung, A.~Delong, B.~Alipanahi, and B.~J. Frey, ``Machine learning in genomic medicine: a review of computational problems and data sets,'' \emph{Proceedings of the IEEE}, vol. 104, no.~1, pp. 176--197, 2015.

\bibitem{schmidt2021deep}
B.~Schmidt and A.~Hildebrandt, ``Deep learning in next-generation sequencing,'' \emph{Drug discovery today}, vol.~26, no.~1, pp. 173--180, 2021.

\bibitem{cortes2022computational}
I.~Cort{\'e}s-Ciriano, D.~C. Gulhan, J.~J.-K. Lee, G.~E. Melloni, and P.~J. Park, ``Computational analysis of cancer genome sequencing data,'' \emph{Nature Reviews Genetics}, vol.~23, no.~5, pp. 298--314, 2022.

\bibitem{dargan2020comprehensive}
S.~Dargan and M.~Kumar, ``A comprehensive survey on the biometric recognition systems based on physiological and behavioral modalities,'' \emph{Expert Systems with Applications}, vol. 143, p. 113114, 2020.

\bibitem{pinto2018evolution}
J.~R. Pinto, J.~S. Cardoso, and A.~Louren{\c{c}}o, ``Evolution, current challenges, and future possibilities in ecg biometrics,'' \emph{IEEE Access}, vol.~6, pp. 34\,746--34\,776, 2018.

\bibitem{paranjape2001electroencephalogram}
R.~Paranjape, J.~Mahovsky, L.~Benedicenti, and Z.~Koles, ``The electroencephalogram as a biometric,'' in \emph{Canadian Conference on Electrical and Computer Engineering 2001. Conference Proceedings (Cat. No. 01TH8555)}, vol.~2.\hskip 1em plus 0.5em minus 0.4em\relax IEEE, 2001, pp. 1363--1366.

\bibitem{abdulrahman2023comprehensive}
S.~A. Abdulrahman and B.~Alhayani, ``A comprehensive survey on the biometric systems based on physiological and behavioural characteristics,'' \emph{Materials Today: Proceedings}, vol.~80, pp. 2642--2646, 2023.

\bibitem{boulgouris2009biometrics}
N.~V. Boulgouris, K.~N. Plataniotis, and E.~Micheli-Tzanakou, \emph{Biometrics: theory, methods, and applications}.\hskip 1em plus 0.5em minus 0.4em\relax John Wiley \& Sons, 2009.

\bibitem{kisku2019design}
D.~R. Kisku, P.~Gupta, and J.~K. Sing, \emph{Design and Implementation of Healthcare Biometric Systems}.\hskip 1em plus 0.5em minus 0.4em\relax IGI Global, 2019.

\bibitem{fatima2019biometric}
K.~Fatima, S.~Nawaz, and S.~Mehrban, ``Biometric authentication in health care sector: A survey,'' in \emph{2019 International Conference on Innovative Computing (ICIC)}.\hskip 1em plus 0.5em minus 0.4em\relax IEEE, 2019, pp. 1--10.

\bibitem{mason2020investigation}
J.~Mason, R.~Dave, P.~Chatterjee, I.~Graham-Allen, A.~Esterline, and K.~Roy, ``An investigation of biometric authentication in the healthcare environment,'' \emph{Array}, vol.~8, p. 100042, 2020.

\bibitem{okoh2015biometrics}
E.~Okoh and A.~I. Awad, ``Biometrics applications in e-health security: A preliminary survey,'' in \emph{Health Information Science: 4th International Conference, HIS 2015, Melbourne, Australia, May 28-30, 2015, Proceedings 4}.\hskip 1em plus 0.5em minus 0.4em\relax Springer, 2015, pp. 92--103.

\bibitem{kaul2020secure}
S.~D. Kaul, V.~K. Murty, and D.~Hatzinakos, ``Secure and privacy preserving biometric based user authentication with data access control system in the healthcare environment,'' in \emph{2020 International Conference on Cyberworlds (CW)}.\hskip 1em plus 0.5em minus 0.4em\relax IEEE, 2020, pp. 249--256.

\bibitem{hei2011biometric}
X.~Hei and X.~Du, ``Biometric-based two-level secure access control for implantable medical devices during emergencies,'' in \emph{2011 Proceedings IEEE INFOCOM}.\hskip 1em plus 0.5em minus 0.4em\relax IEEE, 2011, pp. 346--350.

\bibitem{barka2022implementation}
E.~Barka, M.~Al~Baqari, C.~A. Kerrache, and J.~Herrera-Tapia, ``Implementation of a biometric-based blockchain system for preserving privacy, security, and access control in healthcare records,'' \emph{Journal of Sensor and Actuator Networks}, vol.~11, no.~4, p.~85, 2022.

\bibitem{mohsin2018real}
A.~H. Mohsin, A.~Zaidan, B.~Zaidan, A.~S. Albahri, O.~S. Albahri, M.~Alsalem, and K.~Mohammed, ``Real-time remote health monitoring systems using body sensor information and finger vein biometric verification: A multi-layer systematic review,'' \emph{Journal of medical systems}, vol.~42, pp. 1--36, 2018.

\bibitem{choi2022hybrid}
J.-h. Choi, K.~Khamraev, and D.~Cheriyan, ``Hybrid health risk assessment model using real-time particulate matter, biometrics, and benchmark device,'' \emph{Journal of Cleaner Production}, vol. 350, p. 131443, 2022.

\bibitem{kumar2019deep}
U.~Kumar, E.~Tripathi, S.~P. Tripathi, and K.~K. Gupta, ``Deep learning for healthcare biometrics,'' in \emph{Design and Implementation of Healthcare Biometric Systems}.\hskip 1em plus 0.5em minus 0.4em\relax IGI Global, 2019, pp. 73--108.

\bibitem{sengar2020multimodal}
S.~S. Sengar, U.~Hariharan, and K.~Rajkumar, ``Multimodal biometric authentication system using deep learning method,'' in \emph{2020 International Conference on Emerging Smart Computing and Informatics (ESCI)}.\hskip 1em plus 0.5em minus 0.4em\relax IEEE, 2020, pp. 309--312.

\bibitem{piera2020patient}
J.~Piera-Jim{\'e}nez, J.~Dooling, P.~Ranade-Kharkar, S.~Pollock, D.~Mann, S.~Thornton, R.~Duggal, S.~Khatri, B.~Shukla, R.~Rudin \emph{et~al.}, ``Patient identification techniques--approaches, implications, and findings,'' \emph{Yearbook of medical informatics}, vol.~29, no.~01, pp. 081--086, 2020.

\bibitem{leonard2008realization}
D.~Leonard, A.~P. Pons, and S.~S. Asfour, ``Realization of a universal patient identifier for electronic medical records through biometric technology,'' \emph{IEEE Transactions on Information Technology in Biomedicine}, vol.~13, no.~4, pp. 494--500, 2008.

\bibitem{jonas2014patient}
M.~Jonas, S.~Solangasenathirajan, and D.~Hett, ``Patient identification, a review of the use of biometrics in the icu,'' \emph{Annual Update in Intensive Care and Emergency Medicine 2014}, pp. 679--688, 2014.

\bibitem{sohn2020clinical}
J.~W. Sohn, H.~Kim, S.~B. Park, S.~Lee, J.~I. Monroe, T.~B. Malone, T.~Kinsella, M.~Yao, C.~Kunos, S.~S. Lo \emph{et~al.}, ``Clinical study of using biometrics to identify patient and procedure,'' \emph{Frontiers in Oncology}, vol.~10, p. 586232, 2020.

\bibitem{fatimah2022biometric}
B.~Fatimah, P.~Singh, A.~Singhal, and R.~B. Pachori, ``Biometric identification from ecg signals using fourier decomposition and machine learning,'' \emph{IEEE Transactions on Instrumentation and Measurement}, vol.~71, pp. 1--9, 2022.

\bibitem{labati2019deep}
R.~D. Labati, E.~Mu{\~n}oz, V.~Piuri, R.~Sassi, and F.~Scotti, ``Deep-ecg: Convolutional neural networks for ecg biometric recognition,'' \emph{Pattern Recognition Letters}, vol. 126, pp. 78--85, 2019.

\bibitem{prakash2023deep}
A.~J. Prakash, K.~K. Patro, S.~Samantray, P.~P{\l}awiak, and M.~Hammad, ``A deep learning technique for biometric authentication using ecg beat template matching,'' \emph{Information}, vol.~14, no.~2, p.~65, 2023.

\bibitem{ismail2022recent}
S.~N.~A. Ismail, N.~A. Nayan, R.~Jaafar, and Z.~May, ``Recent advances in non-invasive blood pressure monitoring and prediction using a machine learning approach,'' \emph{Sensors}, vol.~22, no.~16, p. 6195, 2022.

\bibitem{fei2021machine}
C.~Fei, R.~Liu, Z.~Li, T.~Wang, and F.~N. Baig, ``Machine and deep learning algorithms for wearable health monitoring,'' in \emph{Computational intelligence in healthcare}.\hskip 1em plus 0.5em minus 0.4em\relax Springer, 2021, pp. 105--160.

\bibitem{gomes2023survey}
N.~Gomes, M.~Pato, A.~R. Louren{\c{c}}o, and N.~Datia, ``A survey on wearable sensors for mental health monitoring,'' \emph{Sensors}, vol.~23, no.~3, p. 1330, 2023.

\bibitem{xefteris2016behavioral}
S.~Xefteris, N.~Doulamis, V.~Andronikou, T.~Varvarigou, and G.~Cambourakis, ``Behavioral biometrics in assisted living: a methodology for emotion recognition,'' \emph{Engineering, Technology \& Applied Science Research}, vol.~6, no.~4, pp. 1035--1044, 2016.

\bibitem{patel2018mental}
A.~N. Patel, M.~D. Howard, S.~M. Roach, A.~P. Jones, N.~B. Bryant, C.~S. Robinson, V.~P. Clark, and P.~K. Pilly, ``Mental state assessment and validation using personalized physiological biometrics,'' \emph{Frontiers in human neuroscience}, vol.~12, p. 221, 2018.

\bibitem{killoran2023can}
J.~Killoran, Y.~G. Cui, A.~Park, P.~van Esch, and J.~Kietzmann, ``Can behavioral biometrics make everyone happy?'' \emph{Business Horizons}, 2023.

\bibitem{garcia2018mental}
E.~Garcia-Ceja, M.~Riegler, T.~Nordgreen, P.~Jakobsen, K.~J. Oedegaard, and J.~T{\o}rresen, ``Mental health monitoring with multimodal sensing and machine learning: A survey,'' \emph{Pervasive and Mobile Computing}, vol.~51, pp. 1--26, 2018.

\bibitem{arneric2017biometric}
S.~P. Arneri{\'c}, J.~M. Cedarbaum, S.~Khozin, S.~Papapetropoulos, D.~L. Hill, M.~Ropacki, J.~Rhodes, P.~A. Dacks, L.~D. Hudson, M.~F. Gordon \emph{et~al.}, ``Biometric monitoring devices for assessing end points in clinical trials: developing an ecosystem,'' \emph{Nature Reviews Drug Discovery}, vol.~16, no.~10, pp. 736--736, 2017.

\bibitem{harrer2019artificial}
S.~Harrer, P.~Shah, B.~Antony, and J.~Hu, ``Artificial intelligence for clinical trial design,'' \emph{Trends in pharmacological sciences}, vol.~40, no.~8, pp. 577--591, 2019.

\bibitem{weissler2021role}
E.~H. Weissler, T.~Naumann, T.~Andersson, R.~Ranganath, O.~Elemento, Y.~Luo, D.~F. Freitag, J.~Benoit, M.~C. Hughes, F.~Khan \emph{et~al.}, ``The role of machine learning in clinical research: transforming the future of evidence generation,'' \emph{Trials}, vol.~22, no.~1, pp. 1--15, 2021.

\bibitem{lyakso2021voice}
E.~Lyakso, O.~Frolova, and A.~Nikolaev, ``Voice and speech features as diagnostic symptom,'' \emph{C. Pracana \& M. Wang, Psychological applications and trends}, pp. 259--263, 2021.

\bibitem{niebudek2006diagnostic}
E.~Niebudek-Bogusz, M.~Fiszer, P.~Kotylo, and M.~Sliwinska-Kowalska, ``Diagnostic value of voice acoustic analysis in assessment of occupational voice pathologies in teachers,'' \emph{Logopedics Phoniatrics Vocology}, vol.~31, no.~3, pp. 100--106, 2006.

\bibitem{moro2021advances}
L.~Moro-Velazquez, J.~A. Gomez-Garcia, J.~D. Arias-Londo{\~n}o, N.~Dehak, and J.~I. Godino-Llorente, ``Advances in parkinson's disease detection and assessment using voice and speech: A review of the articulatory and phonatory aspects,'' \emph{Biomedical Signal Processing and Control}, vol.~66, p. 102418, 2021.

\bibitem{ngo2022computerized}
Q.~C. Ngo, M.~A. Motin, N.~D. Pah, P.~Drot{\'a}r, P.~Kempster, and D.~Kumar, ``Computerized analysis of speech and voice for parkinson's disease: A systematic review,'' \emph{Computer Methods and Programs in Biomedicine}, p. 107133, 2022.

\bibitem{shahbakhi2014speech}
M.~Shahbakhi, D.~T. Far, and E.~Tahami, ``Speech analysis for diagnosis of parkinson’s disease using genetic algorithm and support vector machine,'' \emph{Journal of Biomedical Science and Engineering}, vol. 2014, 2014.

\bibitem{lella2021automatic}
K.~K. Lella and A.~Pja, ``Automatic covid-19 disease diagnosis using 1d convolutional neural network and augmentation with human respiratory sound based on parameters: Cough, breath, and voice,'' \emph{AIMS public health}, vol.~8, no.~2, p. 240, 2021.

\bibitem{idrisoglu2023applied}
A.~Idrisoglu, A.~L. Dallora, P.~Anderberg, and J.~S. Berglund, ``Applied machine learning techniques to diagnose voice-affecting conditions and disorders: Systematic literature review,'' \emph{Journal of Medical Internet Research}, vol.~25, p. e46105, 2023.

\bibitem{jeon2019facial}
B.~Jeon, B.~Jeong, S.~Jee, Y.~Huang, Y.~Kim, G.~H. Park, J.~Kim, M.~Wufuer, X.~Jin, S.~W. Kim \emph{et~al.}, ``A facial recognition mobile app for patient safety and biometric identification: Design, development, and validation,'' \emph{JMIR mHealth and uHealth}, vol.~7, no.~4, p. e11472, 2019.

\bibitem{onyema2021enhancement}
E.~M. Onyema, P.~K. Shukla, S.~Dalal, M.~N. Mathur, M.~Zakariah, B.~Tiwari \emph{et~al.}, ``Enhancement of patient facial recognition through deep learning algorithm: Convnet,'' \emph{Journal of Healthcare Engineering}, vol. 2021, 2021.

\bibitem{ghazal2021iot}
T.~M. Ghazal, M.~K. Hasan, M.~T. Alshurideh, H.~M. Alzoubi, M.~Ahmad, S.~S. Akbar, B.~Al~Kurdi, and I.~A. Akour, ``Iot for smart cities: Machine learning approaches in smart healthcare—a review,'' \emph{Future Internet}, vol.~13, no.~8, p. 218, 2021.

\bibitem{balakrishna2020iot}
S.~Balakrishna, M.~Thirumaran, and V.~K. Solanki, ``Iot sensor data integration in healthcare using semantics and machine learning approaches,'' \emph{A handbook of internet of things in biomedical and cyber physical system}, pp. 275--300, 2020.

\bibitem{khan2016monitoring}
Y.~Khan, A.~E. Ostfeld, C.~M. Lochner, A.~Pierre, and A.~C. Arias, ``Monitoring of vital signs with flexible and wearable medical devices,'' \emph{Advanced materials}, vol.~28, no.~22, pp. 4373--4395, 2016.

\bibitem{yilmaz2010detecting}
T.~Yilmaz, R.~Foster, and Y.~Hao, ``Detecting vital signs with wearable wireless sensors,'' \emph{Sensors}, vol.~10, no.~12, pp. 10\,837--10\,862, 2010.

\bibitem{wu2021internet}
X.~Wu, C.~Liu, L.~Wang, and M.~Bilal, ``Internet of things-enabled real-time health monitoring system using deep learning,'' \emph{Neural Computing and Applications}, pp. 1--12, 2021.

\bibitem{lv2022wearable}
Z.~Lv and Y.~Li, ``Wearable sensors for vital signs measurement: a survey,'' \emph{Journal of Sensor and Actuator Networks}, vol.~11, no.~1, p.~19, 2022.

\bibitem{da2018internet}
C.~A. Da~Costa, C.~F. Pasluosta, B.~Eskofier, D.~B. Da~Silva, and R.~da~Rosa~Righi, ``Internet of health things: Toward intelligent vital signs monitoring in hospital wards,'' \emph{Artificial intelligence in medicine}, vol.~89, pp. 61--69, 2018.

\bibitem{attal2015physical}
F.~Attal, S.~Mohammed, M.~Dedabrishvili, F.~Chamroukhi, L.~Oukhellou, and Y.~Amirat, ``Physical human activity recognition using wearable sensors,'' \emph{Sensors}, vol.~15, no.~12, pp. 31\,314--31\,338, 2015.

\bibitem{yang2010review}
C.-C. Yang and Y.-L. Hsu, ``A review of accelerometry-based wearable motion detectors for physical activity monitoring,'' \emph{Sensors}, vol.~10, no.~8, pp. 7772--7788, 2010.

\bibitem{cornacchia2016survey}
M.~Cornacchia, K.~Ozcan, Y.~Zheng, and S.~Velipasalar, ``A survey on activity detection and classification using wearable sensors,'' \emph{IEEE Sensors Journal}, vol.~17, no.~2, pp. 386--403, 2016.

\bibitem{zhang2022deep}
S.~Zhang, Y.~Li, S.~Zhang, F.~Shahabi, S.~Xia, Y.~Deng, and N.~Alshurafa, ``Deep learning in human activity recognition with wearable sensors: A review on advances,'' \emph{Sensors}, vol.~22, no.~4, p. 1476, 2022.

\bibitem{nweke2018deep}
H.~F. Nweke, Y.~W. Teh, M.~A. Al-Garadi, and U.~R. Alo, ``Deep learning algorithms for human activity recognition using mobile and wearable sensor networks: State of the art and research challenges,'' \emph{Expert Systems with Applications}, vol. 105, pp. 233--261, 2018.

\bibitem{rodriguez2021mobile}
C.~Rodriguez-Le{\'o}n, C.~Villalonga, M.~Munoz-Torres, J.~R. Ruiz, and O.~Banos, ``Mobile and wearable technology for the monitoring of diabetes-related parameters: Systematic review,'' \emph{JMIR mHealth and uHealth}, vol.~9, no.~6, p. e25138, 2021.

\bibitem{zhu2022enhancing}
T.~Zhu, C.~Uduku, K.~Li, P.~Herrero, N.~Oliver, and P.~Georgiou, ``Enhancing self-management in type 1 diabetes with wearables and deep learning,'' \emph{npj Digital Medicine}, vol.~5, no.~1, p.~78, 2022.

\bibitem{liu2021deep}
X.~Liu, H.~Wang, Z.~Li, and L.~Qin, ``Deep learning in ecg diagnosis: A review,'' \emph{Knowledge-Based Systems}, vol. 227, p. 107187, 2021.

\bibitem{somani2021deep}
S.~Somani, A.~J. Russak, F.~Richter, S.~Zhao, A.~Vaid, F.~Chaudhry, J.~K. De~Freitas, N.~Naik, R.~Miotto, G.~N. Nadkarni \emph{et~al.}, ``Deep learning and the electrocardiogram: review of the current state-of-the-art,'' \emph{EP Europace}, vol.~23, no.~8, pp. 1179--1191, 2021.

\bibitem{minchole2019machine}
A.~Minchol{\'e}, J.~Camps, A.~Lyon, and B.~Rodr{\'\i}guez, ``Machine learning in the electrocardiogram,'' \emph{Journal of electrocardiology}, vol.~57, pp. S61--S64, 2019.

\bibitem{kwon2022flexible}
S.~H. Kwon and L.~Dong, ``Flexible sensors and machine learning for heart monitoring,'' \emph{Nano Energy}, p. 107632, 2022.

\bibitem{sathyanarayana2016sleep}
A.~Sathyanarayana, S.~Joty, L.~Fernandez-Luque, F.~Ofli, J.~Srivastava, A.~Elmagarmid, T.~Arora, S.~Taheri \emph{et~al.}, ``Sleep quality prediction from wearable data using deep learning,'' \emph{JMIR mHealth and uHealth}, vol.~4, no.~4, p. e6562, 2016.

\bibitem{arora2020analysis}
A.~Arora, P.~Chakraborty, and M.~Bhatia, ``Analysis of data from wearable sensors for sleep quality estimation and prediction using deep learning,'' \emph{Arabian Journal for Science and Engineering}, vol.~45, pp. 10\,793--10\,812, 2020.

\bibitem{zhang2020comprehensive}
W.~Zhang and S.~Ram, ``A comprehensive analysis of triggers and risk factors for asthma based on machine learning and large heterogeneous data sources.'' \emph{MIS Quarterly}, vol.~44, no.~1, 2020.

\bibitem{bohlmann2021machine}
A.~Bohlmann, J.~Mostafa, M.~Kumar \emph{et~al.}, ``Machine learning and medication adherence: scoping review,'' \emph{JMIRx Med}, vol.~2, no.~4, p. e26993, 2021.

\bibitem{roh2021deep}
H.~Roh, S.~Shin, J.~Han, and S.~Lim, ``A deep learning-based medication behavior monitoring system,'' \emph{Math. Biosci. Eng}, vol.~18, no.~2, pp. 1513--1528, 2021.

\bibitem{tunca2019deep}
C.~Tunca, G.~Salur, and C.~Ersoy, ``Deep learning for fall risk assessment with inertial sensors: Utilizing domain knowledge in spatio-temporal gait parameters,'' \emph{IEEE journal of biomedical and health informatics}, vol.~24, no.~7, pp. 1994--2005, 2019.

\bibitem{meyer2020wearables}
B.~M. Meyer, L.~J. Tulipani, R.~D. Gurchiek, D.~A. Allen, L.~Adamowicz, D.~Larie, A.~J. Solomon, N.~Cheney, and R.~S. McGinnis, ``Wearables and deep learning classify fall risk from gait in multiple sclerosis,'' \emph{IEEE journal of biomedical and health informatics}, vol.~25, no.~5, pp. 1824--1831, 2020.

\bibitem{al2019deep}
F.~Al~Machot, A.~Elmachot, M.~Ali, E.~Al~Machot, and K.~Kyamakya, ``A deep-learning model for subject-independent human emotion recognition using electrodermal activity sensors,'' \emph{Sensors}, vol.~19, no.~7, p. 1659, 2019.

\bibitem{kyamakya2021emotion}
K.~Kyamakya, F.~Al-Machot, A.~Haj~Mosa, H.~Bouchachia, J.~C. Chedjou, and A.~Bagula, ``Emotion and stress recognition related sensors and machine learning technologies,'' p. 2273, 2021.

\bibitem{gedam2021review}
S.~Gedam and S.~Paul, ``A review on mental stress detection using wearable sensors and machine learning techniques,'' \emph{IEEE Access}, vol.~9, pp. 84\,045--84\,066, 2021.

\bibitem{shapiro2012patient}
M.~Shapiro, D.~Johnston, J.~Wald, and D.~Mon, ``Patient-generated health data,'' \emph{RTI International, April}, vol. 813, p. 814, 2012.

\bibitem{hsueh2017making}
P.-Y.~S. Hsueh, S.~Dey, S.~Das, and T.~Wetter, ``Making sense of patient-generated health data for interpretable patient-centered care: the transition from “more” to “better”,'' in \emph{MEDINFO 2017: Precision Healthcare through Informatics}.\hskip 1em plus 0.5em minus 0.4em\relax IOS Press, 2017, pp. 113--117.

\bibitem{kao2017consumer}
C.-K. Kao and D.~M. Liebovitz, ``Consumer mobile health apps: current state, barriers, and future directions,'' \emph{PM\&R}, vol.~9, no.~5, pp. S106--S115, 2017.

\bibitem{peng2016qualitative}
W.~Peng, S.~Kanthawala, S.~Yuan, and S.~A. Hussain, ``A qualitative study of user perceptions of mobile health apps,'' \emph{BMC public health}, vol.~16, no.~1, pp. 1--11, 2016.

\bibitem{sama2014evaluation}
P.~R. Sama, Z.~J. Eapen, K.~P. Weinfurt, B.~R. Shah, and K.~A. Schulman, ``An evaluation of mobile health application tools,'' \emph{JMIR mHealth and uHealth}, vol.~2, no.~2, p. e3088, 2014.

\bibitem{istepanian2018m}
R.~S. Istepanian and T.~Al-Anzi, ``m-health 2.0: new perspectives on mobile health, machine learning and big data analytics,'' \emph{Methods}, vol. 151, pp. 34--40, 2018.

\bibitem{mendo2021machine}
I.~R. Mendo, G.~Marques, I.~de~la Torre~D{\'\i}ez, M.~L{\'o}pez-Coronado, and F.~Mart{\'\i}n-Rodr{\'\i}guez, ``Machine learning in medical emergencies: a systematic review and analysis,'' \emph{Journal of Medical Systems}, vol.~45, no.~10, p.~88, 2021.

\bibitem{van2013using}
M.~Van~der Eijk, M.~J. Faber, J.~W. Aarts, J.~A. Kremer, M.~Munneke, and B.~R. Bloem, ``Using online health communities to deliver patient-centered care to people with chronic conditions,'' \emph{Journal of medical Internet research}, vol.~15, no.~6, p. e115, 2013.

\bibitem{gupta2020social}
A.~Gupta and R.~Katarya, ``Social media based surveillance systems for healthcare using machine learning: a systematic review,'' \emph{Journal of biomedical informatics}, vol. 108, p. 103500, 2020.

\bibitem{hasib2023depression}
K.~M. Hasib, M.~R. Islam, S.~Sakib, M.~A. Akbar, I.~Razzak, and M.~S. Alam, ``Depression detection from social networks data based on machine learning and deep learning techniques: An interrogative survey,'' \emph{IEEE Transactions on Computational Social Systems}, 2023.

\bibitem{johnson2021precision}
K.~B. Johnson, W.-Q. Wei, D.~Weeraratne, M.~E. Frisse, K.~Misulis, K.~Rhee, J.~Zhao, and J.~L. Snowdon, ``Precision medicine, ai, and the future of personalized health care,'' \emph{Clinical and translational science}, vol.~14, no.~1, pp. 86--93, 2021.

\bibitem{schunke2022rapid}
L.~C. Sch{\"u}nke, B.~Mello, C.~A. da~Costa, R.~S. Antunes, S.~J. Rigo, G.~de~Oliveira~Ramos, R.~da~Rosa~Righi, J.~N. Scherer, and B.~Donida, ``A rapid review of machine learning approaches for telemedicine in the scope of covid-19,'' \emph{Artificial Intelligence in Medicine}, vol. 129, p. 102312, 2022.

\bibitem{verma2022tele}
S.~Verma, R.~Malviya, M.~A. Alam, and B.~D. Tripathi, ``Tele-health monitoring using artificial intelligence deep learning framework,'' \emph{Deep Learning for Targeted Treatments: Transformation in Healthcare}, pp. 199--228, 2022.

\bibitem{national2010prevention}
N.~R. Council \emph{et~al.}, ``The prevention and treatment of missing data in clinical trials,'' 2010.

\bibitem{pettit2021artificial}
R.~W. Pettit, R.~Fullem, C.~Cheng, and C.~I. Amos, ``Artificial intelligence, machine learning, and deep learning for clinical outcome prediction,'' \emph{Emerging topics in life sciences}, vol.~5, no.~6, pp. 729--745, 2021.

\bibitem{liu2019advancing}
F.~Liu, C.~Weng, and H.~Yu, ``Advancing clinical research through natural language processing on electronic health records: traditional machine learning meets deep learning,'' \emph{Clinical Research Informatics}, pp. 357--378, 2019.

\bibitem{cohen2019informed}
I.~G. Cohen, ``Informed consent and medical artificial intelligence: What to tell the patient?'' \emph{Geo. LJ}, vol. 108, p. 1425, 2019.

\bibitem{mckeown2021ethical}
A.~McKeown, M.~Mourby, P.~Harrison, S.~Walker, M.~Sheehan, and I.~Singh, ``Ethical issues in consent for the reuse of data in health data platforms,'' \emph{Science and Engineering Ethics}, vol.~27, pp. 1--21, 2021.

\bibitem{miyoshi2021machine}
J.~Miyoshi, T.~Maeda, K.~Matsuoka, D.~Saito, S.~Miyoshi, M.~Matsuura, S.~Okamoto, S.~Tamura, and T.~Hisamatsu, ``Machine learning using clinical data at baseline predicts the efficacy of vedolizumab at week 22 in patients with ulcerative colitis,'' \emph{Scientific Reports}, vol.~11, no.~1, p. 16440, 2021.

\bibitem{chien2020machine}
I.~Chien, A.~Enrique, J.~Palacios, T.~Regan, D.~Keegan, D.~Carter, S.~Tschiatschek, A.~Nori, A.~Thieme, D.~Richards \emph{et~al.}, ``A machine learning approach to understanding patterns of engagement with internet-delivered mental health interventions,'' \emph{JAMA network open}, vol.~3, no.~7, pp. e2\,010\,791--e2\,010\,791, 2020.

\bibitem{benke2018artificial}
K.~Benke and G.~Benke, ``Artificial intelligence and big data in public health,'' \emph{International journal of environmental research and public health}, vol.~15, no.~12, p. 2796, 2018.

\bibitem{song2023using}
L.~Song, Y.~Li, S.~Nie, Z.~Feng, Y.~Liu, F.~Ding, L.~Gong, L.~Liu, and G.~Yang, ``Using machine learning to predict adverse events in acute coronary syndrome: A retrospective study,'' \emph{Clinical Cardiology}, vol.~46, no.~12, pp. 1594--1602, 2023.

\bibitem{yang2024machine}
J.~Yang, J.~Wan, L.~Feng, S.~Hou, K.~Yv, L.~Xu, and K.~Chen, ``Machine learning algorithms for the prediction of adverse prognosis in patients undergoing peritoneal dialysis,'' \emph{BMC Medical Informatics and Decision Making}, vol.~24, no.~1, p.~8, 2024.

\bibitem{badwan2023machine}
B.~A. Badwan, G.~Liaropoulos, E.~Kyrodimos, D.~Skaltsas, A.~Tsirigos, and V.~G. Gorgoulis, ``Machine learning approaches to predict drug efficacy and toxicity in oncology,'' \emph{Cell Reports Methods}, vol.~3, no.~2, 2023.

\bibitem{gayvert2016data}
K.~M. Gayvert, N.~S. Madhukar, and O.~Elemento, ``A data-driven approach to predicting successes and failures of clinical trials,'' \emph{Cell chemical biology}, vol.~23, no.~10, pp. 1294--1301, 2016.

\bibitem{ezzati2020machine}
A.~Ezzati, R.~B. Lipton, A.~D.~N. Initiative \emph{et~al.}, ``Machine learning predictive models can improve efficacy of clinical trials for alzheimer’s disease,'' \emph{Journal of Alzheimer's Disease}, vol.~74, no.~1, pp. 55--63, 2020.

\bibitem{feijoo2020key}
F.~Feijoo, M.~Palopoli, J.~Bernstein, S.~Siddiqui, and T.~E. Albright, ``Key indicators of phase transition for clinical trials through machine learning,'' \emph{Drug discovery today}, vol.~25, no.~2, pp. 414--421, 2020.

\bibitem{zame2020machine}
W.~R. Zame, I.~Bica, C.~Shen, A.~Curth, H.-S. Lee, S.~Bailey, J.~Weatherall, D.~Wright, F.~Bretz, and M.~van~der Schaar, ``Machine learning for clinical trials in the era of covid-19,'' \emph{Statistics in biopharmaceutical research}, vol.~12, no.~4, pp. 506--517, 2020.

\bibitem{chalasani2023artificial}
S.~H. Chalasani, J.~Syed, M.~Ramesh, V.~Patil, and T.~P. Kumar, ``Artificial intelligence in the field of pharmacy practice: A literature review,'' \emph{Exploratory Research in Clinical and Social Pharmacy}, vol.~12, p. 100346, 2023.

\bibitem{askr2023deep}
H.~Askr, E.~Elgeldawi, H.~Aboul~Ella, Y.~A. Elshaier, M.~M. Gomaa, and A.~E. Hassanien, ``Deep learning in drug discovery: an integrative review and future challenges,'' \emph{Artificial Intelligence Review}, vol.~56, no.~7, pp. 5975--6037, 2023.

\bibitem{nemati2016optimal}
S.~Nemati, M.~M. Ghassemi, and G.~D. Clifford, ``Optimal medication dosing from suboptimal clinical examples: A deep reinforcement learning approach,'' in \emph{2016 38th annual international conference of the IEEE engineering in medicine and biology society (EMBC)}.\hskip 1em plus 0.5em minus 0.4em\relax IEEE, 2016, pp. 2978--2981.

\bibitem{li2023machine}
Q.-Y. Li, B.-H. Tang, Y.-E. Wu, B.-F. Yao, W.~Zhang, Y.~Zheng, Y.~Zhou, J.~van~den Anker, G.-X. Hao, and W.~Zhao, ``Machine learning: a new approach for dose individualization,'' \emph{Clinical Pharmacology \& Therapeutics}, 2023.

\bibitem{lin2022can}
Y.-T. Lin, C.-Y. Chu, K.-S. Hung, C.-H. Lu, E.~M. Bednarczyk, and H.-Y. Chen, ``Can machine learning predict pharmacotherapy outcomes? an application study in osteoporosis,'' \emph{Computer Methods and Programs in Biomedicine}, vol. 225, p. 107028, 2022.

\bibitem{del2023machine}
L.~Del~Fabro, E.~Bondi, F.~Serio, E.~Maggioni, A.~D’Agostino, and P.~Brambilla, ``Machine learning methods to predict outcomes of pharmacological treatment in psychosis,'' \emph{Translational Psychiatry}, vol.~13, no.~1, p.~75, 2023.

\bibitem{wang2018clinical}
Y.~Wang, L.~Wang, M.~Rastegar-Mojarad, S.~Moon, F.~Shen, N.~Afzal, S.~Liu, Y.~Zeng, S.~Mehrabi, S.~Sohn \emph{et~al.}, ``Clinical information extraction applications: a literature review,'' \emph{Journal of biomedical informatics}, vol.~77, pp. 34--49, 2018.

\bibitem{uzuner2010extracting}
{\"O}.~Uzuner, I.~Solti, and E.~Cadag, ``Extracting medication information from clinical text,'' \emph{Journal of the American Medical Informatics Association}, vol.~17, no.~5, pp. 514--518, 2010.

\bibitem{galozy2020prediction}
A.~Galozy and S.~Nowaczyk, ``Prediction and pattern analysis of medication refill adherence through electronic health records and dispensation data,'' \emph{Journal of Biomedical Informatics}, vol. 112, p. 100075, 2020.

\bibitem{hasan2021machine}
M.~M. Hasan, G.~J. Young, J.~Shi, P.~Mohite, L.~D. Young, S.~G. Weiner, and M.~Noor-E-Alam, ``A machine learning based two-stage clinical decision support system for predicting patients’ discontinuation from opioid use disorder treatment: retrospective observational study,'' \emph{BMC Medical Informatics and Decision Making}, vol.~21, pp. 1--21, 2021.

\bibitem{kim2022analyzing}
H.~R. Kim, M.~Sung, J.~A. Park, K.~Jeong, H.~H. Kim, S.~Lee, and Y.~R. Park, ``Analyzing adverse drug reaction using statistical and machine learning methods: A systematic review,'' \emph{Medicine}, vol. 101, no.~25, 2022.

\bibitem{chandak2020using}
P.~Chandak and N.~P. Tatonetti, ``Using machine learning to identify adverse drug effects posing increased risk to women,'' \emph{Patterns}, vol.~1, no.~7, 2020.

\bibitem{zhang2018learning}
S.~Zhang, S.~M.~H. Bamakan, Q.~Qu, and S.~Li, ``Learning for personalized medicine: a comprehensive review from a deep learning perspective,'' \emph{IEEE reviews in biomedical engineering}, vol.~12, pp. 194--208, 2018.

\bibitem{meng2023machine}
W.~Meng, X.~Zhang, B.~Ru, and Y.~Guan, ``A machine learning approach to real-world time to treatment discontinuation prediction,'' \emph{Advanced Intelligent Systems}, vol.~5, no.~4, p. 2200254, 2023.

\bibitem{visweswaran2010identifying}
S.~Visweswaran, J.~Mezger, G.~Clermont, M.~Hauskrecht, and G.~F. Cooper, ``Identifying deviations from usual medical care using a statistical approach,'' in \emph{AMIA Annual Symposium Proceedings}, vol. 2010.\hskip 1em plus 0.5em minus 0.4em\relax American Medical Informatics Association, 2010, p. 827.

\bibitem{gu2021predicting}
Y.~Gu, A.~Zalkikar, M.~Liu, L.~Kelly, A.~Hall, K.~Daly, and T.~Ward, ``Predicting medication adherence using ensemble learning and deep learning models with large scale healthcare data,'' \emph{Scientific Reports}, vol.~11, no.~1, p. 18961, 2021.

\bibitem{wu2020rapid}
J.~Wu, P.~Zhang, L.~Zhang, W.~Meng, J.~Li, C.~Tong, Y.~Li, J.~Cai, Z.~Yang, J.~Zhu \emph{et~al.}, ``Rapid and accurate identification of covid-19 infection through machine learning based on clinical available blood test results,'' \emph{MedRxiv}, pp. 2020--04, 2020.

\bibitem{khan2020review}
S.~Khan, M.~Sajjad, T.~Hussain, A.~Ullah, and A.~S. Imran, ``A review on traditional machine learning and deep learning models for wbcs classification in blood smear images,'' \emph{Ieee Access}, vol.~9, pp. 10\,657--10\,673, 2020.

\bibitem{alam2019machine}
M.~M. Alam and M.~T. Islam, ``Machine learning approach of automatic identification and counting of blood cells,'' \emph{Healthcare technology letters}, vol.~6, no.~4, pp. 103--108, 2019.

\bibitem{zeb2020towards}
B.~Zeb, A.~Khan, Y.~Khan, M.~F. Masood, I.~Tahir, and M.~Asad, ``Towards the selection of the best machine learning techniques and methods for urinalysis,'' in \emph{Proceedings of the 2020 12th International Conference on Machine Learning and Computing}, 2020, pp. 127--133.

\bibitem{chittora2021prediction}
P.~Chittora, S.~Chaurasia, P.~Chakrabarti, G.~Kumawat, T.~Chakrabarti, Z.~Leonowicz, M.~Jasi{\'n}ski, {\L}.~Jasi{\'n}ski, R.~Gono, E.~Jasi{\'n}ska \emph{et~al.}, ``Prediction of chronic kidney disease-a machine learning perspective,'' \emph{IEEE Access}, vol.~9, pp. 17\,312--17\,334, 2021.

\bibitem{de2023applications}
S.~De~Bruyne, P.~De~Kesel, and M.~Oyaert, ``Applications of artificial intelligence in urinalysis: Is the future already here?'' \emph{Clinical Chemistry}, vol.~69, no.~12, pp. 1348--1360, 2023.

\bibitem{peiffer2020machine}
N.~Peiffer-Smadja, S.~Delli{\`e}re, C.~Rodriguez, G.~Birgand, F.-X. Lescure, S.~Fourati, and E.~Rupp{\'e}, ``Machine learning in the clinical microbiology laboratory: has the time come for routine practice?'' \emph{Clinical Microbiology and Infection}, vol.~26, no.~10, pp. 1300--1309, 2020.

\bibitem{goodswen2021machine}
S.~J. Goodswen, J.~L. Barratt, P.~J. Kennedy, A.~Kaufer, L.~Calarco, and J.~T. Ellis, ``Machine learning and applications in microbiology,'' \emph{FEMS microbiology reviews}, vol.~45, no.~5, p. fuab015, 2021.

\bibitem{qu2019application}
K.~Qu, F.~Guo, X.~Liu, Y.~Lin, and Q.~Zou, ``Application of machine learning in microbiology,'' \emph{Frontiers in microbiology}, vol.~10, p. 827, 2019.

\bibitem{ghannam2021machine}
R.~B. Ghannam and S.~M. Techtmann, ``Machine learning applications in microbial ecology, human microbiome studies, and environmental monitoring,'' \emph{Computational and Structural Biotechnology Journal}, vol.~19, pp. 1092--1107, 2021.

\bibitem{madabhushi2016image}
A.~Madabhushi and G.~Lee, ``Image analysis and machine learning in digital pathology: Challenges and opportunities,'' \emph{Medical image analysis}, vol.~33, pp. 170--175, 2016.

\bibitem{janowczyk2016deep}
A.~Janowczyk and A.~Madabhushi, ``Deep learning for digital pathology image analysis: A comprehensive tutorial with selected use cases,'' \emph{Journal of pathology informatics}, vol.~7, no.~1, p.~29, 2016.

\bibitem{obstfeld2023hematology}
A.~E. Obstfeld, ``Hematology and machine learning,'' \emph{The Journal of Applied Laboratory Medicine}, vol.~8, no.~1, pp. 129--144, 2023.

\bibitem{radakovich2020machine}
N.~Radakovich, M.~Nagy, and A.~Nazha, ``Machine learning in haematological malignancies,'' \emph{The Lancet Haematology}, vol.~7, no.~7, pp. e541--e550, 2020.

\bibitem{fang2021using}
K.~Fang, Z.~Dong, X.~Chen, J.~Zhu, B.~Zhang, J.~You, Y.~Xiao, and W.~Xia, ``Using machine learning to identify clotted specimens in coagulation testing,'' \emph{Clinical Chemistry and Laboratory Medicine (CCLM)}, vol.~59, no.~7, pp. 1289--1297, 2021.

\bibitem{guo2021predicting}
K.~Guo, X.~Fu, H.~Zhang, M.~Wang, S.~Hong, and S.~Ma, ``Predicting the postoperative blood coagulation state of children with congenital heart disease by machine learning based on real-world data,'' \emph{Translational Pediatrics}, vol.~10, no.~1, p.~33, 2021.

\bibitem{pertseva2021applications}
M.~Pertseva, B.~Gao, D.~Neumeier, A.~Yermanos, and S.~T. Reddy, ``Applications of machine and deep learning in adaptive immunity,'' \emph{Annual Review of Chemical and Biomolecular Engineering}, vol.~12, pp. 39--62, 2021.

\bibitem{danieli2023machine}
M.~G. Danieli, S.~Brunetto, L.~Gammeri, D.~Palmeri, I.~Claudi, Y.~Shoenfeld, and S.~Gangemi, ``Machine learning application in autoimmune diseases: State of art and future prospectives,'' \emph{Autoimmunity Reviews}, p. 103496, 2023.

\bibitem{usategui2023machine}
I.~Usategui, J.~Barbado, A.~M. Torres, J.~Casc{\'o}n, and J.~Mateo, ``Machine learning, a new tool for the detection of immunodeficiency patterns in systemic lupus erythematosus,'' \emph{Journal of Investigative Medicine}, p. 10815589231171404, 2023.

\bibitem{thomasian2022machine}
N.~M. Thomasian, I.~R. Kamel, and H.~X. Bai, ``Machine intelligence in non-invasive endocrine cancer diagnostics,'' \emph{Nature Reviews Endocrinology}, vol.~18, no.~2, pp. 81--95, 2022.

\bibitem{hong2020machine}
N.~Hong, H.~Park, and Y.~Rhee, ``Machine learning applications in endocrinology and metabolism research: an overview,'' \emph{Endocrinology and metabolism}, vol.~35, no.~1, pp. 71--84, 2020.

\bibitem{gunvcar2018application}
G.~Gun{\v{c}}ar, M.~Kukar, M.~Notar, M.~Brvar, P.~{\v{C}}ernel{\v{c}}, M.~Notar, and M.~Notar, ``An application of machine learning to haematological diagnosis,'' \emph{Scientific reports}, vol.~8, no.~1, p. 411, 2018.

\bibitem{tuckson2017telehealth}
R.~V. Tuckson, M.~Edmunds, and M.~L. Hodgkins, ``Telehealth,'' \emph{New England Journal of Medicine}, vol. 377, no.~16, pp. 1585--1592, 2017.

\bibitem{gajarawala2021telehealth}
S.~N. Gajarawala and J.~N. Pelkowski, ``Telehealth benefits and barriers,'' \emph{The Journal for Nurse Practitioners}, vol.~17, no.~2, pp. 218--221, 2021.

\bibitem{torres2018patient}
A.~D. Torres, H.~Yan, A.~H. Aboutalebi, A.~Das, L.~Duan, and P.~Rad, ``Patient facial emotion recognition and sentiment analysis using secure cloud with hardware acceleration,'' in \emph{Computational Intelligence for Multimedia Big Data on the Cloud with Engineering Applications}.\hskip 1em plus 0.5em minus 0.4em\relax Elsevier, 2018, pp. 61--89.

\bibitem{farzindar2015natural}
A.~Farzindar, D.~Inkpen, and G.~Hirst, \emph{Natural language processing for social media}.\hskip 1em plus 0.5em minus 0.4em\relax Springer, 2015.

\bibitem{fernandez2015machine}
M.~A. Fernandez-Granero, D.~Sanchez-Morillo, M.~A. Lopez-Gordo, and A.~Leon, ``A machine learning approach to prediction of exacerbations of chronic obstructive pulmonary disease,'' in \emph{Artificial Computation in Biology and Medicine: International Work-Conference on the Interplay Between Natural and Artificial Computation, IWINAC 2015, Elche, Spain, June 1-5, 2015, Proceedings, Part I 6}.\hskip 1em plus 0.5em minus 0.4em\relax Springer, 2015, pp. 305--311.

\bibitem{segal2019reducing}
G.~Segal, A.~Segev, A.~Brom, Y.~Lifshitz, Y.~Wasserstrum, and E.~Zimlichman, ``Reducing drug prescription errors and adverse drug events by application of a probabilistic, machine-learning based clinical decision support system in an inpatient setting,'' \emph{Journal of the American Medical Informatics Association}, vol.~26, no.~12, pp. 1560--1565, 2019.

\bibitem{verma2021application}
D.~Verma, K.~Bach, and P.~J. Mork, ``Application of machine learning methods on patient reported outcome measurements for predicting outcomes: a literature review,'' in \emph{Informatics}, vol.~8, no.~3.\hskip 1em plus 0.5em minus 0.4em\relax MDPI, 2021, p.~56.

\bibitem{russell2010artificial}
S.~J. Russell and P.~Norvig, \emph{Artificial intelligence a modern approach}.\hskip 1em plus 0.5em minus 0.4em\relax London, 2010.

\bibitem{weizenbaum1966eliza}
J.~Weizenbaum, ``Eliza—a computer program for the study of natural language communication between man and machine,'' \emph{Communications of the ACM}, vol.~9, no.~1, pp. 36--45, 1966.

\bibitem{shortliffe1974mycin}
E.~H. Shortliffe, ``Mycin: A rule-based computer program for advising physicians regarding antimicrobial therapy selection,'' Ph.D. dissertation, Stanford University Ph. D. dissertation, 1974.

\bibitem{bishop1995neural}
C.~Bishop, ``Neural networks for pattern recognition,'' \emph{Clarendon Press google schola}, vol.~2, pp. 223--228, 1995.

\bibitem{lo1995artificial}
S.-C.~B. Lo, H.-P. Chan, J.-S. Lin, H.~Li, M.~T. Freedman, and S.~K. Mun, ``Artificial convolution neural network for medical image pattern recognition,'' \emph{Neural networks}, vol.~8, no. 7-8, pp. 1201--1214, 1995.

\bibitem{iakovidis1998towards}
I.~Iakovidis, ``Towards personal health record: current situation, obstacles and trends in implementation of electronic healthcare record in europe,'' \emph{International journal of medical informatics}, vol.~52, no. 1-3, pp. 105--115, 1998.

\bibitem{duncan2000medical}
J.~S. Duncan and N.~Ayache, ``Medical image analysis: Progress over two decades and the challenges ahead,'' \emph{IEEE transactions on pattern analysis and machine intelligence}, vol.~22, no.~1, pp. 85--106, 2000.

\bibitem{demner2009can}
D.~Demner-Fushman, W.~W. Chapman, and C.~J. McDonald, ``What can natural language processing do for clinical decision support?'' \emph{Journal of biomedical informatics}, vol.~42, no.~5, pp. 760--772, 2009.

\bibitem{chen2016ibm}
Y.~Chen, J.~E. Argentinis, and G.~Weber, ``Ibm watson: how cognitive computing can be applied to big data challenges in life sciences research,'' \emph{Clinical therapeutics}, vol.~38, no.~4, pp. 688--701, 2016.

\bibitem{younis2024systematic}
H.~A. Younis, T.~A.~E. Eisa, M.~Nasser, T.~M. Sahib, A.~A. Noor, O.~M. Alyasiri, S.~Salisu, I.~M. Hayder, and H.~A. Younis, ``A systematic review and meta-analysis of artificial intelligence tools in medicine and healthcare: Applications, considerations, limitations, motivation and challenges,'' \emph{Diagnostics}, vol.~14, no.~1, p. 109, 2024.

\bibitem{nuseir2024role}
M.~T. Nuseir, I.~A. Akour, H.~M. Alzoubi, B.~Al~Kurdi, M.~T. Alshurideh, and A.~AlHamad, ``Role of big data analytics to empower patient healthcare record management system,'' in \emph{Cyber Security Impact on Digitalization and Business Intelligence: Big Cyber Security for Information Management: Opportunities and Challenges}.\hskip 1em plus 0.5em minus 0.4em\relax Springer, 2024, pp. 39--52.

\bibitem{gowda2024introduction}
D.~Gowda, S.~Shashikala, Y.~Manu, M.~Kaur, and S.~K. Jha, ``Introduction to cloud computing and healthcare 5.0: Transforming the future of healthcare,'' in \emph{Federated Learning and AI for Healthcare 5.0}.\hskip 1em plus 0.5em minus 0.4em\relax IGI Global, 2024, pp. 26--45.

\bibitem{chaddad2023survey}
A.~Chaddad, J.~Peng, J.~Xu, and A.~Bouridane, ``Survey of explainable ai techniques in healthcare,'' \emph{Sensors}, vol.~23, no.~2, p. 634, 2023.

\bibitem{esteva2019guide}
A.~Esteva, A.~Robicquet, B.~Ramsundar, V.~Kuleshov, M.~DePristo, K.~Chou, C.~Cui, G.~Corrado, S.~Thrun, and J.~Dean, ``A guide to deep learning in healthcare,'' \emph{Nature medicine}, vol.~25, no.~1, pp. 24--29, 2019.

\bibitem{jaiswal2019identifying}
A.~K. Jaiswal, P.~Tiwari, S.~Kumar, D.~Gupta, A.~Khanna, and J.~J. Rodrigues, ``Identifying pneumonia in chest x-rays: A deep learning approach,'' \emph{Measurement}, vol. 145, pp. 511--518, 2019.

\bibitem{havaei2017brain}
M.~Havaei, A.~Davy, D.~Warde-Farley, A.~Biard, A.~Courville, Y.~Bengio, C.~Pal, P.-M. Jodoin, and H.~Larochelle, ``Brain tumor segmentation with deep neural networks,'' \emph{Medical image analysis}, vol.~35, pp. 18--31, 2017.

\bibitem{setio2017validation}
A.~A.~A. Setio, A.~Traverso, T.~De~Bel, M.~S. Berens, C.~Van Den~Bogaard, P.~Cerello, H.~Chen, Q.~Dou, M.~E. Fantacci, B.~Geurts \emph{et~al.}, ``Validation, comparison, and combination of algorithms for automatic detection of pulmonary nodules in computed tomography images: the luna16 challenge,'' \emph{Medical image analysis}, vol.~42, pp. 1--13, 2017.

\bibitem{zhavoronkov2019deep}
A.~Zhavoronkov, Y.~A. Ivanenkov, A.~Aliper, M.~S. Veselov, V.~A. Aladinskiy, A.~V. Aladinskaya, V.~A. Terentiev, D.~A. Polykovskiy, M.~D. Kuznetsov, A.~Asadulaev \emph{et~al.}, ``Deep learning enables rapid identification of potent ddr1 kinase inhibitors,'' \emph{Nature biotechnology}, vol.~37, no.~9, pp. 1038--1040, 2019.

\bibitem{chan2019advancing}
H.~S. Chan, H.~Shan, T.~Dahoun, H.~Vogel, and S.~Yuan, ``Advancing drug discovery via artificial intelligence,'' \emph{Trends in pharmacological sciences}, vol.~40, no.~8, pp. 592--604, 2019.

\bibitem{obermeyer2016predicting}
Z.~Obermeyer and E.~J. Emanuel, ``Predicting the future—big data, machine learning, and clinical medicine,'' \emph{The New England journal of medicine}, vol. 375, no.~13, p. 1216, 2016.

\bibitem{rajkomar2018scalable}
A.~Rajkomar, E.~Oren, K.~Chen, A.~M. Dai, N.~Hajaj, M.~Hardt, P.~J. Liu, X.~Liu, J.~Marcus, M.~Sun \emph{et~al.}, ``Scalable and accurate deep learning with electronic health records,'' \emph{NPJ digital medicine}, vol.~1, no.~1, p.~18, 2018.

\bibitem{gulshan2016development}
V.~Gulshan, L.~Peng, M.~Coram, M.~C. Stumpe, D.~Wu, A.~Narayanaswamy, S.~Venugopalan, K.~Widner, T.~Madams, J.~Cuadros \emph{et~al.}, ``Development and validation of a deep learning algorithm for detection of diabetic retinopathy in retinal fundus photographs,'' \emph{jama}, vol. 316, no.~22, pp. 2402--2410, 2016.

\bibitem{futoma2017learning}
J.~Futoma, S.~Hariharan, and K.~Heller, ``Learning to detect sepsis with a multitask gaussian process rnn classifier,'' in \emph{International conference on machine learning}.\hskip 1em plus 0.5em minus 0.4em\relax PMLR, 2017, pp. 1174--1182.

\bibitem{henry2015targeted}
K.~E. Henry, D.~N. Hager, P.~J. Pronovost, and S.~Saria, ``A targeted real-time early warning score (trewscore) for septic shock,'' \emph{Science translational medicine}, vol.~7, no. 299, pp. 299ra122--299ra122, 2015.

\bibitem{kassahun2016surgical}
Y.~Kassahun, B.~Yu, A.~T. Tibebu, D.~Stoyanov, S.~Giannarou, J.~H. Metzen, and E.~Vander~Poorten, ``Surgical robotics beyond enhanced dexterity instrumentation: a survey of machine learning techniques and their role in intelligent and autonomous surgical actions,'' \emph{International journal of computer assisted radiology and surgery}, vol.~11, pp. 553--568, 2016.

\bibitem{biffi2017immersive}
E.~Biffi, E.~Beretta, A.~Cesareo, C.~Maghini, A.~C. Turconi, G.~Reni, and S.~Strazzer, ``An immersive virtual reality platform to enhance walking ability of children with acquired brain injuries,'' \emph{Methods of information in medicine}, vol.~56, no.~02, pp. 119--126, 2017.

\bibitem{miner2020chatbots}
A.~S. Miner, L.~Laranjo, and A.~B. Kocaballi, ``Chatbots in the fight against the covid-19 pandemic,'' \emph{NPJ digital medicine}, vol.~3, no.~1, p.~65, 2020.

\bibitem{dorsey2016state}
E.~R. Dorsey and E.~J. Topol, ``State of telehealth,'' \emph{New England journal of medicine}, vol. 375, no.~2, pp. 154--161, 2016.

\bibitem{mesko2018will}
B.~Mesk{\'o}, G.~Het{\'e}nyi, and Z.~Gy{\H{o}}rffy, ``Will artificial intelligence solve the human resource crisis in healthcare?'' \emph{BMC health services research}, vol.~18, no.~1, pp. 1--4, 2018.

\bibitem{harper2002framework}
P.~R. Harper, ``A framework for operational modelling of hospital resources,'' \emph{Health care management science}, vol.~5, pp. 165--173, 2002.

\bibitem{wong2023leveraging}
F.~Wong, C.~de~la Fuente-Nunez, and J.~J. Collins, ``Leveraging artificial intelligence in the fight against infectious diseases,'' \emph{Science}, vol. 381, no. 6654, pp. 164--170, 2023.

\bibitem{li2020artificial}
L.~Li, L.~Qin, Z.~Xu, Y.~Yin, X.~Wang, B.~Kong, J.~Bai, Y.~Lu, Z.~Fang, Q.~Song \emph{et~al.}, ``Artificial intelligence distinguishes covid-19 from community acquired pneumonia on chest ct,'' \emph{Radiology}, 2020.

\bibitem{zhavoronkov2020potential}
A.~Zhavoronkov, V.~Aladinskiy, A.~Zhebrak, B.~Zagribelnyy, V.~Terentiev, D.~S. Bezrukov, D.~Polykovskiy, R.~Shayakhmetov, A.~Filimonov, P.~Orekhov \emph{et~al.}, ``Potential covid-2019 3c-like protease inhibitors designed using generative deep learning approaches. chemrxiv,'' \emph{Preprint. https://doi. org/10.26434/chemrxiv}, vol. 11829102, p.~v2, 2020.

\bibitem{chen2019artificial}
M.~Chen, U.~Challita, W.~Saad, C.~Yin, and M.~Debbah, ``Artificial neural networks-based machine learning for wireless networks: A tutorial,'' \emph{IEEE Communications Surveys \& Tutorials}, vol.~21, no.~4, pp. 3039--3071, 2019.

\bibitem{pereira2009machine}
F.~Pereira, T.~Mitchell, and M.~Botvinick, ``Machine learning classifiers and fmri: a tutorial overview,'' \emph{Neuroimage}, vol.~45, no.~1, pp. S199--S209, 2009.

\bibitem{vercio2020supervised}
L.~L. Vercio, K.~Amador, J.~J. Bannister, S.~Crites, A.~Gutierrez, M.~E. MacDonald, J.~Moore, P.~Mouches, D.~Rajashekar, S.~Schimert \emph{et~al.}, ``Supervised machine learning tools: a tutorial for clinicians,'' \emph{Journal of Neural Engineering}, vol.~17, no.~6, p. 062001, 2020.

\bibitem{erickson2017machine}
B.~J. Erickson, P.~Korfiatis, Z.~Akkus, and T.~L. Kline, ``Machine learning for medical imaging,'' \emph{Radiographics}, vol.~37, no.~2, pp. 505--515, 2017.

\bibitem{kukreja2024review}
S.~Kukreja, A.~Kumar, and G.~A. Khan, ``A review paper on the diagnosis of lung cancer using machine learning,'' \emph{Artificial Intelligence, Blockchain, Computing and Security Volume 2}, pp. 15--19, 2024.

\bibitem{bertsimas2020machine}
D.~Bertsimas and H.~Wiberg, ``Machine learning in oncology: methods, applications, and challenges,'' \emph{JCO Clinical Cancer Informatics}, vol.~4, 2020.

\bibitem{asri2016using}
H.~Asri, H.~Mousannif, H.~Al~Moatassime, and T.~Noel, ``Using machine learning algorithms for breast cancer risk prediction and diagnosis,'' \emph{Procedia Computer Science}, vol.~83, pp. 1064--1069, 2016.

\bibitem{talwar2023performance}
A.~Talwar, M.~A. Lopez-Olivo, Y.~Huang, L.~Ying, and R.~R. Aparasu, ``Performance of advanced machine learning algorithms overlogistic regression in predicting hospital readmissions: A meta-analysis,'' \emph{Exploratory Research in Clinical and Social Pharmacy}, vol.~11, p. 100317, 2023.

\bibitem{colace2024unsupervised}
F.~Colace, B.~B. Gupta, A.~Lorusso, A.~Troiano, D.~Santaniello, and C.~Valentino, ``Unsupervised learning techniques for vibration-based structural health monitoring systems driven by data: A general overview,'' \emph{Handbook of Research on AI and ML for Intelligent Machines and Systems}, pp. 305--347, 2024.

\bibitem{eisen1998cluster}
M.~B. Eisen, P.~T. Spellman, P.~O. Brown, and D.~Botstein, ``Cluster analysis and display of genome-wide expression patterns,'' \emph{Proceedings of the National Academy of Sciences}, vol.~95, no.~25, pp. 14\,863--14\,868, 1998.

\bibitem{monti2003consensus}
S.~Monti, P.~Tamayo, J.~Mesirov, and T.~Golub, ``Consensus clustering: a resampling-based method for class discovery and visualization of gene expression microarray data,'' \emph{Machine learning}, vol.~52, pp. 91--118, 2003.

\bibitem{wu2017unsupervised}
J.~Wu, Y.~Cui, X.~Sun, G.~Cao, B.~Li, D.~M. Ikeda, A.~W. Kurian, and R.~Li, ``Unsupervised clustering of quantitative image phenotypes reveals breast cancer subtypes with distinct prognoses and molecular pathways,'' \emph{Clinical Cancer Research}, vol.~23, no.~13, pp. 3334--3342, 2017.

\bibitem{ringner2008principal}
M.~Ringn{\'e}r, ``What is principal component analysis?'' \emph{Nature biotechnology}, vol.~26, no.~3, pp. 303--304, 2008.

\bibitem{van2008visualizing}
L.~Van~der Maaten and G.~Hinton, ``Visualizing data using t-sne.'' \emph{Journal of machine learning research}, vol.~9, no.~11, 2008.

\bibitem{smith2002fast}
S.~M. Smith, ``Fast robust automated brain extraction,'' \emph{Human brain mapping}, vol.~17, no.~3, pp. 143--155, 2002.

\bibitem{menze2014multimodal}
B.~H. Menze, A.~Jakab, S.~Bauer, J.~Kalpathy-Cramer, K.~Farahani, J.~Kirby, Y.~Burren, N.~Porz, J.~Slotboom, R.~Wiest \emph{et~al.}, ``The multimodal brain tumor image segmentation benchmark (brats),'' \emph{IEEE transactions on medical imaging}, vol.~34, no.~10, pp. 1993--2024, 2014.

\bibitem{chapelle2009semi}
O.~Chapelle, B.~Scholkopf, and A.~Zien, ``Semi-supervised learning (chapelle, o. et al., eds.; 2006)[book reviews],'' \emph{IEEE Transactions on Neural Networks}, vol.~20, no.~3, pp. 542--542, 2009.

\bibitem{qiu2023federated}
L.~Qiu, J.~Cheng, H.~Gao, W.~Xiong, and H.~Ren, ``Federated semi-supervised learning for medical image segmentation via pseudo-label denoising,'' \emph{IEEE Journal of Biomedical and Health Informatics}, 2023.

\bibitem{ren2020not}
Z.~Ren, R.~Yeh, and A.~Schwing, ``Not all unlabeled data are equal: Learning to weight data in semi-supervised learning,'' \emph{Advances in Neural Information Processing Systems}, vol.~33, pp. 21\,786--21\,797, 2020.

\bibitem{IBM_semi_supervised_learning}
\BIBentryALTinterwordspacing
IBM, ``What is semi-supervised learning?'' [Online]. Available: \url{https://www.ibm.com/topics/semi-supervised-learning}
\BIBentrySTDinterwordspacing

\bibitem{eckardt2022semi}
J.-N. Eckardt, M.~Bornh{\"a}user, K.~Wendt, and J.~M. Middeke, ``Semi-supervised learning in cancer diagnostics,'' \emph{Frontiers in oncology}, vol.~12, p. 960984, 2022.

\bibitem{jiao2023learning}
R.~Jiao, Y.~Zhang, L.~Ding, B.~Xue, J.~Zhang, R.~Cai, and C.~Jin, ``Learning with limited annotations: a survey on deep semi-supervised learning for medical image segmentation,'' \emph{Computers in Biology and Medicine}, p. 107840, 2023.

\bibitem{gosavi2009reinforcement}
A.~Gosavi, ``Reinforcement learning: A tutorial survey and recent advances,'' \emph{INFORMS Journal on Computing}, vol.~21, no.~2, pp. 178--192, 2009.

\bibitem{yu2021reinforcement}
C.~Yu, J.~Liu, S.~Nemati, and G.~Yin, ``Reinforcement learning in healthcare: A survey,'' \emph{ACM Computing Surveys (CSUR)}, vol.~55, no.~1, pp. 1--36, 2021.

\bibitem{komorowski2018artificial}
M.~Komorowski, L.~A. Celi, O.~Badawi, A.~C. Gordon, and A.~A. Faisal, ``The artificial intelligence clinician learns optimal treatment strategies for sepsis in intensive care,'' \emph{Nature medicine}, vol.~24, no.~11, pp. 1716--1720, 2018.

\bibitem{zhao2009reinforcement}
Y.~Zhao, M.~R. Kosorok, and D.~Zeng, ``Reinforcement learning design for cancer clinical trials,'' \emph{Statistics in medicine}, vol.~28, no.~26, pp. 3294--3315, 2009.

\bibitem{chi2019context}
W.~Chi, ``Context-aware learning for robot-assisted endovascular catheterization,'' 2019.

\bibitem{wang2022predicting}
Y.~Wang, Y.~Zhao, and L.~Petzold, ``Predicting the need for blood transfusion in intensive care units with reinforcement learning,'' in \emph{Proceedings of the 13th ACM International Conference on Bioinformatics, Computational Biology and Health Informatics}, 2022, pp. 1--10.

\bibitem{mclaverty2023unifying}
B.~McLaverty, ``Unifying data-driven modeling with machine learning to improve personalized treatment of critical care patients,'' Ph.D. dissertation, University of Pittsburgh, 2023.

\bibitem{almagrabi2022reinforcement}
A.~O. Almagrabi, R.~Ali, D.~Alghazzawi, A.~AlBarakati, and T.~Khurshaid, ``A reinforcement learning-based framework for crowdsourcing in massive health care internet of things,'' \emph{Big data}, vol.~10, no.~2, pp. 161--170, 2022.

\bibitem{esteva2017dermatologist}
A.~Esteva, B.~Kuprel, R.~A. Novoa, J.~Ko, S.~M. Swetter, H.~M. Blau, and S.~Thrun, ``Dermatologist-level classification of skin cancer with deep neural networks,'' \emph{nature}, vol. 542, no. 7639, pp. 115--118, 2017.

\bibitem{shah2020heart}
D.~Shah, S.~Patel, and S.~K. Bharti, ``Heart disease prediction using machine learning techniques,'' \emph{SN Computer Science}, vol.~1, pp. 1--6, 2020.

\bibitem{levin2018machine}
S.~Levin, M.~Toerper, E.~Hamrock, J.~S. Hinson, S.~Barnes, H.~Gardner, A.~Dugas, B.~Linton, T.~Kirsch, and G.~Kelen, ``Machine-learning-based electronic triage more accurately differentiates patients with respect to clinical outcomes compared with the emergency severity index,'' \emph{Annals of emergency medicine}, vol.~71, no.~5, pp. 565--574, 2018.

\bibitem{zou2018predicting}
Q.~Zou, K.~Qu, Y.~Luo, D.~Yin, Y.~Ju, and H.~Tang, ``Predicting diabetes mellitus with machine learning techniques,'' \emph{Frontiers in genetics}, vol.~9, p. 515, 2018.

\bibitem{tatonetti2012data}
N.~P. Tatonetti, P.~P. Ye, R.~Daneshjou, and R.~B. Altman, ``Data-driven prediction of drug effects and interactions,'' \emph{Science translational medicine}, vol.~4, no. 125, pp. 125ra31--125ra31, 2012.

\bibitem{mucaki2019predicting}
E.~J. Mucaki, J.~Z. Zhao, D.~J. Lizotte, and P.~K. Rogan, ``Predicting responses to platin chemotherapy agents with biochemically-inspired machine learning,'' \emph{Signal transduction and targeted therapy}, vol.~4, no.~1, p.~1, 2019.

\bibitem{dhiman2022methodological}
P.~Dhiman, J.~Ma, C.~L. Andaur~Navarro, B.~Speich, G.~Bullock, J.~A. Damen, L.~Hooft, S.~Kirtley, R.~D. Riley, B.~Van~Calster \emph{et~al.}, ``Methodological conduct of prognostic prediction models developed using machine learning in oncology: a systematic review,'' \emph{BMC medical research methodology}, vol.~22, no.~1, pp. 1--16, 2022.

\bibitem{sinha2023semi}
A.~Sinha, T.~Aljrees, S.~K. Pandey, A.~Kumar, P.~Banerjee, B.~Kumar, K.~U. Singh, T.~Singh, and P.~Jha, ``Semi-supervised clustering-based dana algorithm for data gathering and disease detection in healthcare wireless sensor networks (wsn),'' \emph{Sensors}, vol.~24, no.~1, p.~18, 2023.

\bibitem{qiu2023review}
S.~Qiu, Y.~Chen, Y.~Yang, P.~Wang, Z.~Wang, H.~Zhao, Y.~Kang, and R.~Nie, ``A review on semi-supervised learning for eeg-based emotion recognition,'' \emph{Information Fusion}, p. 102190, 2023.

\bibitem{christopoulou2024machine}
S.~C. Christopoulou, ``Machine learning models and technologies for evidence-based telehealth and smart care: A review,'' \emph{BioMedInformatics}, vol.~4, no.~1, pp. 754--779, 2024.

\bibitem{mckinney2020international}
S.~M. McKinney, M.~Sieniek, V.~Godbole, J.~Godwin, N.~Antropova, H.~Ashrafian, T.~Back, M.~Chesus, G.~S. Corrado, A.~Darzi \emph{et~al.}, ``International evaluation of an ai system for breast cancer screening,'' \emph{Nature}, vol. 577, no. 7788, pp. 89--94, 2020.

\bibitem{piening2023improved}
B.~Piening, B.~Bapat, R.~K. Weerasinghe, R.~Meng, A.~K. Dowdell, S.-C. Chang, A.~Vita, C.~Wong, R.~Tinn, L.~Harold \emph{et~al.}, ``Improved outcomes from reflex comprehensive genomic profiling-guided precision therapeutic selection across a major us healthcare system.'' 2023.

\bibitem{iyortsuun2023review}
N.~K. Iyortsuun, S.-H. Kim, M.~Jhon, H.-J. Yang, and S.~Pant, ``A review of machine learning and deep learning approaches on mental health diagnosis,'' in \emph{Healthcare}, vol.~11, no.~3.\hskip 1em plus 0.5em minus 0.4em\relax MDPI, 2023, p. 285.

\bibitem{konevcny2016federated}
J.~Kone{\v{c}}n{\`y}, H.~B. McMahan, D.~Ramage, and P.~Richt{\'a}rik, ``Federated optimization: Distributed machine learning for on-device intelligence,'' \emph{arXiv preprint arXiv:1610.02527}, 2016.

\bibitem{yang2019federated}
Q.~Yang, Y.~Liu, T.~Chen, and Y.~Tong, ``Federated machine learning: Concept and applications,'' \emph{ACM Transactions on Intelligent Systems and Technology (TIST)}, vol.~10, no.~2, pp. 1--19, 2019.

\bibitem{zhang2023federated}
X.~Zhang, A.~Mavromatics, A.~Vafeas, R.~Nejabati, and D.~Simeonidou, ``Federated feature selection for horizontal federated learning in iot networks,'' \emph{IEEE Internet of Things Journal}, 2023.

\bibitem{gao2019hhhfl}
D.~Gao, C.~Ju, X.~Wei, Y.~Liu, T.~Chen, and Q.~Yang, ``Hhhfl: Hierarchical heterogeneous horizontal federated learning for electroencephalography,'' \emph{arXiv preprint arXiv:1909.05784}, 2019.

\bibitem{huang2022fairness}
W.~Huang, T.~Li, D.~Wang, S.~Du, J.~Zhang, and T.~Huang, ``Fairness and accuracy in horizontal federated learning,'' \emph{Information Sciences}, vol. 589, pp. 170--185, 2022.

\bibitem{liu2024vertical}
Y.~Liu, Y.~Kang, T.~Zou, Y.~Pu, Y.~He, X.~Ye, Y.~Ouyang, Y.-Q. Zhang, and Q.~Yang, ``Vertical federated learning: Concepts, advances, and challenges,'' \emph{IEEE Transactions on Knowledge and Data Engineering}, 2024.

\bibitem{feng2024mmvfl}
S.~Feng, H.~Yu, and Y.~Zhu, ``Mmvfl: A simple vertical federated learning framework for multi-class multi-participant scenarios,'' \emph{Sensors}, vol.~24, no.~2, p. 619, 2024.

\bibitem{gupta2024federated}
M.~Gupta, P.~Sharma, and R.~Kalra, ``Federated learning and artificial intelligence in e-healthcare,'' in \emph{Federated Learning and AI for Healthcare 5.0}.\hskip 1em plus 0.5em minus 0.4em\relax IGI Global, 2024, pp. 104--118.

\bibitem{liu2020secure}
Y.~Liu, Y.~Kang, C.~Xing, T.~Chen, and Q.~Yang, ``A secure federated transfer learning framework,'' \emph{IEEE Intelligent Systems}, vol.~35, no.~4, pp. 70--82, 2020.

\bibitem{chen2020fedhealth}
Y.~Chen, X.~Qin, J.~Wang, C.~Yu, and W.~Gao, ``Fedhealth: A federated transfer learning framework for wearable healthcare,'' \emph{IEEE Intelligent Systems}, vol.~35, no.~4, pp. 83--93, 2020.

\bibitem{chorney2024towards}
W.~Chorney and H.~Wang, ``Towards federated transfer learning in electrocardiogram signal analysis,'' \emph{Computers in Biology and Medicine}, vol. 170, p. 107984, 2024.

\bibitem{nguyen2022federated}
D.~C. Nguyen, Q.-V. Pham, P.~N. Pathirana, M.~Ding, A.~Seneviratne, Z.~Lin, O.~Dobre, and W.-J. Hwang, ``Federated learning for smart healthcare: A survey,'' \emph{ACM Computing Surveys (CSUR)}, vol.~55, no.~3, pp. 1--37, 2022.

\bibitem{li2020federated}
T.~Li, A.~K. Sahu, A.~Talwalkar, and V.~Smith, ``Federated learning: Challenges, methods, and future directions,'' \emph{IEEE signal processing magazine}, vol.~37, no.~3, pp. 50--60, 2020.

\bibitem{heyndrickx2023melloddy}
W.~Heyndrickx, L.~Mervin, T.~Morawietz, N.~Sturm, L.~Friedrich, A.~Zalewski, A.~Pentina, L.~Humbeck, M.~Oldenhof, R.~Niwayama \emph{et~al.}, ``Melloddy: Cross-pharma federated learning at unprecedented scale unlocks benefits in qsar without compromising proprietary information,'' \emph{Journal of chemical information and modeling}, 2023.

\bibitem{foley2022openfl}
P.~Foley, M.~J. Sheller, B.~Edwards, S.~Pati, W.~Riviera, M.~Sharma, P.~N. Moorthy, S.-h. Wang, J.~Martin, P.~Mirhaji \emph{et~al.}, ``Openfl: the open federated learning library,'' \emph{Physics in Medicine \& Biology}, vol.~67, no.~21, p. 214001, 2022.

\bibitem{malik2024federated}
H.~Malik and T.~Anees, ``Federated learning with deep convolutional neural networks for the detection of multiple chest diseases using chest x-rays,'' \emph{Multimedia Tools and Applications}, pp. 1--29, 2024.

\bibitem{dasaradharami2023comprehensive}
K.~Dasaradharami~Reddy, T.~R. Gadekallu \emph{et~al.}, ``A comprehensive survey on federated learning techniques for healthcare informatics,'' \emph{Computational Intelligence and Neuroscience}, vol. 2023, 2023.

\bibitem{kaissis2020secure}
G.~A. Kaissis, M.~R. Makowski, D.~R{\"u}ckert, and R.~F. Braren, ``Secure, privacy-preserving and federated machine learning in medical imaging,'' \emph{Nature Machine Intelligence}, vol.~2, no.~6, pp. 305--311, 2020.

\bibitem{rafi2024fairness}
T.~H. Rafi, F.~A. Noor, T.~Hussain, and D.-K. Chae, ``Fairness and privacy preserving in federated learning: A survey,'' \emph{Information Fusion}, vol. 105, p. 102198, 2024.

\bibitem{djebrouni2024bias}
Y.~Djebrouni, N.~Benarba, O.~Touat, P.~De~Rosa, S.~Bouchenak, A.~Bonifati, P.~Felber, V.~Marangozova, and V.~Schiavoni, ``Bias mitigation in federated learning for edge computing,'' \emph{Proceedings of the ACM on Interactive, Mobile, Wearable and Ubiquitous Technologies}, vol.~7, no.~4, pp. 1--35, 2024.

\bibitem{chen2024credible}
L.~Chen, D.~Zhao, L.~Tao, K.~Wang, S.~Qiao, X.~Zeng, and C.~W. Tan, ``A credible and fair federated learning framework based on blockchain,'' \emph{IEEE Transactions on Artificial Intelligence}, 2024.

\bibitem{you2024slmfed}
L.~You, Z.~Guo, B.~Zuo, Y.~Chang, and C.~Yuen, ``Slmfed: A stage-based and layer-wise mechanism for incremental federated learning to assist dynamic and ubiquitous iot,'' \emph{IEEE Internet of Things Journal}, 2024.

\bibitem{mazzocca2024enabling}
C.~Mazzocca, N.~Romandini, R.~Montanari, and P.~Bellavista, ``Enabling federated learning at the edge through the iota tangle,'' \emph{Future Generation Computer Systems}, vol. 152, pp. 17--29, 2024.

\bibitem{ji2024edge}
J.~Ji, Z.~Shu, H.~Li, K.~X. Lai, M.~Lu, G.~Jiang, W.~Wang, Y.~Zheng, and X.~Jiang, ``Edge-computing based knowledge distillation and multi-task learning for partial discharge recognition,'' \emph{IEEE Transactions on Instrumentation and Measurement}, 2024.

\bibitem{elhattab2024pastel}
F.~Elhattab, S.~Bouchenak, and C.~Boscher, ``Pastel: Privacy-preserving federated learning in edge computing,'' \emph{Proceedings of the ACM on Interactive, Mobile, Wearable and Ubiquitous Technologies}, vol.~7, no.~4, pp. 1--29, 2024.

\bibitem{myrzashova2023blockchain}
R.~Myrzashova, S.~H. Alsamhi, A.~V. Shvetsov, A.~Hawbani, and X.~Wei, ``Blockchain meets federated learning in healthcare: A systematic review with challenges and opportunities,'' \emph{IEEE Internet of Things Journal}, 2023.

\bibitem{bitcoin2008bitcoin}
N.~S. Bitcoin, ``Bitcoin: A peer-to-peer electronic cash system,'' 2008.

\bibitem{wood2014ethereum}
G.~Wood \emph{et~al.}, ``Ethereum: A secure decentralised generalised transaction ledger,'' \emph{Ethereum project yellow paper}, vol. 151, no. 2014, pp. 1--32, 2014.

\bibitem{azzi2019power}
R.~Azzi, R.~K. Chamoun, and M.~Sokhn, ``The power of a blockchain-based supply chain,'' \emph{Computers \& industrial engineering}, vol. 135, pp. 582--592, 2019.

\bibitem{dutta2020blockchain}
P.~Dutta, T.-M. Choi, S.~Somani, and R.~Butala, ``Blockchain technology in supply chain operations: Applications, challenges and research opportunities,'' \emph{Transportation research part e: Logistics and transportation review}, vol. 142, p. 102067, 2020.

\bibitem{casado2018blockchain}
R.~Casado-Vara, J.~Prieto, F.~De~la Prieta, and J.~M. Corchado, ``How blockchain improves the supply chain: Case study alimentary supply chain,'' \emph{Procedia computer science}, vol. 134, pp. 393--398, 2018.

\bibitem{korpela2017digital}
K.~Korpela, J.~Hallikas, and T.~Dahlberg, ``Digital supply chain transformation toward blockchain integration,'' 2017.

\bibitem{queiroz2020blockchain}
M.~M. Queiroz, R.~Telles, and S.~H. Bonilla, ``Blockchain and supply chain management integration: a systematic review of the literature,'' \emph{Supply chain management: An international journal}, vol.~25, no.~2, pp. 241--254, 2020.

\bibitem{raval2016decentralized}
S.~Raval, \emph{Decentralized applications: harnessing Bitcoin's blockchain technology}.\hskip 1em plus 0.5em minus 0.4em\relax " O'Reilly Media, Inc.", 2016.

\bibitem{joshi2018survey}
A.~P. Joshi, M.~Han, and Y.~Wang, ``A survey on security and privacy issues of blockchain technology.'' \emph{Mathematical foundations of computing}, vol.~1, no.~2, 2018.

\bibitem{mohanta2019blockchain}
B.~K. Mohanta, D.~Jena, S.~S. Panda, and S.~Sobhanayak, ``Blockchain technology: A survey on applications and security privacy challenges,'' \emph{Internet of Things}, vol.~8, p. 100107, 2019.

\bibitem{hassan2022anomaly}
M.~U. Hassan, M.~H. Rehmani, and J.~Chen, ``Anomaly detection in blockchain networks: A comprehensive survey,'' \emph{IEEE Communications Surveys \& Tutorials}, vol.~25, no.~1, pp. 289--318, 2022.

\bibitem{shahsavari2020theoretical}
Y.~Shahsavari, K.~Zhang, and C.~Talhi, ``A theoretical model for block propagation analysis in bitcoin network,'' \emph{IEEE Transactions on Engineering Management}, vol.~69, no.~4, pp. 1459--1476, 2020.

\bibitem{belchior2021survey}
R.~Belchior, A.~Vasconcelos, S.~Guerreiro, and M.~Correia, ``A survey on blockchain interoperability: Past, present, and future trends,'' \emph{ACM Computing Surveys (CSUR)}, vol.~54, no.~8, pp. 1--41, 2021.

\bibitem{lafourcade2020blockchain}
P.~Lafourcade and M.~Lombard-Platet, ``About blockchain interoperability,'' \emph{Information Processing Letters}, vol. 161, p. 105976, 2020.

\bibitem{schulte2019towards}
S.~Schulte, M.~Sigwart, P.~Frauenthaler, and M.~Borkowski, ``Towards blockchain interoperability,'' in \emph{Business Process Management: Blockchain and Central and Eastern Europe Forum: BPM 2019 Blockchain and CEE Forum, Vienna, Austria, September 1--6, 2019, Proceedings 17}.\hskip 1em plus 0.5em minus 0.4em\relax Springer, 2019, pp. 3--10.

\bibitem{hardjono2019toward}
T.~Hardjono, A.~Lipton, and A.~Pentland, ``Toward an interoperability architecture for blockchain autonomous systems,'' \emph{IEEE Transactions on Engineering Management}, vol.~67, no.~4, pp. 1298--1309, 2019.

\bibitem{zhang2017applying}
P.~Zhang, J.~White, D.~C. Schmidt, and G.~Lenz, ``Applying software patterns to address interoperability in blockchain-based healthcare apps,'' \emph{arXiv preprint arXiv:1706.03700}, 2017.

\bibitem{qasse2019inter}
I.~A. Qasse, M.~Abu~Talib, and Q.~Nasir, ``Inter blockchain communication: A survey,'' in \emph{Proceedings of the ArabWIC 6th Annual International Conference Research Track}, 2019, pp. 1--6.

\bibitem{zou2019smart}
W.~Zou, D.~Lo, P.~S. Kochhar, X.-B.~D. Le, X.~Xia, Y.~Feng, Z.~Chen, and B.~Xu, ``Smart contract development: Challenges and opportunities,'' \emph{IEEE Transactions on Software Engineering}, vol.~47, no.~10, pp. 2084--2106, 2019.

\bibitem{delmolino2016step}
K.~Delmolino, M.~Arnett, A.~Kosba, A.~Miller, and E.~Shi, ``Step by step towards creating a safe smart contract: Lessons and insights from a cryptocurrency lab,'' in \emph{Financial Cryptography and Data Security: FC 2016 International Workshops, BITCOIN, VOTING, and WAHC, Christ Church, Barbados, February 26, 2016, Revised Selected Papers 20}.\hskip 1em plus 0.5em minus 0.4em\relax Springer, 2016, pp. 79--94.

\bibitem{xu2023survey}
J.~Xu, C.~Wang, and X.~Jia, ``A survey of blockchain consensus protocols,'' \emph{ACM Computing Surveys}, 2023.

\bibitem{nguyen2018survey}
G.-T. Nguyen and K.~Kim, ``A survey about consensus algorithms used in blockchain.'' \emph{Journal of Information processing systems}, vol.~14, no.~1, 2018.

\bibitem{monrat2019survey}
A.~A. Monrat, O.~Schel{\'e}n, and K.~Andersson, ``A survey of blockchain from the perspectives of applications, challenges, and opportunities,'' \emph{IEEE Access}, vol.~7, pp. 117\,134--117\,151, 2019.

\bibitem{gervais2016security}
A.~Gervais, G.~O. Karame, K.~W{\"u}st, V.~Glykantzis, H.~Ritzdorf, and S.~Capkun, ``On the security and performance of proof of work blockchains,'' in \emph{Proceedings of the 2016 ACM SIGSAC conference on computer and communications security}, 2016, pp. 3--16.

\bibitem{shahsavari2019theoretical}
Y.~Shahsavari, K.~Zhang, and C.~Talhi, ``A theoretical model for fork analysis in the bitcoin network,'' in \emph{2019 IEEE international conference on Blockchain (Blockchain)}.\hskip 1em plus 0.5em minus 0.4em\relax IEEE, 2019, pp. 237--244.

\bibitem{nguyen2019proof}
C.~T. Nguyen, D.~T. Hoang, D.~N. Nguyen, D.~Niyato, H.~T. Nguyen, and E.~Dutkiewicz, ``Proof-of-stake consensus mechanisms for future blockchain networks: fundamentals, applications and opportunities,'' \emph{IEEE access}, vol.~7, pp. 85\,727--85\,745, 2019.

\bibitem{buterin2014next}
V.~Buterin \emph{et~al.}, ``A next-generation smart contract and decentralized application platform,'' \emph{white paper}, vol.~3, no.~37, pp. 2--1, 2014.

\bibitem{cardanodocs}
\BIBentryALTinterwordspacing
{Cardano Foundation, IOHK, and EMURGO}. (2023) Cardano documentation. Cardano Foundation, IOHK, and EMURGO. Accessed: February 22, 2024. [Online]. Available: \url{https://docs.cardano.org/introduction/}
\BIBentrySTDinterwordspacing

\bibitem{gilad2017algorand}
Y.~Gilad, R.~Hemo, S.~Micali, G.~Vlachos, and N.~Zeldovich, ``Algorand: Scaling byzantine agreements for cryptocurrencies,'' in \emph{Proceedings of the 26th symposium on operating systems principles}, 2017, pp. 51--68.

\bibitem{wang2020revisiting}
Q.~Wang, M.~Xu, X.~Li, and H.~Qian, ``Revisiting the fairness and randomness of delegated proof of stake consensus algorithm,'' in \emph{2020 IEEE Intl Conf on Parallel \& Distributed Processing with Applications, Big Data \& Cloud Computing, Sustainable Computing \& Communications, Social Computing \& Networking (ISPA/BDCloud/SocialCom/SustainCom)}.\hskip 1em plus 0.5em minus 0.4em\relax IEEE, 2020, pp. 305--312.

\bibitem{lamport2019byzantine}
L.~Lamport, R.~Shostak, and M.~Pease, ``The byzantine generals problem,'' in \emph{Concurrency: the works of leslie lamport}, 2019, pp. 203--226.

\bibitem{castro1999practical}
M.~Castro, B.~Liskov \emph{et~al.}, ``Practical byzantine fault tolerance,'' in \emph{OsDI}, vol.~99, no. 1999, 1999, pp. 173--186.

\bibitem{androulaki2018hyperledger}
E.~Androulaki, A.~Barger, V.~Bortnikov, C.~Cachin, K.~Christidis, A.~De~Caro, D.~Enyeart, C.~Ferris, G.~Laventman, Y.~Manevich \emph{et~al.}, ``Hyperledger fabric: a distributed operating system for permissioned blockchains,'' in \emph{Proceedings of the thirteenth EuroSys conference}, 2018, pp. 1--15.

\bibitem{cohen2019chia}
B.~Cohen and K.~Pietrzak, ``The chia network blockchain,'' \emph{White Paper, Chia. net}, vol.~9, 2019.

\bibitem{li2020direct}
Y.~Li, B.~Cao, M.~Peng, L.~Zhang, L.~Zhang, D.~Feng, and J.~Yu, ``Direct acyclic graph-based ledger for internet of things: Performance and security analysis,'' \emph{IEEE/ACM Transactions on Networking}, vol.~28, no.~4, pp. 1643--1656, 2020.

\bibitem{popov2018tangle}
S.~Popov, ``The tangle,'' \emph{White paper}, vol.~1, no.~3, p.~30, 2018.

\bibitem{saa2023iota}
O.~Saa, A.~Cullen, and L.~Vigneri, ``Iota 2.0 incentives and tokenomics whitepaper,'' 2023.

\bibitem{karantias2020proof}
K.~Karantias, A.~Kiayias, and D.~Zindros, ``Proof-of-burn,'' in \emph{Financial Cryptography and Data Security: 24th International Conference, FC 2020, Kota Kinabalu, Malaysia, February 10--14, 2020 Revised Selected Papers 24}.\hskip 1em plus 0.5em minus 0.4em\relax Springer, 2020, pp. 523--540.

\bibitem{douceur2002sybil}
J.~R. Douceur, ``The sybil attack,'' in \emph{International workshop on peer-to-peer systems}.\hskip 1em plus 0.5em minus 0.4em\relax Springer, 2002, pp. 251--260.

\bibitem{pass2016hybrid}
R.~Pass and E.~Shi, ``Hybrid consensus: Efficient consensus in the permissionless model,'' \emph{Cryptology ePrint Archive}, 2016.

\bibitem{chepurnoy2017twinscoin}
A.~Chepurnoy, T.~Duong, L.~Fan, and H.-S. Zhou, ``Twinscoin: A cryptocurrency via proof-of-work and proof-of-stake,'' \emph{Cryptology ePrint Archive}, 2017.

\bibitem{kwon2014tendermint}
J.~Kwon, ``Tendermint: Consensus without mining,'' \emph{Draft v. 0.6, fall}, vol.~1, no.~11, pp. 1--11, 2014.

\bibitem{cheng2018new}
Z.~Cheng, G.~Wu, H.~Wu, M.~Zhao, L.~Zhao, and Q.~Cai, ``A new hybrid consensus protocol: Deterministic proof of work,'' \emph{arXiv preprint arXiv:1808.04142}, 2018.

\bibitem{dinh2018untangling}
T.~T.~A. Dinh, R.~Liu, M.~Zhang, G.~Chen, B.~C. Ooi, and J.~Wang, ``Untangling blockchain: A data processing view of blockchain systems,'' \emph{IEEE transactions on knowledge and data engineering}, vol.~30, no.~7, pp. 1366--1385, 2018.

\bibitem{decker2013information}
C.~Decker and R.~Wattenhofer, ``Information propagation in the bitcoin network,'' in \emph{IEEE P2P 2013 Proceedings}.\hskip 1em plus 0.5em minus 0.4em\relax IEEE, 2013, pp. 1--10.

\bibitem{kan2018multiple}
L.~Kan, Y.~Wei, A.~H. Muhammad, W.~Siyuan, L.~C. Gao, and H.~Kai, ``A multiple blockchains architecture on inter-blockchain communication,'' in \emph{2018 IEEE international conference on software quality, reliability and security companion (QRS-C)}.\hskip 1em plus 0.5em minus 0.4em\relax IEEE, 2018, pp. 139--145.

\bibitem{chen2017inter}
Z.~Chen, Y.~Zhuo, Z.-B. Duan, and H.~Kai, ``Inter-blockchain communication,'' \emph{DEStech Transactions on Computer Science and Engineering http://dx. doi. org/10.12783/dtcse/cst2017/12539}, 2017.

\bibitem{vo2018internet}
H.~T. Vo, Z.~Wang, D.~Karunamoorthy, J.~Wagner, E.~Abebe, and M.~Mohania, ``Internet of blockchains: Techniques and challenges ahead,'' in \emph{2018 IEEE international conference on internet of things (iThings) and IEEE green computing and communications (GreenCom) and IEEE cyber, physical and social computing (CPSCom) and IEEE smart data (SmartData)}.\hskip 1em plus 0.5em minus 0.4em\relax IEEE, 2018, pp. 1574--1581.

\bibitem{zarrin2021blockchain}
J.~Zarrin, H.~Wen~Phang, L.~Babu~Saheer, and B.~Zarrin, ``Blockchain for decentralization of internet: prospects, trends, and challenges,'' \emph{Cluster Computing}, vol.~24, no.~4, pp. 2841--2866, 2021.

\bibitem{kwon2019cosmos}
J.~Kwon and E.~Buchman, ``Cosmos whitepaper,'' \emph{A Netw. Distrib. Ledgers}, vol.~27, pp. 1--32, 2019.

\bibitem{wood2016polkadot}
G.~Wood, ``Polkadot: Vision for a heterogeneous multi-chain framework,'' \emph{White paper}, vol.~21, no. 2327, p. 4662, 2016.

\bibitem{wanchain}
``Wanchain,'' \url{https://www.wanchain.org/}, accessed: March 11, 2024.

\bibitem{dinh2017blockbench}
T.~T.~A. Dinh, J.~Wang, G.~Chen, R.~Liu, B.~C. Ooi, and K.-L. Tan, ``Blockbench: A framework for analyzing private blockchains,'' in \emph{Proceedings of the 2017 ACM international conference on management of data}, 2017, pp. 1085--1100.

\bibitem{pahlajani2019survey}
S.~Pahlajani, A.~Kshirsagar, and V.~Pachghare, ``Survey on private blockchain consensus algorithms,'' in \emph{2019 1st International Conference on Innovations in Information and Communication Technology (ICIICT)}.\hskip 1em plus 0.5em minus 0.4em\relax IEEE, 2019, pp. 1--6.

\bibitem{li2017towards}
W.~Li, A.~Sforzin, S.~Fedorov, and G.~O. Karame, ``Towards scalable and private industrial blockchains,'' in \emph{Proceedings of the ACM workshop on blockchain, cryptocurrencies and contracts}, 2017, pp. 9--14.

\bibitem{gramoli2016danger}
V.~Gramoli, ``On the danger of private blockchains,'' in \emph{Workshop on Distributed Cryptocurrencies and Consensus Ledgers (DCCL’16)}, 2016, pp. 1--4.

\bibitem{cachin2016architecture}
C.~Cachin \emph{et~al.}, ``Architecture of the hyperledger blockchain fabric,'' in \emph{Workshop on distributed cryptocurrencies and consensus ledgers}, vol. 310, no.~4.\hskip 1em plus 0.5em minus 0.4em\relax Chicago, IL, 2016, pp. 1--4.

\bibitem{irresberger2021public}
F.~Irresberger, K.~John, P.~Mueller, and F.~Saleh, ``The public blockchain ecosystem: An empirical analysis,'' \emph{NYU Stern School of Business}, 2021.

\bibitem{ferdous2021survey}
M.~S. Ferdous, M.~J.~M. Chowdhury, and M.~A. Hoque, ``A survey of consensus algorithms in public blockchain systems for crypto-currencies,'' \emph{Journal of Network and Computer Applications}, vol. 182, p. 103035, 2021.

\bibitem{benhamouda2020can}
F.~Benhamouda, C.~Gentry, S.~Gorbunov, S.~Halevi, H.~Krawczyk, C.~Lin, T.~Rabin, and L.~Reyzin, ``Can a public blockchain keep a secret?'' in \emph{Theory of Cryptography: 18th International Conference, TCC 2020, Durham, NC, USA, November 16--19, 2020, Proceedings, Part I 18}.\hskip 1em plus 0.5em minus 0.4em\relax Springer, 2020, pp. 260--290.

\bibitem{rebello2024survey}
G.~A.~F. Rebello, G.~F. Camilo, L.~A.~C. de~Souza, M.~Potop-Butucaru, M.~D. de~Amorim, M.~E.~M. Campista, and L.~H.~M. Costa, ``A survey on blockchain scalability: From hardware to layer-two protocols,'' \emph{IEEE Communications Surveys \& Tutorials}, 2024.

\bibitem{dib2018consortium}
O.~Dib, K.-L. Brousmiche, A.~Durand, E.~Thea, and E.~B. Hamida, ``Consortium blockchains: Overview, applications and challenges,'' \emph{Int. J. Adv. Telecommun}, vol.~11, no.~1, pp. 51--64, 2018.

\bibitem{li2017consortium}
Z.~Li, J.~Kang, R.~Yu, D.~Ye, Q.~Deng, and Y.~Zhang, ``Consortium blockchain for secure energy trading in industrial internet of things,'' \emph{IEEE transactions on industrial informatics}, vol.~14, no.~8, pp. 3690--3700, 2017.

\bibitem{yao2021survey}
W.~Yao, J.~Ye, R.~Murimi, and G.~Wang, ``A survey on consortium blockchain consensus mechanisms,'' \emph{arXiv preprint arXiv:2102.12058}, 2021.

\bibitem{zhang2018towards}
A.~Zhang and X.~Lin, ``Towards secure and privacy-preserving data sharing in e-health systems via consortium blockchain,'' \emph{Journal of medical systems}, vol.~42, no.~8, p. 140, 2018.

\bibitem{qammar2023securing}
A.~Qammar, A.~Karim, H.~Ning, and J.~Ding, ``Securing federated learning with blockchain: a systematic literature review,'' \emph{Artificial Intelligence Review}, vol.~56, no.~5, pp. 3951--3985, 2023.

\bibitem{li2022blockchain}
D.~Li, D.~Han, T.-H. Weng, Z.~Zheng, H.~Li, H.~Liu, A.~Castiglione, and K.-C. Li, ``Blockchain for federated learning toward secure distributed machine learning systems: a systemic survey,'' \emph{Soft Computing}, vol.~26, no.~9, pp. 4423--4440, 2022.

\bibitem{qu2022blockchain}
Y.~Qu, M.~P. Uddin, C.~Gan, Y.~Xiang, L.~Gao, and J.~Yearwood, ``Blockchain-enabled federated learning: A survey,'' \emph{ACM Computing Surveys}, vol.~55, no.~4, pp. 1--35, 2022.

\bibitem{issa2023blockchain}
W.~Issa, N.~Moustafa, B.~Turnbull, N.~Sohrabi, and Z.~Tari, ``Blockchain-based federated learning for securing internet of things: A comprehensive survey,'' \emph{ACM Computing Surveys}, vol.~55, no.~9, pp. 1--43, 2023.

\bibitem{passerat2020blockchain}
J.~Passerat-Palmbach, T.~Farnan, M.~McCoy, J.~D. Harris, S.~T. Manion, H.~L. Flannery, and B.~Gleim, ``Blockchain-orchestrated machine learning for privacy preserving federated learning in electronic health data,'' in \emph{2020 IEEE international conference on blockchain (Blockchain)}.\hskip 1em plus 0.5em minus 0.4em\relax IEEE, 2020, pp. 550--555.

\bibitem{aich2022protecting}
S.~Aich, N.~K. Sinai, S.~Kumar, M.~Ali, Y.~R. Choi, M.-I. Joo, and H.-C. Kim, ``Protecting personal healthcare record using blockchain \& federated learning technologies,'' in \emph{2022 24th international conference on advanced communication technology (ICACT)}.\hskip 1em plus 0.5em minus 0.4em\relax Ieee, 2022, pp. 109--112.

\bibitem{el2020blockchain}
O.~El~Rifai, M.~Biotteau, X.~de~Boissezon, I.~Megdiche, F.~Ravat, and O.~Teste, ``Blockchain-based federated learning in medicine,'' in \emph{Artificial Intelligence in Medicine: 18th International Conference on Artificial Intelligence in Medicine, AIME 2020, Minneapolis, MN, USA, August 25--28, 2020, Proceedings 18}.\hskip 1em plus 0.5em minus 0.4em\relax Springer, 2020, pp. 214--224.

\bibitem{kim2019blockchain}
Y.~J. Kim and C.~S. Hong, ``Blockchain-based node-aware dynamic weighting methods for improving federated learning performance,'' in \emph{2019 20th Asia-pacific network operations and management symposium (APNOMS)}.\hskip 1em plus 0.5em minus 0.4em\relax IEEE, 2019, pp. 1--4.

\bibitem{lu2020communication}
Y.~Lu, X.~Huang, K.~Zhang, S.~Maharjan, and Y.~Zhang, ``Communication-efficient federated learning and permissioned blockchain for digital twin edge networks,'' \emph{IEEE Internet of Things Journal}, vol.~8, no.~4, pp. 2276--2288, 2020.

\bibitem{pandey2209fedtoken}
S.~Pandey, L.~Nguyen, and P.~Popovski, ``Fedtoken: Tokenized incentives for data contribution in federated learning. arxiv 2022,'' \emph{arXiv preprint arXiv:2209.09775}.

\bibitem{behera2021federated}
M.~R. Behera, S.~Upadhyay, and S.~Shetty, ``Federated learning using smart contracts on blockchains, based on reward driven approach,'' \emph{arXiv preprint arXiv:2107.10243}, 2021.

\bibitem{ma2021transparent}
S.~Ma, Y.~Cao, and L.~Xiong, ``Transparent contribution evaluation for secure federated learning on blockchain,'' in \emph{2021 IEEE 37th International Conference on Data Engineering Workshops (ICDEW)}.\hskip 1em plus 0.5em minus 0.4em\relax IEEE, 2021, pp. 88--91.

\bibitem{wang2021blockchain}
Z.~Wang and Q.~Hu, ``Blockchain-based federated learning: A comprehensive survey,'' \emph{arXiv preprint arXiv:2110.02182}, 2021.

\bibitem{liang2023architectural}
X.~Liang, J.~Zhao, Y.~Chen, E.~Bandara, and S.~Shetty, ``Architectural design of a blockchain-enabled, federated learning platform for algorithmic fairness in predictive health care: Design science study,'' \emph{Journal of medical Internet research}, vol.~25, p. e46547, 2023.

\bibitem{om2023securing}
C.~Om~Kumar, S.~Gajendran, V.~Balaji, A.~Nhaveen, and S.~Sai~Balakrishnan, ``Securing health care data through blockchain enabled collaborative machine learning,'' \emph{Soft Computing}, vol.~27, no.~14, pp. 9941--9954, 2023.

\bibitem{moulahi2023blockchain}
W.~Moulahi, I.~Jdey, T.~Moulahi, M.~Alawida, and A.~Alabdulatif, ``A blockchain-based federated learning mechanism for privacy preservation of healthcare iot data,'' \emph{Computers in Biology and Medicine}, vol. 167, p. 107630, 2023.

\bibitem{ali2023empowering}
A.~Ali, B.~A.~S. Al-Rimy, T.~T. Tin, S.~N. Altamimi, S.~N. Qasem, and F.~Saeed, ``Empowering precision medicine: Unlocking revolutionary insights through blockchain-enabled federated learning and electronic medical records,'' \emph{Sensors}, vol.~23, no.~17, p. 7476, 2023.

\bibitem{chang2021blockchain}
Y.~Chang, C.~Fang, and W.~Sun, ``A blockchain-based federated learning method for smart healthcare,'' \emph{Computational Intelligence and Neuroscience}, vol. 2021, 2021.

\bibitem{lian2023blockchain}
Z.~Lian, W.~Wang, Z.~Han, and C.~Su, ``Blockchain-based personalized federated learning for internet of medical things,'' \emph{IEEE Transactions on Sustainable Computing}, 2023.

\bibitem{farooq2022blockchain}
K.~Farooq, H.~J. Syed, S.~O. Alqahtani, W.~Nagmeldin, A.~O. Ibrahim, and A.~Gani, ``Blockchain federated learning for in-home health monitoring,'' \emph{Electronics}, vol.~12, no.~1, p. 136, 2022.

\bibitem{zhang2021blockchain}
H.~Zhang, G.~Li, Y.~Zhang, K.~Gai, and M.~Qiu, ``Blockchain-based privacy-preserving medical data sharing scheme using federated learning,'' in \emph{Knowledge Science, Engineering and Management: 14th International Conference, KSEM 2021, Tokyo, Japan, August 14--16, 2021, Proceedings, Part III 14}.\hskip 1em plus 0.5em minus 0.4em\relax Springer, 2021, pp. 634--646.

\bibitem{passerat2019blockchain}
J.~Passerat-Palmbach, T.~Farnan, R.~Miller, M.~S. Gross, H.~L. Flannery, and B.~Gleim, ``A blockchain-orchestrated federated learning architecture for healthcare consortia,'' \emph{arXiv preprint arXiv:1910.12603}, 2019.

\bibitem{rahman2020secure}
M.~A. Rahman, M.~S. Hossain, M.~S. Islam, N.~A. Alrajeh, and G.~Muhammad, ``Secure and provenance enhanced internet of health things framework: A blockchain managed federated learning approach,'' \emph{Ieee Access}, vol.~8, pp. 205\,071--205\,087, 2020.

\bibitem{singh2022framework}
S.~Singh, S.~Rathore, O.~Alfarraj, A.~Tolba, and B.~Yoon, ``A framework for privacy-preservation of iot healthcare data using federated learning and blockchain technology,'' \emph{Future Generation Computer Systems}, vol. 129, pp. 380--388, 2022.

\bibitem{liu2022blockchain}
Y.~Liu, W.~Yu, Z.~Ai, G.~Xu, L.~Zhao, and Z.~Tian, ``A blockchain-empowered federated learning in healthcare-based cyber physical systems,'' \emph{IEEE Transactions on Network Science and Engineering}, 2022.

\bibitem{otoum2021preventing}
S.~Otoum, I.~Al~Ridhawi, and H.~T. Mouftah, ``Preventing and controlling epidemics through blockchain-assisted ai-enabled networks,'' \emph{Ieee Network}, vol.~35, no.~3, pp. 34--41, 2021.

\bibitem{kumar2021blockchain}
R.~Kumar, A.~A. Khan, J.~Kumar, N.~A. Golilarz, S.~Zhang, Y.~Ting, C.~Zheng, W.~Wang \emph{et~al.}, ``Blockchain-federated-learning and deep learning models for covid-19 detection using ct imaging,'' \emph{IEEE Sensors Journal}, vol.~21, no.~14, pp. 16\,301--16\,314, 2021.

\bibitem{durga2021federated}
R.~Durga and E.~Poovammal, ``Federated learning model for healthchain system,'' in \emph{2021 6th IEEE International Conference on Recent Advances and Innovations in Engineering (ICRAIE)}, vol.~6.\hskip 1em plus 0.5em minus 0.4em\relax IEEE, 2021, pp. 1--6.

\bibitem{lakhan2022federated}
A.~Lakhan, M.~A. Mohammed, J.~Nedoma, R.~Martinek, P.~Tiwari, A.~Vidyarthi, A.~Alkhayyat, and W.~Wang, ``Federated-learning based privacy preservation and fraud-enabled blockchain iomt system for healthcare,'' \emph{IEEE journal of biomedical and health informatics}, vol.~27, no.~2, pp. 664--672, 2022.

\bibitem{samuel2022iomt}
O.~Samuel, A.~B. Omojo, A.~M. Onuja, Y.~Sunday, P.~Tiwari, D.~Gupta, G.~Hafeez, A.~S. Yahaya, O.~J. Fatoba, and S.~Shamshirband, ``Iomt: A covid-19 healthcare system driven by federated learning and blockchain,'' \emph{IEEE Journal of Biomedical and Health Informatics}, vol.~27, no.~2, pp. 823--834, 2022.

\bibitem{yang2024federated}
X.~Yang, C.~Xing \emph{et~al.}, ``Federated medical learning framework based on blockchain and homomorphic encryption,'' \emph{Wireless Communications and Mobile Computing}, vol. 2024, 2024.

\bibitem{lo2022toward}
S.~K. Lo, Y.~Liu, Q.~Lu, C.~Wang, X.~Xu, H.-Y. Paik, and L.~Zhu, ``Toward trustworthy ai: Blockchain-based architecture design for accountability and fairness of federated learning systems,'' \emph{IEEE Internet of Things Journal}, vol.~10, no.~4, pp. 3276--3284, 2022.

\bibitem{nguyen2021blockchain}
D.~C. Nguyen, M.~Ding, P.~N. Pathirana, and A.~Seneviratne, ``Blockchain and ai-based solutions to combat coronavirus (covid-19)-like epidemics: A survey,'' \emph{Ieee Access}, vol.~9, pp. 95\,730--95\,753, 2021.

\end{thebibliography}


\end{document}